\documentclass[12pt]{article}
\pdfoutput=1
\usepackage{epsf,amsfonts,amssymb,epsfig,amsmath,graphics,cite}
\usepackage[dvipsnames]{xcolor}
\usepackage{graphicx,subfig}
\usepackage{booktabs}
\usepackage{cancel}
\usepackage{float}
\usepackage{verbatim}
\usepackage{mathrsfs}
\usepackage{empheq}
\usepackage{enumerate}
\usepackage{bbm}
\usepackage{soul}
\usepackage{slashed}
\usepackage{tikz}
\usetikzlibrary{arrows,decorations.markings,positioning}
\usepackage{tikz-3dplot}
\usetikzlibrary{scopes,fadings,shadings}

\tikzset{
    canvas/.style={draw,left color=blue!35!black,right color=white},
    sign/.style={align=center,fill=white,fill opacity=0.2,text opacity=1,text=white},
    axis label/.style={midway,below,sloped}
}

\usepackage[colorlinks=true
,urlcolor=blue
,anchorcolor=blue
,citecolor=blue
,filecolor=blue
,linkcolor=blue
,menucolor=blue
,linktocpage=true
]{hyperref}
 
\makeatletter

\newcommand{\Rmnum}[1]{\expandafter\@slowromancap\romannumeral #1@}
\makeatother

\newcommand{\nn}{\notag }
\def\be{\begin{equation}}
\def\ee{\end{equation}}

\newcommand{\ii}{\mathrm{i}}
\newcommand{\ex}{\mathrm{e}}
\newcommand{\me}{\mathrm{e}}
\newcommand{\diff}{\mathrm{d}}
\newcommand{\dd}{\mathrm{d}}
\newcommand{\R}{\mathbb{R}}
\newcommand{\Z}{\mathbb{Z}}
\newcommand{\vol}{\mathrm{vol}}

\newcommand{\C}{\mathbb{C}}

\newcommand{\mt}{\mathrm{T}}

\newcommand{\hook}{\mathbin{\rule[.2ex]{.4em}{.03em}\rule[.2ex]{.03em}{.9ex}}}

\newcommand{\del}{\partial}

\newcommand{\Phib}{\Phi^{I_{\mathrm{bulk}}}}

\newsavebox{\ns}
\newsavebox{\dbrane}
\newsavebox{\dbshort}

\def\be{\begin{equation}}
	\def\ee{\end{equation}}
\def\bea{\begin{eqnarray}}
	\def\eea{\end{eqnarray}}

\newlength{\sswidth}

\newcommand{\xinew}{\zeta}

\newcommand{\vv}{\mathrm{v}}

\newcommand{\epsft}{\tilde{\varepsilon}}

\newcommand{\Bc}{\mathfrak{b}}

\newcommand{\sigmai}{\sigma^{(i)}}
\newcommand{\sigmaone}{\sigma^{(1)}}
\newcommand{\sigmatwo}{\sigma^{(2)}}
\newcommand{\sigmathree}{\sigma^{(3)}}

\newcommand{\KSone}{K_{\Sigma_{g_1}}}
\newcommand{\KStwo}{K_{\Sigma_{g_2}}}

\numberwithin{equation}{section}       

\begin{document}

\begin{titlepage}

\vskip 1cm

\begin{center}

{\Large \bf Localizing Romans supergravity}

\vskip 1cm
{Christopher Couzens$^{\mathrm{a}}$,  Carolina Matte Gregory$^{\mathrm{b,c}}$, Davide Muniz$^{\mathrm{b}}$, 
 Tabea Sieper$^{\mathrm{a}}$  and James Sparks$^{\mathrm{a}}$}

\vskip 1cm

${}^{\mathrm{a}}$\textit{Mathematical Institute, University of Oxford,\\
Andrew Wiles Building, Radcliffe Observatory Quarter,\\
Woodstock Road, Oxford, OX2 6GG, U.K.\\}

\vskip 0.2cm

${}^{\mathrm{b}}$\textit{Universidade de Bras\'ilia, Instituto de F\'isica, 
70910-900, Bras\'ilia, DF, Brasil }\\

\vskip 0.2cm

${}^{\mathrm{c}}$\textit{International Center of Physics, 
C.P. 04667, Bras\'ilia, DF, Brasil }\\

\vskip 0.2 cm

\end{center}

\vskip 0.5 cm

\begin{abstract}
\noindent  
We derive a general formula for the on-shell action of six-dimensional Euclidean Romans supergravity using equivariant localization. Our results are obtained without the need for solving any of the equations of motion, instead working on the assumption of the existence of a supersymmetric solution. We show that the on-shell action is completely determined in terms of the R-symmetry Killing vector and topological data. We easily recover known results in the literature, make predictions for hitherto unknown solutions, and also match to holographic field theory duals. 

\end{abstract}

\end{titlepage}

\pagestyle{plain}
\setcounter{page}{1}
\newcounter{bean}
\baselineskip18pt

\tableofcontents

\newpage

\section{Introduction}\label{sec:intro}

Supersymmetric solutions to supergravity theories continue to play a central role in the AdS/CFT correspondence. The supergravity 
limit is typically a strong coupling limit in the dual superconformal 
field theory (SCFT), and over the last two decades a number 
of approaches have been developed for computing various protected physical observables exactly in field theory. The results may then be compared to supergravity. However, a key problem is that while 
solving first order Killing spinor  equations in supergravity is conceptually straightforward, 
in practice these are coupled non-linear PDEs, and finding 
solutions in closed form usually relies on a high degree of symmetry. 
Ultimately, one is in any case not interested in the solution 
{\it per se}, but rather the corresponding protected physical observables 
for that solution, given that it exists. 

Starting with \cite{BenettiGenolini:2023kxp}, it has recently been 
appreciated that supersymmetric solutions to supergravity theories 
are canonically equipped with an ``R-symmetry Killing vector'' 
together with a set of equivariantly closed forms. 
This Killing vector is constructed as a 
bilinear in the Killing
spinor of the solution, while the equivariantly closed forms are constructed as polynomials in 
Killing spinor bilinears and supergravity fields. 
Remarkably, various physical 
observables of interest arise as integrals of 
these forms, and these 
may in turn be evaluated 
using the fixed point formula of \cite{BV:1982, Atiyah:1984px}, crucially  
without solving any PDEs, but 
instead assuming a corresponding solution exists.
While the precise details depend on
which particular supergravity theory one is studying, 
in 
\cite{BenettiGenolini:2023kxp}, together with many subsequent works \cite{BenettiGenolini:2023yfe, BenettiGenolini:2023ndb, BenettiGenolini:2024kyy, BenettiGenolini:2024xeo, Cassani:2024kjn, BenettiGenolini:2024hyd, BenettiGenolini:2024lbj, Couzens:2024vbn, Suh:2024asy, Colombo:2025ihp}, this method has been 
used to not only reproduce a swathe of results in the 
supergravity literature with just a few lines of calculation, 
but also to significantly generalize them.

In the present paper we continue this line of research, focusing on six-dimensional Romans $F(4)$  gauged supergravity \cite{Romans:1985tw}. 
This is a consistent 
truncation of massive type IIA supergravity on $S^4$ \cite{Cvetic:1999un}, meaning that any solution to the Romans theory can be uplifted to a solution of type IIA string theory. 
An asymptotically locally AdS Romans solution on a manifold $M_6$, 
with conformal boundary $M_5=\partial M_6$, is conjectured to be dual to a particular five-dimensional superconformal gauge theory on $M_5$, with gauge group
USp$(2N)$ and arising from a D4-D8-brane system \cite{Ferrara:1998gv, Brandhuber:1999np, Bergman:2012kr}. 
This SCFT has $N_f<8$ matter fields in the fundamental  representation of the gauge group, and a single hypermultiplet in the anti-symmetric representation.
The relation between the Newton 
constant $G_N$ of the Romans theory 
and the gauge theory parameters is
\begin{equation}\label{parameters}
 -\frac{27\pi^2}{4G_N} = F_{S^5}=-\frac{9 \sqrt{2}\pi}{5}\frac{N^{5/2}}{\sqrt{8-N_f}} \, ,    
\end{equation}
where $F_{S^5}$ 
is the holographically renormalized on-shell action of Euclidean AdS$_6$, and \eqref{parameters} holds in the $N\rightarrow \infty$ limit.

Euclidean signature supersymmetric 
solutions to the $D=6$ Romans theory were 
analyzed in \cite{Alday:2015jsa}. 
Using these results we construct a number of equivariantly closed forms -- essentially one for each field strength and its Hodge dual, together with an equivariantly closed form for the on-shell action. 
Our main result is  the following general expression for the holographically renormalized on-shell action $I$, generalizing the result for isolated fixed points presented in \cite{BenettiGenolini:2023kxp}:
\begin{align}\label{main}
     I = \ &   \bigg\{\sum_{\text{dim  }  0} \sigmaone\sigmatwo\sigmathree\frac{(\sigmaone\epsilon_1+\sigmatwo\epsilon_2+\sigmathree\epsilon_3)^3}{\epsilon_1\epsilon_2\epsilon_3} \\
+&\sum_{\text{dim  }  2} \sigmaone\sigmatwo\sigmathree\frac{
(\sigmaone\epsilon_1+\sigmatwo\epsilon_2)^2}{\epsilon_1\epsilon_2} \bigg[3\sigmathree \chi(\Sigma_g)\nonumber\\ & \qquad +
 \int_{\Sigma_g}2\big(\sigmaone c_1(L_1)+\sigmatwo c_1(L_2)\big)-\sigmaone \frac{\epsilon_1}{\epsilon_2}c_1(L_2)
 -\sigmatwo\frac{\epsilon_2}{\epsilon_1}c_1(L_1)\bigg]\nonumber\\
 +& \sum_{\text{dim  }  4} \bigg[  6 \chi(B_4)+9 \tau (B_4)+\int_{B_4}c_1(L_1)\wedge\big( c_1(L_1)\mp 3 \sigmaone c_1(K_{B_4})\big)\bigg]\bigg\} \frac{F_{S^5}}{27}\, .\nonumber
 \end{align}
Although this looks a little unwieldy at first sight, this is largely because it is a very general formula. Here $I$ is 
the holographically 
renormalized on-shell action for a solution with topology $M_6$. 
This six-manifold is equipped with an R-symmetry Killing vector $\xi$, and the sums in \eqref{main} are over the various connected components of the 
fixed point set, where $\xi=0$. These may either be zero-dimensional (isolated fixed points), two-dimensional (fixed 
Riemann surfaces $\Sigma_g$), 
or four-dimensional (fixed four-manifolds $B_4$). The notation 
in \eqref{main} is then as follows:
\begin{itemize}
\item zero-dimensional fixed points: the tangent space at the fixed point splits as $\R^6=\R^2_1\oplus \R^2_2\oplus \R^2_3$, and we correspondingly write 
$\xi=\sum_{i=1}^3 \epsilon_i\partial_{\varphi_i}$, where $\varphi_i$ is an angular (polar) coordinate on the $i^{\mathrm{th}}$ plane. 
The $\epsilon_i\in\R$ are 
referred to as the \emph{weights} of $\xi$ 
at the fixed point.
The 
$\sigmai\in\{\pm1\}$ are 
certain signs associated to the 
Killing spinor (more precisely, $\sigmai$ is twice the charge of the Killing spinor under $\partial_{\varphi_i}$).
\item two-dimensional 
fixed points: here a fixed 
Riemann surface $\Sigma_g$, 
with Euler number $\chi(\Sigma_g)=2-2g$, has 
normal bundle $\mathcal{N}=L_1\oplus L_2$. Here 
the fibre of the complex line bundle $L_i$ is $\mathbb{C}_i\cong \R^2_i$, and we again 
write $\xi=\sum_{i=1}^2\epsilon_i\partial_{\varphi_i}$, 
with the weights $\epsilon_i$ parametrizing the action of $\xi$ on the normal directions to $\Sigma_g$ in $M_6$. The $\int_{\Sigma_g}c_1(L_i)\in\Z$ 
denote first Chern numbers, and $\sigmathree\in\{\pm1\}$ is a choice of sign (associated with the spinor chirality).
\item four-dimensional 
fixed points: here the fixed 
four-manifold $B_4$ has Euler number
and signature $\chi(B_4),\tau(B_4)\in\Z$, respectively. 
In this case the normal 
bundle is a complex line bundle $\mathcal{N}=L_1$, and supersymmetry equips $B_4$ with 
a U$(2)$ structure with first Chern 
class $c_1(K_{B_4})\in H^2(B_4,\Z)$. 
\end{itemize}

We illustrate the general formula \eqref{main} by applying it to a variety of examples. These include various black hole-type solutions, and other explicitly known supergravity solutions, but 
the formalism also allows us to write down results for 
solutions that are
unlikely to ever be found in closed form. 
To highlight two particular (not unrelated) examples: 
(i) we straightforwardly derive the conjectured
 entropy functions for  AdS$_2 \times M_4$ toric orbifold solutions of \cite{Faedo:2022rqx}, (ii)
 we can conjecture the existence of general families of solutions with topology $M_6=\R^2\times B_4$, which may be thought of as Euclidean black holes with ``horizon'' a four-manifold $B_4$. The on-shell action for the latter should be compared to the free energy (minus the logarithm of the partition function) of the USp$(2N)$ gauge theories on $S^1\times B_4$ in the large $N$ limit, and \eqref{main} immediately predicts the simple result
\begin{align}\label{B4intro}
F_{S^1\times B_4} = [2\chi(B_4)+3 \tau(B_4)]\frac{F_{S^5}}{9}\,.
\end{align}
It is clearly of interest to compute this directly 
in field theory. We note that in the special case that $B_4$ is a complex surface we have $2\chi(B_4)+3\tau(B_4)=\int_{B_4}c_1^2(K_{B_4})$, and \eqref{B4intro} agrees with minus the Bekenstein--Hawking entropy of the black holes constructed in \cite{Hosseini:2018usu}, where 
$B_4$ is K\"ahler--Einstein. 

The plan of the rest of the paper is as follows. 
In section \ref{sec:Romans} we introduce the Euclidean Romans  theory, summarizing the classification of supersymmetric solutions obtained in \cite{Alday:2015jsa}. Section \ref{sec:localizeRomans} identifies a set of equivariantly closed forms, and, after a careful analysis of the local fixed point contributions, derives the main formula \eqref{main} using the Berline--Vergne--Atiyah--Bott fixed point theorem \cite{BV:1982, Atiyah:1984px}. In sections~\ref{sec:examplesI}, \ref{sec:examplesII} and~\ref{sec:examplesIII} we study a variety of examples, both where there is a known supergravity solution, but also where there is not, and compare to dual field theory results where these are available. We briefly conclude in section~\ref{sec:discussion}. A number of technical details have been included in five appendices.

\section{Romans supergravity}\label{sec:Romans}

In this section we begin by
summarizing the (Euclidean) Romans $F(4)$ gauged supergravity theory and 
its uplift to massive IIA. 
This will fix the conventions that we will employ in the remainder of the paper, before then discussing the torsion conditions for the existence of a supersymmetric solution, as first computed in \cite{Alday:2015jsa}. 

\subsection{Uplift to massive type IIA}

A particular class of five-dimensional superconformal gauge theories,
with gauge group USp$(2N)$ and arising from a D4-D8-brane
system, is expected to have a large $N$ description in terms
of the AdS$_6\times S^4$ solution of massive type IIA supergravity \cite{Ferrara:1998gv, Brandhuber:1999np, Bergman:2012kr}.
To find gravity duals for these superconformal theories on different background five-manifolds, it is natural to work within the six-dimensional Romans $F(4)$ gauged supergravity theory \cite{Romans:1985tw}. The key point, as demonstrated in \cite{Cvetic:1999un}, is that Romans theory is a consistent truncation of massive type IIA supergravity on $S^4$. Here, we briefly review its uplift to ten dimensions before presenting the theory in Euclidean signature in the following subsection.

The Romans theory \cite{Romans:1985tw} is a six-dimensional gauged supergravity that admits a unique supersymmetric AdS$_6$ vacuum. Its bosonic field content 
comprises the metric, a scalar field $X=\exp(-\tfrac{\phi}{2\sqrt{2}})$, where $\phi$ is the dilaton, 
a one-form potential $\mathcal{A}$, 
a two-form potential $B$, and a $\mathtt{so}(3)_R= \mathtt{su}(2)_R$ R-symmetry gauge field $A^i$ with field strength $F^i=\diff A^i-\frac{1}{2}g\varepsilon_{ijk}A^j\wedge A^k$, where $i=1,2,3$.  Notably, $B$ appears in the field strength of $\mathcal{A}$ via 
 $\mathcal{F}=\diff\mathcal{A}+\frac{2}{3}gB$, 
 while the three-form field strength is $H= \diff B$. It follows that one can work in a gauge in which the Stueckelberg one-form $\mathcal{A}$ has been set to zero, rendering the $B$-field massive, and we employ this gauge throughout. We will later fix the gauge coupling constant $g$ to unity, but, in this subsection, it will remain explicit.

Since the Romans theory is a consistent truncation of massive type IIA supergravity on $S^4$, any solution automatically uplifts, via the non-linear Kaluza-Klein ansatz of \cite{Cvetic:1999un}, to a solution of massive type IIA supergravity. Moreover, the AdS$_6\times S^4$ solution of the latter is precisely the uplift of the AdS$_6$ vacuum of the Romans theory.
The gauge coupling constant $g$ is related to the ten-dimensional mass parameter by $m_{\mathrm{IIA}}=\frac{\sqrt{2}}{3}g$, while the remaining fields uplift as follows
\bea
\diff s^2_{10}&=& (\sin\xinew)^{\frac{1}{12}}X^{\frac{1}{8}}\Big[\Delta^{\frac{3}{8}}\diff s^2_6 + 
2g^{-2}\Delta^{\frac{3}{8}}X^2\diff\xinew^2 \nn\\ 
& & \qquad \qquad \qquad +\tfrac{1}{2}g^{-2}\Delta^{-\frac{5}{8}}X^{-1}\cos^2\xinew\sum_{i=1}^3
(\hat\tau^i-g A^i)^2\Big]\, ,\nn\\
F_{(4)} &=&   -\tfrac{\sqrt2}{6} g^{-3} s^{1/3} c^3 \Delta^{-2}
U \, \dd\xinew\wedge\vol_3 -\sqrt2 g^{-3} s^{4/3} c^4 \Delta^{-2}
X^{-3} \, \dd X\wedge \vol_3 \nn\\
&&+\sqrt2 g^{-1} 
s^{1/3} c X^4\, {*H}\wedge \dd\xinew
-\tfrac{1}{\sqrt2} s^{4/3} X^{-2} {*\mathcal{F}}+\tfrac{1}{\sqrt2} g^{-2}
s^{1/3} c\, F^i  h^i\wedge \dd\xinew  \nn\\
&&  -\tfrac{1}{4\sqrt2} g^{-2}
s^{4/3} c^2 \Delta^{-1} X^{-3}  F^i \wedge
h^j\wedge h^k\, \varepsilon_{ijk}\, ,\nn\\
F_{(3)} &=& s^{2/3} H + g^{-1} s^{-1/3} c \, \mathcal{F}\wedge \dd\xinew~,\nn\\
F_{(2)} &=& \tfrac{1}{\sqrt2} s^{2/3} \mathcal{F}\, , \qquad
\ex^{\Phi} \ = \  s^{-5/6} \Delta^{1/4} X^{-5/4}\, ,
\label{uplift}
\eea
where
\be
\begin{split}
\Delta &\equiv  X c^2 +X^{-3} s^2\, ,\\
U &\equiv  X^{-6} s^2 - 3 X^2 c^2 + 4 X^{-2} c^2 - 6 X^{-2}\, .
\end{split}
\ee
Here, $\diff s^2_{10}$ is the ten-dimensional metric in the Einstein frame, $\Phi$ is the ten-dimensional dilaton, $F_{(3)}$ is the NS--NS three-form field strength, and $F_{(4)}$ and $F_{(2)}$ are the RR four-form and two-form field strengths, respectively. The forms $\hat\tau^i$ ($i=1,2,3$) are left-invariant one-forms on SU$(2)\cong S^3$, defined as in \cite{Cvetic:1999un}. Additionally, we define $h^i\equiv \hat\tau^i - g A^i$ and $\vol_3\equiv h^1\wedge h^2\wedge h^3$, while $s\equiv\sin\xinew$ and $c\equiv\cos\xinew$. The Hodge duals in the uplift equations are taken with respect to the six-dimensional metric $\diff s^2_6$.

The ten-dimensional metric in (\ref{uplift}) describes a warped product $M_6\times S^4$. More precisely, the solution only covers one half of a four-sphere, where the coordinate $\xinew \in (0,\frac{\pi}{2}]$ acts as a polar coordinate. For constant $\xinew \in (0,\frac{\pi}{2})$, slices are three-spheres parametrized by Euler angles on $S^3$. The solution remains smooth at the north pole ($\xinew=\frac{\pi}{2}$), where the $S^3$ slices collapse to zero size, but it exhibits a singularity at the equator ($\xinew=0$) due to the presence of a D8/O8 stack. Nevertheless, as argued in \cite{Brandhuber:1999np, Bergman:2012kr}, the supergravity solution can be trusted away from this singularity.

Finally, we emphasize that our focus is on a specific class of theories with gauge group $G=\mathrm{USp}(2N)$, which arise from a system of $N$ D4-branes and $0<N_f<8$ D8-branes on top of an O8 orientifold plane in massive type IIA string theory. These are sometimes referred to as \emph{Seiberg theories} \cite{Seiberg:1996bd}. The $S^5$ free energy of these solutions is given by equation \eqref{parameters}, 
where $G_N$ is the Newton constant of Romans supergravity. 
These theories are expected to have a well-defined large $N$ limit, which, in turn, has a dual description in massive type IIA supergravity \cite{Ferrara:1998gv, Brandhuber:1999np, Bergman:2012kr}.


\subsection{Euclidean Romans supergravity}\label{sec:SUGRA}

  The Euclidean signature equations of motion of Euclidean Romans supergravity are (now setting $g\equiv1$ without loss of generality) given by \cite{Alday:2014rxa, Alday:2014bta}
  \begin{equation}\begin{split}\label{FullEOM}
  \diff\left(X^{-1}*\dd X\right) = & -  \left(\tfrac{1}{6}X^{-6}-\tfrac{2}{3}X^{-2}+\tfrac{1}{2}X^2\right)*1  \\
&-\tfrac{1}{8}X^{-2}\left(\tfrac{4}{9}B\wedge *B+F^i\wedge *F^i\right) + \tfrac{1}{4}X^4H\wedge *H \,,\\
\diff\left(X^4 * H\right) = & \ \tfrac{2\, \ii}{9}B\wedge B + \tfrac{\ii}{2}F^i\wedge F^i +\tfrac{4}{9}  X^{-2}*B\,,\\
D(X^{-2}*F^i)  = & - \ii F^i\wedge H\, .
\end{split}\end{equation}
Here 
$*$ denotes the Hodge duality operator 
on differential forms, and $D$ is the $\mathtt{so}(3)$ covariant derivative defined to be $D\omega^i\equiv\dd\omega^i - \varepsilon_{ijk}A^j\wedge \omega^k$. Notice that the theory 
contains Chern--Simons-type couplings that become purely imaginary in Euclidean signature.
The Einstein equation is
\begin{equation}\begin{split}\label{Einstein}
R_{\mu\nu} = & \ 4X^{-2}\partial_\mu X\partial_\nu X + \left(\tfrac{1}{18}X^{-6}-\tfrac{2}{3}X^{-2}-\tfrac{1}{2}X^2\right) g_{\mu\nu} + \tfrac{1}{4}X^4\left(H^2_{\mu\nu}-\tfrac{1}{6}H^2g_{\mu\nu}\right) \\
& +  \tfrac{2}{9}X^{-2}\left(B^2_{\mu\nu}-\tfrac{1}{8}B^2g_{\mu\nu}\right) +  \tfrac{1}{2}X^{-2}\left((F^i)^2_{\mu\nu}-\tfrac{1}{8}(F^i)^2g_{\mu\nu}\right)\, ,
\end{split}
\end{equation}
where $B^2_{\mu\nu} \equiv B_{\mu\rho} B_\nu{}^\rho$, $H^2_{\mu\nu}\equiv H_{\mu\rho\sigma}H_{\nu}^{\ \rho\sigma}$. The  bulk 
action that leads to these equations of motion is
\begin{equation}\begin{split}
I_{\mathrm{bulk}}  =   -\frac{1}{16\pi G_N}\int_{M_6}& \Big[R*1-4X^{-2}\diff X\wedge *\diff X  -\left(\tfrac{2}{9}X^{-6}-\tfrac{8}{3}X^{-2}-2X^2\right)*1\\
&   -\tfrac{1}{2}X^{-2} \left(\tfrac{4}{9}B\wedge *B+ F^i \wedge * F^i\right) -\tfrac{1}{2}X^4H\wedge *H \\
&  - \ii B \wedge \big( \tfrac{2}{27}  B\wedge B + \tfrac{1}{2} F^i \wedge F^i \big)\Big]\, .\label{Fullaction}
\end{split}
\end{equation}

A  solution is supersymmetric provided there 
exists a non-trivial SU$(2)_R$ doublet of Dirac spinors $\epsilon_I$, $I=1,2$, satisfying 
the following Killing spinor  and dilatino equations, respectively
\begin{align}
D_\mu \epsilon_I  =  & \ \tfrac{\ii}{4\sqrt{2}}  ( X + \tfrac{1}{3} X^{-3} ) \gamma_\mu \gamma_7 \epsilon_I - \tfrac{\ii}{24\sqrt{2}} X^{-1} B_{\nu\rho} ( \gamma_\mu{}^{\nu\rho} - 6 \delta_\mu{}^\nu \gamma^\rho ) \epsilon_I \label{KSE} \\
& - \tfrac{1}{48} X^2 H_{\nu\rho\sigma} \gamma^{\nu\rho\sigma} \gamma_\mu \gamma_7 \epsilon_I
+ \tfrac{1}{16\sqrt{2}}X^{-1} F_{\nu\rho}^i ( \gamma_\mu{}^{\nu\rho} - 6 \delta_\mu{}^\nu \gamma^\rho ) \gamma_7 ( \sigma_i )_I{}^J \epsilon_J \,,\nn\\
0  = &  - \ii X^{-1} \partial_\mu X \gamma^\mu \epsilon_ I + \tfrac{1}{2\sqrt{2}}  \left( X - X^{-3} \right) \gamma_7 \epsilon_I + \tfrac{\ii}{24} X^2 H_{\mu\nu\rho} \gamma^{\mu\nu\rho} \gamma_7 \epsilon_I \label{dilatino} \\
&- \tfrac{1}{12\sqrt{2}} X^{-1} B_{\mu\nu} \gamma^{\mu\nu} \epsilon_I - \tfrac{\ii}{8\sqrt{2}} X^{-1} F^i_{\mu\nu} \gamma^{\mu\nu} \gamma_7 ( \sigma_i )_I{}^J \epsilon_J\, .\nonumber
\end{align}
Here the covariant derivative acting on the spinor is $D_\mu\epsilon_I\equiv {\nabla}_\mu\epsilon_I+\frac{\ii}{2} A^i_\mu(\sigma_i)_I{}^J\epsilon_J$, where ${\nabla}_\mu=\partial_\mu+\frac{1}{4}\omega_{\mu}^{\ \, \nu\rho}\gamma_{\nu\rho}$ denotes the usual Levi--Civita 
spin connection, while $\sigma_i$, for $i=1,2,3$, are the Pauli matrices that convert between 
$3\times 3$ matrices of $\mathtt{so}(3)_R$ and $2\times 2$ matrices of $\mathtt{su}(2)_R$. More precisely, 
(\ref{KSE}),  (\ref{dilatino}) are the Lorentzian signature equations, 
and as usual we need to be careful when defining fermions and supersymmetry in 
Euclidean signature. We clarify the conditions we shall impose momentarily, 
but note that in Euclidean signature $\gamma_\mu$, $\mu=1,\ldots,6$ are taken to be Hermitian and generate the Clifford 
algebra $\mathrm{Cliff}(6,0)$ in an orthonormal frame, and the chirality 
operator is 
 $\gamma_7 \equiv \ii \gamma^{123456}$, which satisfies $\gamma_7^2=1$. 

As in \cite{Alday:2015jsa}, for simplicity, we shall consider an Abelian truncation of this supergravity theory in which 
we set $A^1_\mu\equiv A^2_\mu\equiv 0$, and $A^3_\mu\equiv {A}_\mu$, with field strength ${F}\equiv \diff {A}$. 
Further, as in \cite{Alday:2015jsa}, we consider a ``real'' Euclidean class of solutions
 for which $\epsilon_I$ satisfies the symplectic Majorana condition 
 $\varepsilon_{I}^{\ J}\epsilon_J = \mathcal{C}\epsilon_I^*\equiv \epsilon_I^c$, where 
$\varepsilon_{IJ}$ is the two-dimensional Levi--Civita symbol and
 $\mathcal{C}$ denotes the charge conjugation matrix, satisfying $\gamma_\mu^\mt=\mathcal{C}^{-1}\gamma_\mu \mathcal{C}$.
This is consistent with taking all the bosonic fields to be real, 
with the exception of the $B$-field which is purely imaginary. 
 With these reality properties, one can show that the Killing spinor equation (\ref{KSE}) and dilatino equation 
 (\ref{dilatino}) for $\epsilon_2$  are simply the charge conjugates of the corresponding 
 equations for $\epsilon_1$. 
The conditions for Euclidean supersymmetry therefore reduce to the existence of a single Killing spinor $\epsilon\equiv \epsilon_1$, with SU$(2)_R$ doublet $(\epsilon_1,\epsilon_2)=(\epsilon,\epsilon^c)$.  

As discussed in Appendix \ref{app:chargeconjugate}, the charge conjugate spinor $\epsilon^c$ satisfies the same Killing spinor equation and dilatino equation as $\epsilon$, after replacing $A$ by $-A$. There is thus a charge conjugation symmetry which exchanges $(\epsilon,A)\leftrightarrow(\epsilon^c,A^c)$, where $A^c\equiv -A$. 


\subsection{Supersymmetry and bilinear forms}\label{sec:SUSY}

Following \cite{Alday:2015jsa}, given a Dirac spinor $\epsilon$  such that
$(\epsilon_1,\epsilon_2)=(\epsilon,\epsilon^c)$ and solves \eqref{KSE}, \eqref{dilatino}, we may construct the following real bilinear forms\footnote{The 
one-form $\xi^\flat$ was denoted by $K$ in \cite{Alday:2015jsa}, while 
$P$ was denoted $\tilde{S}$. Note that one could also have used the $\epsilon^c$ spinor in the following. As discussed in Appendix~\ref{app:chargeconjugate}, this leads to some overall minus sign differences in the identifications since the bilinear forms we will construct momentarily satisfy $\bar{\epsilon}^c \gamma_{(n)}\epsilon^c=(-1)^{\tfrac{n(n-1)}{2}}\bar{\epsilon}\gamma_{(n)}\epsilon$ and $\bar{\epsilon}^c \gamma_7\gamma_{(n)}\epsilon^c=(-1)^{\tfrac{n(n+1)}{2}+1}\bar{\epsilon}\gamma_{(n)}\epsilon$.}
\begin{align}\label{bilineardefs}
& S\equiv \bar\epsilon\epsilon\, , \quad &&P\equiv \bar\epsilon \gamma_7\epsilon\, , 
\quad &&\xi^\flat \equiv \bar\epsilon \gamma_{(1)}\epsilon\, , \quad &&\tilde{K}\equiv 
\ii\bar\epsilon \gamma_{(1)}\gamma_7\epsilon\, ,\nn\\
& Y \equiv \ii\bar\epsilon \gamma_{(2)}\epsilon\, , \quad &&\tilde{Y}\equiv 
\ii\bar\epsilon \gamma_{(2)}\gamma_7\epsilon\, , \quad &&V \equiv \ii\bar\epsilon \gamma_{(3)}\epsilon\, , \quad&&
\tilde{V}  \equiv \bar\epsilon \gamma_{(3)}\gamma_7\epsilon\, ,
\end{align}
where we have defined $\gamma_{(r)}\equiv\frac{1}{r!}\gamma_{\mu_1\cdots 
\mu_r}\diff x^{\mu_1}\wedge\cdots\wedge\diff x^{\mu_r}$, 
and $\bar\epsilon=\epsilon^\dagger$ is the Hermitian conjugate of $\epsilon$.  
We may 
furthermore choose a basis of the gamma matrices $\gamma_\mu$ which are purely imaginary and 
anti-symmetric, with charge conjugation matrix $\mathcal{C}=-\ii\gamma_7$. 
The equations \eqref{KSE}, \eqref{dilatino} imply
that $\xi^\flat$ is a Killing one-form, with dual Killing vector field $\xi$  \cite{Alday:2015jsa}. 
In \cite{Alday:2015jsa},  it  is further shown that 
all of the bosonic supergravity fields are annihilated by the Lie derivative $\mathcal{L}_\xi$, 
and moreover so are the real bilinear forms in \eqref{bilineardefs}, so that 
$\xi$ generates a symmetry of the full solution.

A generic Dirac spinor in $D=6$ defines an SU$(2)$ structure, and 
making this manifest will be helpful for subsequent calculations. 
We first decompose $\epsilon$ into its positive and negative chirality\footnote{Compared 
to  \cite{Alday:2015jsa}, notice we have swapped $\epsilon_\pm^{\mathrm{there}}=
\epsilon_\mp^{\mathrm{here}}$.} parts $\epsilon_\pm$ under 
$\gamma_7$
\begin{align}
\epsilon_\pm \equiv \tfrac{1}{2}(1\pm \gamma_7)\epsilon\, ,
\end{align}
and then write
\begin{align}\label{epsilonpm}
\epsilon_+ = \sqrt{S}\sin\vartheta\, \eta_2^*\, , \quad \epsilon_- = \sqrt{S}\cos
\vartheta\, \eta_1\, .
\end{align}
Here $\vartheta$ is a function, while $\eta_1,\eta_2$ are two orthogonal unit norm chiral spinors, so that $\bar\eta_1\eta_1=\bar\eta_2\eta_2=1$ and $\bar\eta_2\eta_1=0$. 
The stabilizer group of each $\eta_i$ is a different SU$(3)$ subgroup 
of the double cover $\mathrm{Spin}(6)$ of the rotation group SO$(6)$, and these 
intersect to give an SU$(2)$ stabilizer group of $\epsilon$. 
From this one can define two real orthonormal one-forms $K_1, K_2$ and a triplet 
of real two-forms $J_i$, $i=1,2,3$ (orthogonal to each of $K_1$, $K_2$), via
\begin{align}
K_1 - \ii K_2 \equiv -\tfrac{1}{2}\varepsilon^{\alpha\beta}\, \eta_\alpha^\mt\gamma_{(1)}\eta_\beta\,, \quad J_i \equiv  - \tfrac{\ii}{2}\sigma_i^{\alpha\beta}\bar{\eta}_\alpha\gamma_{(2)}\eta_\beta\, .
\end{align}
The Riemannian volume form on $M_6$ is then $\vol_6=K_1\wedge K_2\wedge\tfrac{1}{2}J_i\wedge J_i$, where here there is no sum on $i$ and this holds for any of $i=1,2,3$.  
We then further distinguish $J\equiv J_3$, while the complex combination 
$\Omega\equiv J_2+\ii J_1$ that appears in certain bilinear equations in 
\cite{Alday:2015jsa} will play only a small role in this paper. 
One can then verify \cite{Alday:2015jsa} that the bilinear forms \eqref{bilineardefs} 
may be expressed in terms of this canonically normalized SU$(2)$ structure as 
\begin{align}\label{SU2}
& P = - S\cos2\vartheta\, , \quad \xi^\flat =  S \sin 2\vartheta K_1\, , \quad 
\tilde{K} = -S \sin2\vartheta K_2\, , \nn\\
& Y = S \left(\cos2\vartheta K_1\wedge K_2 - J\right)\, , \quad \tilde{Y} = S \left(- K_1\wedge K_2 +\cos2\vartheta J\right)\, , \nn\\
& V = -S \sin 2\vartheta K_1\wedge J\, , \quad \tilde{V} = -S \sin 2\vartheta K_2\wedge J\, .
\end{align}
The $G$-structure analysis fixes the metric on $M_6$ to take the form
\begin{equation}
	g_{M_6} = K_1^2 + K_2^2 + g_{\mathrm{SU}(2)}\, ,
\end{equation}
in an orthonormal frame such that $\{\mathrm{e}^a_{\mathrm{SUSY}}, \mathrm{e}^5_{\mathrm{SUSY}}\equiv K_1, \mathrm{e}^6_{\mathrm{SUSY}} \equiv K_2\}, a=1, ..., 4$, and $g_{\mathrm{SU}(2)} = \Sigma_{a=1}^4 (\mathrm{e}^a_{\mathrm{SUSY}})^2$.\footnote{The subscript on $\mathrm{e}^A_{\mathrm{SUSY}}$ is to distinguish this canonical frame defined by supersymmetry from other (local) frames used in the following section. }

Finally, we record the following differential constraints  \cite{Alday:2015jsa} on the bilinear forms 
\eqref{bilineardefs}, which follow from imposing  \eqref{KSE}, \eqref{dilatino}:
\begin{align}
\dd ( X S )  = & \  \tfrac{\sqrt{2}}{3} ( X^{-2} \tilde{K} - \ii  \xi\hook B ) \, , \label{biS} \\
\dd ( X P)  =&  - \tfrac{1}{\sqrt{2}} \xi\hook {F} \, , \label{biP} \\
\dd ( X^2 \xi^\flat )  =&  - \tfrac{2\sqrt{2}}{3} X^{-1} \tilde{Y} - \ii  X^4 \xi\hook *H - 
\sqrt{2} X (P{F} -\tfrac{2}{3} \ii S B)\, , \label{biK} \\
\dd ( X^{-2} \tilde{K} )  =&  - \ii \xi\hook H \, , \label{bitK} \\
\dd ( X^{-1} Y )  =&  - \sqrt{2} \tilde{V} + \ii ( X P ) H + \tfrac{1}{\sqrt{2}}  X^{-2} ( \xi \hook  *  {F}  + {F} \wedge \tilde{K} ) \, , \label{biY} \\
\dd ( X^{-1} \tilde{Y} )  =&  \ \ii  ( X S ) H + \ii\tfrac{\sqrt{2}}{3}X^{-2}(\xi \hook  * B +B \wedge \tilde{K} ) \, , \label{bitY}\\
\dd V  =& \  \sqrt{2} ( X + \tfrac{1}{3} X^{-3} ) *   Y   + \ii\tfrac{\sqrt{2}}{3}X^{-1}(P  * B + B\wedge Y)\nn \\ 
& \  - \tfrac{1}{\sqrt{2}}X^{-1}(S*{F} + {F}\wedge \tilde{Y})\, , \label{biV}
\end{align}
where we also have $\diff \tilde{V}=0$.


\section{Localization of Romans supergravity}\label{sec:localizeRomans}


The aim of this section is to derive the main formula \eqref{main} for the holographically renormalized on-shell action. We begin by introducing a set of equivariantly closed forms in section \ref{sec:equivariant}, then apply the Berline--Vergne--Atiyah--Bott fixed point theorem. The
final form of \eqref{main} is obtained after analyzing and simplifying the various fixed point contributions. 

\subsection{Equivariantly closed forms}\label{sec:equivariant}

Following \cite{BenettiGenolini:2023kxp}, and subsequent 
developments in \cite{BenettiGenolini:2023yfe, BenettiGenolini:2023ndb, BenettiGenolini:2024kyy, BenettiGenolini:2024xeo, Cassani:2024kjn, BenettiGenolini:2024hyd, BenettiGenolini:2024lbj, Couzens:2024vbn, Suh:2024asy, Colombo:2025ihp},  we introduce 
the equivariant exterior derivative
\begin{align}
\diff_\xi \equiv \diff - \xi \hook\, .
\end{align}
A polyform $\Phi$, constructed as sums of differential forms of different degrees, is by definition \emph{equivariantly closed} if $\diff_\xi \Phi=0$. 
Such a polyform (given in \eqref{Phib} below) that is related to the bulk on-shell action \eqref{Fullaction}
 was written down in  \cite{BenettiGenolini:2023kxp}; 
we shall see here that there exists a plethora of $\diff_\xi$-closed polyforms 
in the Euclidean Romans supergravity theory.

We may immediately write down polyforms whose top degree components are field strengths:
\begin{align}\label{PhiF}
\Phi^F & \equiv F - {\sqrt{2}}\, (XP)\, ,\\
\Phi^H & \equiv H + \ii X^{-2}\tilde{K}\, .\label{eq:PhiH}
\end{align}
These are equivariantly closed by  virtue of the Bianchi identities, together with equations \eqref{biP}, \eqref{biK}, respectively. 

We next write down the following equivariantly closed polyform associated to $*B$:
\begin{align}\label{eq:Phi*B}
\Phi^{*B}\equiv \left[X^{-2}* B + \tfrac{\ii}{2}B\wedge B\right] -\tfrac{3}{\sqrt{2}}[\ii 
X^{-1}\tilde{Y} + (XS)B] -\tfrac{9}{4}\ii (XS)^2\, ,
\end{align}
where this is a sum of a four-form, two-form and zero-form. The four-form 
is closed by virtue of 
the middle equation in \eqref{FullEOM} (the equation of motion for $B$). 
We have then used \eqref{biS}, \eqref{bitY} and the requirement 
$\diff_\xi \Phi^{*B}=0$ 
to obtain the two-form component, 
and the zero-form component similarly follows from using the algebraic relation $\xi\hook  \tilde{Y} = 
S\tilde{K}$ (obtained from \eqref{SU2}), together with \eqref{biS} again. Integrating the top form over a compact four-cycle $M_4$ gives us the conserved charge $\Bc$,
\begin{equation}\label{eq:starBcharge}
    \Bc\equiv \frac{1}{(2\pi)^2}\int_{M_4}\Phi^{\star B}\,.
\end{equation}
We may similarly construct an equivariantly closed polyform associated to $*F$: 
\begin{align}
\Phi^{*F} \equiv & \ [X^{-2}*F + \ii F\wedge B]+[\sqrt{2}X^{-1}Y-\sqrt{2}\ii (XP)B
-\tfrac{3}{\sqrt{2}}(XS)F+2C]\nn\\
& +3(XS)(XP)\, .\label{eq:Phi*F}
\end{align}
The four-form is closed by virtue of the last equation in \eqref{FullEOM} (the Maxwell equation of motion for the gauge field $A$). The electric charge, $q$, is defined by
\begin{equation}\label{eq:electricdef}
    q\equiv \frac{1}{(2\pi)^2}\int_{M_4}\Phi^{*F}\,  ,
\end{equation}
with $M_4$ a compact four-cycle.
The two-form component part of $\Phi^{*F}$ follows upon using equations \eqref{biS}, \eqref{biP},
\eqref{biY}, and we have introduced the two-form $C$ defined via $\tilde{V}=\diff C$, 
whose (local) existence follows from the equations $\diff\tilde{V}=0=\xi\hook \tilde{V}$. 
Finally, the zero-form component  follows from using the algebraic relation $\xi\hook  {Y} = 
P\tilde{K}$ (obtained from \eqref{SU2}), together with \eqref{biS}, \eqref{biP} again. 

Finally, we have the equivariantly closed polyform constructed in 
\cite{BenettiGenolini:2023kxp}, where we have corrected the overall 
signs of the four-form and zero-form components\footnote{The form 
written  in \cite{BenettiGenolini:2023kxp} is closed under 
$\diff+\xi\hook\, $, rather than $\diff_\xi = \diff-\xi\hook\, $.}
\begin{align}\label{Phib}
  \Phib \equiv \Phib_6+\Phib_4+\Phib_2+\Phib_0\, ,
\end{align}
where
\begin{equation}
\begin{split}
\Phib_6 \equiv & \ \tfrac{4}{9}(2+3X^4)X^{-2}\, \vol_6 + \tfrac{1}{3}
X^{-2}F\wedge *F+\tfrac{\ii}{3}B\wedge F\wedge F\, , \\
\Phib_4\equiv & \ -\tfrac{\sqrt{2}}{3}(XP)X^{-2}*F +\tfrac{2\sqrt{2}}{3}X*\tilde{Y}+ \tfrac{\sqrt{2}}{3}F \wedge X^{-1}Y \\
& -\tfrac{1}{\sqrt{2}}
(XS){F}\wedge{F} -\tfrac{2\sqrt{2}\ii}{3}(XP)B\wedge F\, , \\
\Phib_2 \equiv & \ -\tfrac{2}{3}P Y + \tfrac{2\ii}{3}(XP)^2 B + 2(XS)(XP)F\, ,\\
\Phib_0 \equiv & \ -\sqrt{2}(XS)(XP)^2\, .
\end{split}
\end{equation}
The action \eqref{Fullaction}, evaluated on a solution to 
equations of motion (i.e. on-shell) and including the boundary Gibbons--Hawking--York and holographic counterterms, is
\begin{align}\label{action}
I = \frac{\pi^2}{2G_N}\frac{1}{(2\pi)^3}\int_{M_6} \Phib + \mbox{boundary terms}\, .
\end{align}
The full set of boundary terms were constructed in \cite{Alday:2014rxa, Alday:2014bta}; however, in the present context, as argued in \cite{BenettiGenolini:2023kxp}, they contribute zero to our final expressions and we therefore do not consider them further.\footnote{Since the proof does not fit in the margin we leave it to the interested reader.} A similar result for general four-dimensional gauged supergravity coupled to matter was shown to hold true in  \cite{BenettiGenolini:2024lbj}.


\subsection{Fixed point analysis}\label{sec:fps}

Suppose we have a supersymmetric solution to this supergravity theory, defined on a manifold $M_6$ 
equipped with an R-symmetry Killing vector field $\xi$. In this subsection 
we deduce some properties of the fixed point set of $\xi$, which 
we denote 
\begin{align}
\mathscr{F}\equiv \{\xi=0\}\subset M_6\, .
\end{align} Notice 
that generically this may not be connected, and that a connected component 
will be an even-dimensional submanifold of $M_6$. We thus have a partition of the fixed point set,
$\mathscr{F}=\mathscr{F}_4\cup\mathscr{F}_2\cup\mathscr{F}_0$, 
where the subscript denotes dimension, and some of the $\mathscr{F}_i$ may be empty.

Notice first from 
 \eqref{SU2} that we may write down the norm of the Killing vector field $\|\xi\| = S |\sin 2\vartheta|$. 
The spinor $\epsilon$ is zero if and only if $S\equiv\bar\epsilon\epsilon=0$, but since 
$\epsilon$ solves a first order Killing spinor equation, a standard argument\footnote{See 
\cite{Closset:2012ru}, or the summary of this argument in section 2.5 of 
\cite{Ferrero:2021etw}.} shows 
that, if $\epsilon$ vanishes at a point, it will then be identically zero in a neighbourhood of that point. 
Thus  a supersymmetric solution with a non-trivial Killing spinor necessarily has $S>0$ everywhere. 
The R-symmetry Killing vector field $\xi$ is then zero 
precisely where $\sin 2\vartheta=0$, i.e., where $\vartheta=\frac{\pi}{2}, \frac{3\pi}{2}$ or 
$\vartheta=0,\pi$. In turn, from \eqref{epsilonpm} we see that\footnote{Notice that if one uses $\epsilon^c$ instead of $\epsilon$, one finds instead that $P=\mp S$ on the zeroes of $\xi$. }
\begin{align}\label{pm}
\epsilon = \epsilon_\pm \quad \Leftrightarrow\quad P = \pm S \quad \Leftrightarrow
\quad \xi = 0\, .
\end{align}
This allows us to further refine the fixed point set as
$\mathscr{F}=\mathscr{F}^+\cup \mathscr{F}^-$, 
where from \eqref{pm} we may also identify 
\begin{align}
\mathscr{F}^\pm = 
\{\epsilon=\epsilon_\pm\}=\{P=\pm S\}\, .
\end{align}
The Killing spinor in particular has definite chirality on a connected component of $\mathscr{F}$. We may then further delineate $\mathscr{F}^\pm_i$, 
where the subscript $i$ labels dimension, and in general a given $\mathscr{F}^\pm_i$ 
may have a number of connected components (including being empty).
From \eqref{SU2} we can also immediately deduce that $\tilde{K}|_{\mathscr{F}}=0$, and  
using $\cos2\vartheta=\mp 1$ or equivalently $P=\pm S$ on $\mathscr{F}^\pm$, we also immediately have
\begin{align}\label{YYtilde}
Y|_{\mathscr{F}^\pm}   = \pm \tilde{Y}|_{\mathscr{F}^\pm} ,
\end{align}
where the vertical bar implies a restriction of the forms to the relevant subspace of $M_6$, rather than a pull-back.\footnote{A pull-back would be a differential form on $\mathscr{F}$, while 
the restriction is still a differential form on $M_6$.} 

Next we turn to analysing the differential equations and equivariant forms. 
Since on $\mathscr{F}$ the right hand sides of both \eqref{biS} and \eqref{biP} are 
zero, we have
\begin{align}\label{XSXP}
\diff(XS)|_{\mathscr{F}} = 0 = \diff(XP)|_{\mathscr{F}}\, .
\end{align}
Therefore, the fixed point set is a critical point set of both
functions $XS$ and $XP$. In more detail, taking a contraction of 
\eqref{XSXP} with a tangent vector to a connected component of 
$\mathscr{F}$ shows 
that $XS$ and $XP$ are constant on that connected component, 
while taking a contraction with a normal vector shows that 
this is also a critical point set, with the first order change in $XS$ and $XP$ 
as one moves away from the connected component being zero. 

Turning to the equivariant forms $\Phi$ constructed in section 
\ref{sec:equivariant}, note that it is immediate from 
$\diff_\xi\Phi=0$ that, restricted to $\mathscr{F}$, each 
fixed degree component $\Phi_i$ is separately closed, since 
$\diff \Phi_{i} = \xi\hook \Phi_{i+2}$ holds for each $i$. 
In particular, notice that the zero-form components $\Phi_0$ of all the even-degree polyforms are monomials in $XS$, $XP$, consistent with \eqref{XSXP}. Thus, the fact that
$\diff \Phi_0|_{\mathscr{F}}=0$ 
for all the polyforms in section \ref{sec:equivariant} follows 
from \eqref{XSXP}.

We next turn to the two-form components. From $\Phi^{*B}_2$ we deduce 
that
\begin{align}
\diff [X^{-1}\tilde{Y}-\ii (XS)B]|_{\mathscr{F}} = 0 \, .
\end{align} 
Notice that the quantity $X^{-1}\tilde{Y}-\ii (XS)B$ defines a closed two-form 
when pulled back to the fixed point set, where it is already known that 
$XS$ is locally constant. One can similarly examine 
$\Phi^{*F}_2$, but, upon using  \eqref{pm}  and \eqref{YYtilde},
one sees that the same linear combination of 
$\tilde{Y}$ and $B$ again appears. There is no new information coming from this, as we already know that $F$ is closed 
and, from \eqref{SU2} $\tilde{V}|_{\mathscr{F}}=0$, it implies 
that $\diff C |_{\mathscr{F}}=0$.   

On the other hand, further information about two-forms restricted to 
$\mathscr{F}$ may be obtained by looking directly at \eqref{biK}. Setting $\xi=0$ in this equation  gives 
\begin{align}\label{twoformrelation}
\left.\left[X^{-1}\tilde{Y}- \ii (XS)B + \tfrac{3}{2}(XP)F\right]\right|_{\mathscr{F}} = \mbox{exact}\, .
\end{align}
Notice here that although $\xi^\flat$ restricts to zero on $\mathscr{F}$, this 
is not necessarily the case for $\diff (X^2\xi^\flat)$. However, the latter 
is a globally exact form, and that will be sufficient for our purposes.
Remarkably, we then find 
\begin{alignat}{2}
\left.\Phi^F_2\right|_{\mathscr{F}} &=  F\, , \quad 
&\left.\Phi^{*F}_2\right|_{\mathscr{F}} &=  -3\sqrt{2}(XS)F + 2C + \mbox{exact}\, ,\nonumber \\
\left.\Phi^{*B}_2\right|_{\mathscr{F}} &=  \tfrac{9 }{2\sqrt{2}} \ii(XP)F+\mbox{exact}\, ,\,~ 
&\left.\Phib_2\right|_{\mathscr{F}} &=  3(XS) (XP) F+\mbox{exact}\, ,\label{Phi2}
\end{alignat}
where we have used \eqref{YYtilde} and $P|_{\mathscr{F}^\pm}=\pm S|_{\mathscr{F}^\pm}$. 
The two-form components of all four equivariantly closed forms 
are thus closely related when evaluated on a fixed point set. 
We have no information about the two-form $C$, other than that it is closed 
on $\mathscr{F}$, and thus defines some cohomology class $[C]\in H^2(\mathscr{F},\R)$. 

Finally we turn to the four-form components. The four-form component $\Phib_4$ remarkably reduces to being proportional to $F\wedge F$
when evaluated on the fixed point sets. 
To derive this, one needs to make use of both \eqref{twoformrelation} and the relation between four-forms when
 evaluated on fixed point sets, equation \eqref{biV}. In addition, $ *\tilde{Y}\big|_{\mathscr{F}}$ can be written in terms of 
$Y\wedge Y\big|_{\mathscr{F}} =\tilde{Y}\wedge \tilde{Y}\big|_{\mathscr{F}}$ when restricted 
to a fixed point set $\mathscr{F}$, namely
\begin{align}
    Y\wedge Y\Big|_{\mathscr{F}} = \left.\tilde{Y}\wedge \tilde{Y}\right|_{\mathscr{F}} = \left. -2\,  S  *\tilde{Y}\right|_{\mathscr{F}} \, .
\end{align}
This results in a relation between $\Phib_4$ and $\Phi_4^{*B}$ and $F\wedge F$ on fixed point sets, which is
\begin{align}
  \left.\Phib_4\right|_{\mathscr{F}}=-\frac{ 2 \sqrt{2} }{9} \ii(XS) \Phi_4^{*B} - \frac{5}{2\sqrt{2}} (XS) F \wedge F \, .
\end{align}
Using the second  of the equations of motion in \eqref{FullEOM}, one sees that $\Phi_4^{*B}$ is proportional to $F\wedge F$ when evaluated on a closed four-dimensional fixed point set, and the final expression for $\Phib_4$ on such a set then simplifies to
\begin{align}\label{Phi4}
  \left.\Phib_4\right|_{\mathscr{F}} = - \frac{3}{\sqrt{2}}(XS) F \wedge F \, .
\end{align}


\subsection{Localization}\label{sec:localize}

Given a compact Riemannian manifold $M$ of dimension $2n$, equipped with a 
vector field $\xi$ and an equivariantly closed form $\Phi$, we can express the integral of $\Phi$ over $M$ via the BV--AB fixed point formula \cite{BV:1982, Atiyah:1984px}
(see also \cite{BenettiGenolini:2023ndb} for a review)
\begin{equation}\label{BV-AB}
\begin{split}
    \int_{M}\Phi =& \sum_{\text{dim  }  0}\frac{1}{d_{\mathscr{F}_0}}\frac{(2\pi)^{n}}{\epsilon_1 ... \epsilon_n}\Phi_0 + \sum_{\text{dim  } 2 } \frac{1}{d_{\mathscr{F}_2}}\frac{(2\pi)^{n-1}}{\epsilon_1 ... \epsilon_{n-1}}\int \left [ \Phi_2 -\Phi_0 \sum_{1\leq i \leq n-1}\frac{2\pi}{\epsilon_i}c_{1}(L_i)\right] \\
    ~& +\sum_{\text{dim  } 4 }\frac{1}{d_{\mathscr{F}_4}}\frac{(2\pi)^{n-2}}{\epsilon_1 ... \epsilon_{n-2}}\int \left [ \Phi_4 -\Phi_2 \wedge \sum_{1\leq i\leq n-2}\frac{2\pi}{\epsilon_i}c_1 (L_i) \right. \\ 
    ~& \left. + ~ \Phi_0 \sum_{1\leq i \leq j \leq n-2}\frac{(2\pi)^2}{\epsilon_i \epsilon_j}c_1 (L_i)\wedge c_1 (L_j) \right ] + \cdots\, .
   \end{split}
\end{equation}
Recall that $\mathscr{F}_j$  is the $j$-dimensional fixed point set under $\xi$, 
and that the sums are over connected components of $\mathscr{F}_j$. 
$\Phi_i$ is the degree $i$ component of  the polyform $\Phi$, and we have allowed for the possibility 
of orbifold singularities on $M$, where 
 $d_{\mathscr{F}_j}\in\mathbb{N}$ is the order of the orbifold structure group of (a connected component of) ${\mathscr{F}_j}$. 
The $\epsilon_i$'s are the weights of the action of $\xi$ on the normal bundle $\mathcal{N}\mathscr{F}_j$  of each connected component of ${\mathscr{F}}$. In more detail, 
the normal space to a point in $\mathscr{F}_j$ will, for generic\footnote{More generally 
it is the (equivariant) Euler class of the normal bundle that appears in \eqref{BV-AB}, 
rather than the Chern classes $c_1(L_i)$ written here, but this formula will be sufficient 
for the present paper.} $\xi$, split as $\R^{2n-j}=\oplus_{i=1}^k\R^2$, where $2k=2n-j$ is the rank 
of the normal bundle, 
with $\xi$ inducing a rotation of each two-plane $\R^2$. 
Fixing an orientation on each $\R^2\cong \C$, 
this then correspondingly splits the normal bundle into a sum of 
complex line bundles $\mathcal{N}\mathscr{F}_j\cong \oplus_{i=1}^k L_i$
whose Chern classes appear in \eqref{BV-AB}. 
 The vector field 
$\xi$ on this normal space can then be written as 
\begin{equation}\label{xiweights}
    \xi = \sum_{i=1}^k \epsilon_i \mskip2mu\partial_{\varphi_i},
\end{equation}
where $\varphi_i$ is an angular coordinate of the $i^{\text{th}}$ plane;
that is, $z_i=|z_i|\ex^{\ii \varphi_i}$ is the corresponding complex coordinate 
on the  $i^{\text{th}}$ complex plane. 
 Notice that in general the weights $\epsilon_i$ will vary from fixed point set to fixed point set (which is suppressed in the notation in \eqref{BV-AB}), 
and that a fixed point  of codimension $2k$ will have only $k$ weights associated to it. 

We may now apply the  general formula \eqref{BV-AB} to the integral of $\Phib$ over $M_6$. The expressions \eqref{Phi2}, \eqref{Phi4}  may be used to simplify the 
final formula. We note specifically that the exact part of the two-form $\Phib_2\big|_{\mathscr{F}} $ is proportional to $\text{d}((XS)(X^2 \xi^\flat))$ and thus vanishes when integrated over the fixed point set, upon using Stokes' Theorem. Hence we obtain\footnote{Note that the $\pm$ and $\mp$ would be interchanged if we used $\epsilon^c$ instead of $\epsilon$. }
\begin{align}\label{Phiblocalize}
    \int_{M_6} \Phib& = \sum_{\text{dim  }  0}-\sqrt{2}(2\pi)^3\frac{(XS)^3}{\epsilon_1 \epsilon_2 \epsilon_3}\nonumber \\
    ~& +\sum_{\text{dim  } 2 } \frac{(2\pi)^{2}}{\epsilon_1  \epsilon_{2}}\int_{\mathscr{F}_2} \left [ \pm 3(XS)^2 F  +2\pi \sqrt{2}(XS)^3 \left ( \frac{c_{1}(L_1)}{\epsilon_1} + \frac{c_{1}(L_2)}{\epsilon_2} \right )\right] \nonumber\\
    ~ & +\sum_{\text{dim  } 4 }\frac{2\pi}{\epsilon_1 }\int_{\mathscr{F}_4} \left [ -\frac{3}{\sqrt{2}}(XS)F\wedge F \mp\frac{6\pi}{\epsilon_1}  (XS)^2 F \wedge c_1 (L_1) \right . \nonumber \\
    ~& \left . \qquad \qquad  -\frac{(2\pi)^2 \sqrt{2}(XS)^3 }{\epsilon_1^2}c_1 (L_1)\wedge c_1 (L_1) \right ] + \mbox{boundary terms} .
\end{align}
A few comments are in order. Firstly, $M_6$ has a UV boundary five-manifold $M_5\equiv \partial M_6$, 
and in applying the BV--AB formula there will be a boundary term integrated over $M_5$
after applying Stokes' Theorem to the top form part of 
$ \Phib$. However, as explained in \cite{BenettiGenolini:2023kxp}, this boundary 
term will cancel the boundary terms in the holographically renormalized action 
\eqref{action}, and thus we will not need to keep track of its precise form. 
Secondly, the plus and minus signs in \eqref{Phiblocalize} are those associated to the 
chirality of the fixed point set $\mathscr{F}^\pm$, as explained in section \ref{sec:fps}, 
while $\mathscr{F}_2$ has a normal bundle $L_1\oplus L_2$ and 
$\mathscr{F}_4$ has a normal bundle $L_1$.

The formula \eqref{Phiblocalize} contains two types of terms: 
(i) the weights $\epsilon_i$, Chern classes $c_1(L_i)$ and $\pm$ signs 
are global topological information on $M_6$, requiring one to only know 
the topology of $M_6$, the R-symmetry vector $\xi$, and chirality data; (ii) the values of 
the constant scalar $(XS)$ on a fixed point set and the integrals of the 
gauge field flux terms involving $F$ \emph{a priori} would require one to know more about the explicit solution 
in order to evaluate them. We turn to further analyzing the second type of terms in the next subsections.


\subsection{Further evaluating the action}\label{sec:charges}

We claim that at a general fixed point locus $\mathscr{F}_{6-2k}$ of 
codimension $2k$ we have the formula
\begin{equation}\label{XP}
	(XP)|_{\text{$\mathscr{F}_{6-2k}$}} = \frac{\sum_{i=1}^k \sigmai \epsilon_i}{\sqrt{2}}\, ,
\end{equation}
where $\epsilon_i$ are the weights, and $\sigmai\in \{\pm 1\}$ are 
signs associated to certain spinor projections (see below). 
A version of equation \eqref{XP} already appeared in \cite{Alday:2015jsa}, 
where it was derived for a solution with topology $M_6=\R^6$. 
Our proof here takes a different approach, following \cite{BenettiGenolini:2024xeo}, 
and is more general.

To show \eqref{XP}, we begin by locally writing $F=\diff A$, where we note that we may always choose a gauge for the gauge field $A$ such that 
$\mathcal{L}_\xi A=0$. Recall that the Lie derivative acting on forms is 
$\mathcal{L}_\xi  A= \xi\hook (\diff A) + \diff (\xi\hook A)$. The 
 equivariant closure of \eqref{PhiF} then implies that
\begin{align}\label{Asusy}
\xi \hook A = \sqrt{2}(XP) + c\, ,
\end{align}
where $c$ is an integration constant. We define a \emph{supersymmetric gauge} for 
$A$ to be the choice $c=0$.\footnote{This will not fix the gauge choice completely, but 
this will not matter for the argument that follows.} 

In appendix 
\ref{app:charge}, it is shown that in this gauge the Killing spinor has 
charge zero under the R-symmetry vector: $\mathcal{L}_\xi \epsilon =0$. 
Here the Lie derivative acting on a 
spinor is 
\begin{align}\label{Lieepsilon}
\mathcal{L}_\xi \epsilon = \xi^\mu\nabla_\mu \epsilon + \frac{1}{8}\diff\xi^\flat_{\mu\nu}\gamma^{\mu \nu}\epsilon\, .
\end{align}
On the other hand, we may substitute for $\nabla_\mu\epsilon$ in this equation using the Killing spinor equation \eqref{KSE} (with $I=1$). Contracting the latter with $\xi^\mu$, $\xi^\mu\nabla_\mu\epsilon$ will trivially give zero at a fixed point where $\xi=0$ (assuming the fields of the solution 
do not diverge, making the solution singular), where 
we only need to be careful with the gauge field contribution in the covariant derivative $D_\mu = \nabla_\mu + \frac{\ii}{2}A_\mu \epsilon$. 
Specifically, in the supersymmetric gauge \eqref{Asusy} with $c=0$, the 
gauge field $A$ is \emph{precisely} singular at the fixed point locus $\xi=0$, when
$\xi^\mu A_\mu =\sqrt{2}(XP)\neq 0$. 
Combining the above formulae, from \eqref{Lieepsilon}
we immediately deduce
\begin{align}\label{Lieepsilonagain}
 0 & = -\frac{\ii}{2}\sqrt{2}(XP) + \left.\frac{1}{2}\sum_{i=1}^k \epsilon_i\gamma^{(2i-1)2 i} \epsilon\right|_{\mathscr{F}_{6-2k}}\, ,
\end{align}
where we have used that, normal to a fixed point set of codimension $2k$, in a local orthonormal frame, one has
\begin{align}\label{dxiframe}
\diff \xi^\flat\mid_{\mathcal{N}\mathscr{F}_{6-2k}} = 2\, \bigoplus_{i=1}^k \begin{pmatrix} 0 & \epsilon_i \\ -\epsilon_i & 0\end{pmatrix}\, .
\end{align}
Hence,
\begin{align}\label{spinorchargethingy}
\left.\sum_{i=1}^k \epsilon_i\gamma^{(2i-1)2i} \epsilon = \ii \sqrt{2}(XP)\epsilon\right|_{\mathscr{F}_{6-2k}}\, 
\end{align}
holds at a given fixed point. 
The operators $-\ii \gamma^{(2i-1)2i}$ square to 1, 
and one can check from \eqref{spinorchargethingy} 
that, for $\epsilon$ to obey this equation, it must 
be an eigenspinor of each of these operators.\footnote{To be more precise, this is 
true for \emph{generic} weights 
where $|\epsilon_i|\neq |\epsilon_j|$ for all $i\neq j$. If some 
$\epsilon_i=\pm \epsilon_j$ then 
the normal bundle to the fixed point set does not necessarily split into a sum of complex line bundles, which recall we 
have assumed to be the case. This non-generic 
case will potentially have weaker projection conditions and a different normal bundle contribution,
but we leave this for the interested reader to pursue.}
At a given fixed point, we may thus write\footnote{Notice that the chiralities would reverse sign if we used the charge conjugate spinor $\epsilon^c$, as opposed to $\epsilon$. }
\begin{align}\label{sigmai}
-\ii \gamma^{(2i-1)2i} \epsilon = \sigmai\mskip1mu\epsilon\, ,
\end{align}
where necessarily $\sigmai\in\{\pm 1\}$. Combining with \eqref{spinorchargethingy}
then proves \eqref{XP}.

At first sight one might think that the signs $\sigmai$ in \eqref{XP} 
can simply be absorbed into the definition of the weights 
$\epsilon_i$: defining the latter above equation \eqref{xiweights} required one to pick orientations 
on the normal two-planes $\R^2\cong \C$ to the fixed point set, and 
changing the orientation changes the sign of the corresponding 
weight $\epsilon_i$ (effectively replacing $\C$ by its complex conjugate).  
While this is true, there are two subtleties that become important 
when there is more than one connected component of 
$\mathscr{F}$:

\begin{itemize}
\item 
First, the overall orientation on $M_6$ is fixed, and 
the BV--AB formula requires us to use this orientation. 
This means that the local orthonormal frame used in \eqref{Lieepsilonagain}, \eqref{dxiframe} 
must have orientation $\mathrm{e}^{123456}=\vol_6$, and, 
since $\gamma_7=\ii\gamma^{123456}=(-\ii\gamma^{12})(-\ii\gamma^{34})(-\ii\gamma^{56})$, the signs $\sigmai$ in \eqref{sigmai} are correlated with the 
chirality of the fixed point locus, with $\gamma_7\epsilon=\pm \epsilon$ on $\mathscr{F}^\pm$. The analysis is then slightly different depending on the 
dimension of $\mathscr{F}_j$. 
At an isolated fixed point $\mathscr{F}_0$, we have
\begin{align}\label{XS0dim}
(XS)|_{\mathscr{F}_0^{\pm}} = \pm (XP) |_{\mathscr{F}_0^{\pm}} = \chi \frac{\sum_{i=1}^3\sigmai\epsilon_i}{\sqrt{2}}\, ,
\end{align} 
where the argument above implies that we may identify the chirality of the spinor as
$ \chi=\chi|_{\mathscr{F}_0^\pm}=\sigmaone\sigmatwo\sigmathree=\pm 1$.
For a two-dimensional fixed point locus $\mathscr{F}_2$, we may 
essentially, without loss of generality, choose this to be 
along the $\mathrm{e}^5$--$\mathrm{e}^6$ directions of the frame. 
Then the projections \eqref{sigmai} hold for $i=1,2$, and, 
since the spinor is necessarily chiral, this implies that the same projection must
hold  also for $i=3$, where the chirality is again $ \chi=\chi|_{\mathscr{F}_2^\pm}=\sigmaone\sigmatwo\sigmathree=\pm 1$, and where 
 $\sigmathree$ is precisely the chirality of the spinor 
along $\mathscr{F}_2$. Finally, for a four-dimensional 
fixed point locus $\mathscr{F}_4$, there is only one 
normal plane, say the $\mathrm{e}^1$--$\mathrm{e}^{2}$ plane, 
with projection \eqref{sigmai} holding with $i=1$.\footnote{Comparing 
to the supersymmetric frame introduced in section \ref{sec:SUSY}, notice that 
the $\mathrm{e}^1$--$\mathrm{e}^{2}$ plane here is the same as the $\mathrm{e}^5_{\mathrm{SUSY}}$--$\mathrm{e}^{6}_{\mathrm{SUSY}}$ plane.} 
We now deduce that $\chi=\chi|_{\mathscr{F}_4^\pm}=\sigmaone \eta =\pm 1$, 
where necessarily $-\gamma^{3456}\epsilon=\eta\mskip1mu \epsilon$, with $\eta\in\{\pm 1\}$ 
now denoting the   chirality of the spinor 
along $\mathscr{F}_4$; in general there will be no particular 
decomposition of the tangent space of $\mathscr{F}_4$ into 
two two-planes, and correspondingly we cannot in general simply write $\eta=\sigmatwo
\sigmathree$.

\item Secondly, as soon as there is more than one connected component 
of $\mathscr{F}$,
one needs to specify how the relative orientations of $\R^2\cong \C$ are 
chosen at different points, and the corresponding relative signs (specified by the $\sigmai$) 
can then distinguish different solutions. We illustrate 
this in more detail in sections \ref{sec:examplesI}-\ref{sec:examplesIII}, when we look at various 
examples.
\end{itemize}

We may now use this together with \eqref{Phiblocalize} to give a 
localization formula for the on-shell action \eqref{action}.
In our normalization of the AdS radius we can take the following formula for the Newton constant \cite{Alday:2014bta}:
\begin{align}
F_{S^5}\equiv I_{AdS_6} = -\frac{27\pi^2}{4G_N}\, .
\end{align}
Here the free energy $F_{S^5}$ of the dual field theory on the round $S^5$ is 
identified holographically with the renormalized on-shell action $I_{AdS_6}$ of Euclidean AdS$_6$. Then we find
\begin{align}\label{actionwithsigns}
& I =  \frac{F_{S^5}}{27}  \bigg\{\sum_{\text{dim  }  0} \frac{\chi(\sigmaone\epsilon_1+\sigmatwo\epsilon_2+\sigmathree\epsilon_3)^3}{\epsilon_1\epsilon_2\epsilon_3} \\
&-\sum_{\text{dim  }  2} \frac{\chi
(\sigmaone\epsilon_1+\sigmatwo\epsilon_2)^2}{\epsilon_1\epsilon_2}\int_{\mathscr{F}_2} \bigg[3\mskip2mu c_1(F) 
+(\sigmaone\epsilon_1+\sigmatwo\epsilon_2)\left(\frac{c_1(L_1)}{\epsilon_1}+\frac{c_1(L_2)}{\epsilon_2}\right)\bigg]\nonumber\\
& + \sum_{\text{dim  }  4} \chi\sigmaone\int_{\mathscr{F}_4}\bigg[ 3 c_1(F)\wedge c_1(F) 
+ 3 \mskip1mu \sigmaone c_1(F)\wedge c_1(L_1)+c_1(L_1)\wedge c_1(L_1)\bigg]\bigg\} \, ,\nonumber
\end{align}
where we have defined
\begin{align}
c_1(F) \equiv \frac{F}{2\pi}\, .
\end{align}
As explained above, for $\mathscr{F}_0^\pm$ and $\mathscr{F}_2^\pm$, 
we may identify $\chi=\sigmaone\sigmatwo\sigmathree=\pm 1$ 
in  \eqref{actionwithsigns}, where for $\mathscr{F}_2$ 
the two-dimensional chirality of the Killing spinor along $\mathscr{F}_2$ is $\sigmathree$, 
while for $\mathscr{F}_4$ instead $\chi\sigmaone=\eta$ may be 
interpreted as the four-dimensional  chirality of the Killing spinor along 
$\mathscr{F}_4$. 

We emphasize that, if we had begun with the spinor $\epsilon^c$ rather than $\epsilon$, the above formula would differ in the overall signs  of all $\sigma^{(i)}$, and there would be additional minus signs in front of the $c_1(F)$ terms. The final result is then 
independent of this choice, as must be the case.


\subsection{Gauge field flux}\label{sec:gaugefield}

The formula \eqref{actionwithsigns} for the renormalized on-shell action is now manifestly purely topological, but to evaluate it for a solution we still need to know 
the integrals involving the gauge field flux $c_1(F)\equiv F/2\pi$. Physically, these are 
magnetic fluxes for the R-symmetry gauge field $A$. In this section, we will  show that for a two-dimensional fixed point set $\mathscr{F}_2$ we may determine the contribution by generalizing an argument that first appeared in 
\cite{BenettiGenolini:2019jdz}. For four-dimensional fixed point sets $\mathscr{F}_4$ we will obtain a similar result, but the argument is a little more involved.

As used earlier, the spinor $\epsilon$ transforms covariantly under the charged covariant derivative $D_\mu=\nabla_\mu + \frac{\ii}{2}A_\mu$. This means that globally $\epsilon$ is a section of $\mathcal{S}M_6\otimes \mathcal{L}^{1/2}$, where $\mathcal{S}M_6$ is the spin bundle of $M_6$, and $A$ is a connection on the complex line bundle $\mathcal{L}$. This makes $\epsilon$ into a ``spin$^c$''-type spinor, which may exist globally even when the spin bundle of $\mathcal{S}M_6$ does not exist.\footnote{One needs $w_2(M_6)=w_2(\mathcal{L})\in H^2(M_6,\Z_2)$, where these are the second Stiefel-Whitney classes.}

\paragraph{Two-dimensional fixed point set}
Consider restricting $\epsilon$ to a connected component of the two-dimensional fixed point set, which we simply denote by  $\mathscr{F}_2$. Choosing an orientation on this makes $\mathscr{F}_2$ into a Riemann surface, and these are classified: $\mathscr{F}_2\cong \Sigma_g$, where $g\in\Z_{\geq 0}$ is the genus. Isomorphism classes of 
complex line bundles $L$ over a compact Riemann surface are in one-to-one correspondence with $H^2(\Sigma_g,\Z) \cong \Z$, and the (group) isomorphism is given by the first Chern number $\int_{\Sigma_g}c_1(L)$ of the bundle, also known as the \emph{degree}. Thus, we may unambiguously use the notation $\mathcal{O}(n)$ for the line bundle over $\Sigma_g$ with degree $n$. In particular, having fixed a choice of orientation for $\mathscr{F}_2$, in order to agree with the given orientation on $M_6$, this also fixes an orientation for the normal bundle $\mathcal{N}\mathscr{F}_2=L_1\oplus L_2$, as discussed in section \ref{sec:localize}. We may then write $L_i=\mathcal{O}(-p_i)$, with $-p_i=\int_{\Sigma_g} c_1(L_i)$ for $i=1,2$. Similarly, the tangent bundle of $\mathscr{F}_2\cong \Sigma_g$ is $T\Sigma_g= \mathcal{O}(2-2g)$. 

It follows that the tangent bundle splits as $TM_6|_{\Sigma_g} 
\cong \mathcal{O}(2-2g)\oplus \mathcal{O}(-p_1)\oplus \mathcal{O}(-p_2)$. 
The chiral spinor bundles $\mathcal{S}^\pm\equiv \mathcal{S}^\pm M_6 |_{\Sigma_g}$ restricted 
to $\Sigma_g$ are then 
\begin{equation}\label{spinors}
\begin{split}
\begin{array}{lllll}
\mathcal{S}^+ \cong &&  \mathcal{O}(\tfrac{1}{2}p_1+\tfrac{1}{2}p_2+(1-g))&\oplus &
\mathcal{O}(-\tfrac{1}{2}p_1-\tfrac{1}{2}p_2+(1-g))\\
& \oplus&  
\mathcal{O}(-\tfrac{1}{2}p_1+\tfrac{1}{2}p_2-(1-g))&\oplus &
\mathcal{O}(\tfrac{1}{2}p_1-\tfrac{1}{2}p_2-(1-g))\, ,\\
\mathcal{S}^- \cong & &  \mathcal{O}(-\tfrac{1}{2}p_1+\tfrac{1}{2}p_2+(1-g))&\oplus &
\mathcal{O}(\tfrac{1}{2}p_1-\tfrac{1}{2}p_2+(1-g))\\
& \oplus&   
\mathcal{O}(\tfrac{1}{2}p_1+\tfrac{1}{2}p_2-(1-g))&\oplus &
\mathcal{O}(-\tfrac{1}{2}p_1-\tfrac{1}{2}p_2-(1-g))\, .
\end{array}
\end{split}
\end{equation}
Altogether these make up the 8 components of a spinor in  six dimensions, 4 of each chirality. Notice that these are precisely the $8=2^3$ different sign choices in $\sigmai$, with the chirality $ \chi=\chi|_{\mathscr{F}_2^\pm}=\sigmaone\sigmatwo\sigmathree=\pm 1$ discussed in the previous subsection. That is, we may identify each of the 8 factors in \eqref{spinors} with the bundle
\begin{align}
\mathcal{O}\left(-\tfrac{1}{2}\sigmaone p_1 - \tfrac{1}{2}\sigmatwo p_2+ 
\sigmathree (1-g)\right)\, ,
\end{align}
for all possible choices of $\sigmai\in\{\pm 1\}$, where as claimed in section \ref{sec:charges}, $\sigmathree$ is effectively the two-dimensional chirality of the spinor along $\mathscr{F}_2\cong \Sigma_g$ with the two chiral spin bundles on $\Sigma_g$ being $\mathcal{O}(\pm (1-g))$. 

Reinstating the complex line bundle $\mathcal{L}$, the full Killing spinor $\epsilon$ is then a section of $\mathcal{S}M_6\otimes\mathcal{L}^{1/2}$. Defining the magnetic flux as
\begin{align}
m \equiv \int_{\Sigma_g} c_1(F)\, ,
\end{align}
it follows that the 8 components of $\epsilon$ are bundles of
\begin{align}\label{epsilonbundle}
\mathcal{O}\big(-\tfrac{1}{2}\sigmaone p_1 - \tfrac{1}{2}\sigmatwo p_2+ 
\sigmathree (1-g)+\tfrac{1}{2}m\big)\, .
\end{align}
On the other hand, the analysis in section \ref{sec:charges} also showed that 
the spinor is in \emph{precisely one} of these 8 components, due to the projection 
conditions \eqref{sigmai}. The final step of the argument is now that 
we know that $\epsilon$ is everywhere nowhere-zero, and, 
if a complex line bundle \eqref{epsilonbundle} has a nowhere-zero section, 
it must be a trivial line bundle. That is,
\begin{align}\label{linebundlezero}
\mathcal{O}\big(-\tfrac{1}{2}\sigmaone p_1 - \tfrac{1}{2}\sigmatwo p_2+ 
\sigmathree (1-g)+\tfrac{1}{2}m\big) \cong \mathcal{O}(0)\, ,
\end{align}
and therefore
\begin{align}\label{mformula}
\int_{\Sigma_g}c_1(F) = m =\sigmaone p_1 + \sigmatwo p_2 - \sigmathree (2-2g)\, .
\end{align}
This may now be inserted into the middle line of the on-shell action formula \eqref{actionwithsigns}, 
leading to a purely topological formula for the isolated fixed points and 
two-dimensional fixed points.\footnote{Equation 
\eqref{mformula} should be compared to equation (3) of \cite{BenettiGenolini:2024xeo}, where it was shown that the on-shell action for four-dimensional gauged supergravity can also be written in terms of purely topological terms.} 

\paragraph{Four-dimensional fixed point set} Consider now the four-dimensional fixed point sets, $\mathscr{F}_4$.  
Let us denote by $\mathscr{F}_4=B_4$, a connected four-manifold, and note that now the spin bundle on $M_6$ restricted to $B_4$ splits as
\begin{equation}
\begin{array}{lllll}
\mathcal{S}^+M_6|_{B_4} &\cong & \left(\mathcal{S}^+_{B_4}\otimes L_1^{1/2}\right)&\oplus &
\left(\mathcal{S}^-_{B_4}\otimes L_1^{-1/2}\right)\, ,\nonumber\\
\mathcal{S}^-M_6|_{B_4}& \cong & \left(\mathcal{S}^+_{B_4}\otimes L_1^{-1/2}\right)&\oplus &
\left(\mathcal{S}^-_{B_4}\otimes L_1^{1/2}\right)\, .
\end{array}
\end{equation}
Here $\mathcal{S}^\pm_{B_4}$ are now the rank two chiral spin bundles of $B_4$ (which 
recall we distinguished by $\eta\in\{\pm 1\}$), while the sign of $L_1^{\pm 1/2}$ is 
fixed by $\sigmaone\in\{\pm 1\}$, where $L_1$ is the normal bundle of $B_4$ inside $M_6$. The projections in section~\ref{sec:charges} then say that $\epsilon\mskip1mu |_{B_4}$ is precisely one of these four factors (tensored with the gauge bundle $\mathcal{L}^{1/2}$), determined by the $4=2^2$ choices of $\eta,\sigmaone\in\{\pm 1\}$. Thus we find that the spinor is a section of the rank two bundle
\begin{align}\label{curlyE}
\mathcal{E}\equiv\mathcal{S}^\eta_{B_4}\otimes L_1^{\sigmaone/2} \otimes \mathcal{L}^{1/2}\, . 
\end{align}
Again, $\epsilon$ is a nowhere-zero section of $\mathcal{E}$, which being rank two implies that its second Chern class must be zero, $\int_{B_4} c_2(\mathcal{E})=0$. Moreover, 
$\mathcal{E}$ must then split as a sum of two complex line bundles $\mathcal{E}\cong L_{\epsilon}\oplus L_{\perp}$, where the trivial line bundle
$L_{\epsilon}$ consists of multiples of $\epsilon$. 
This is quite a strong topological constraint on $\mathcal{E}$,
which moreover is a spin$^c$ bundle on $B_4$ 
with structure group U$(2)$.
The SU$(2)\subset \mathrm{U}(2)$ subgroup 
may be identified with supersymmetric SU$(2)$-structure described in 
section \ref{sec:SUSY}. 
It is a standard result that the spinor bundles on $B_4$ are, in terms of this almost complex structure,
\begin{align}\label{4dspinors}
\mathcal{S}^+_{B_4} = K^{1/2}_{B_4}\oplus K^{-1/2}_{B_4}\, , \qquad \mathcal{S}^-_{B_4} = K^{1/2}_{B_4}\otimes \Lambda^{0,1}\, ,    
\end{align}
where $K_{B_4}$ is the canonical line bundle and $\Lambda^{0,1}$ denotes the bundle of  $(0,1)$-forms.\footnote{If $B_4$ is not spin, $K_{B_4}$ will not have a global square root, but it is only the total tensor product in \eqref{curlyE} that needs to exist globally for a spin$^c$ structure.}
We see that the condition 
that $\mathcal{E}$ splits as a sum of complex line bundles 
is automatically true for $\eta=1$, but for $\eta=-1$ this implies that necessarily
$\Lambda^{0,1}$ is a direct sum, and generically this will 
not be true.\footnote{For example, this is not true for $B_4=\mathbb{CP}^2$, discussed in section \ref{sec:examplesIII}. If instead $B_4$ is the direct product of two Riemann surfaces, there is a natural decomposition of $\Lambda^{0,1}$ into the sum of two complex line bundles and $\eta=-1$ is a possibility, see section \ref{sec:OpSg1Sg2}. }

Rather than analyse the latter as a potential special case, we thus henceforth set $\eta=1$. 
The above condition that $c_1(L_\epsilon)=0\in H^2(B_4,\Z)$ then implies that
\begin{equation}\label{mformulaB4}
    \pm c_1(K_{B_4})= \sigmaone c_1(L_1)+c_1(F)\, ,
\end{equation}
where the $\pm$ sign is the choice of canonical or anti-canonical bundle factor in $\mathcal{S}^+_{B_4}$ in \eqref{4dspinors}. 
Equation \eqref{mformulaB4} then allows us to 
eliminate $c_1(F)$ in terms of purely topological data. 
In particular, using 
the standard formula
\begin{align}\label{eulersig}
\int_{B_4} c_2(\mathcal{S}^\pm_{B_4}) = - \frac{1}{4}\left[\pm 2\chi(B_4)+ 3\tau(B_4)\right]\, ,
\end{align}
where $\chi(B_4),\tau(B_4)\in\Z$ are the Euler number and signature, respectively, of the
oriented four-manifold $B_4$,\footnote{$\chi(B_4)$ here is not to be confused with the chirality 
$\chi$ of the spinor $\epsilon$. The right hand side of \eqref{eulersig} is only an integer when $B_4$ is a spin manifold, but, as with $M_6$, more generally, one can take $B_4$ to be spin$^c$ and the derivation we have presented still remains correct. } 
$\int_{B_4}c_2(\mathcal{E})=0$ is equivalent to\footnote{Note that if we did not set $\eta=1$ above the left-hand side of the following equation would have an additional factor of $\eta$ multiplying the Euler characteristic and the right-hand side would be absent.} 
\begin{align}\label{identity}
2 \chi(B_4)+3 \tau(B_4) = & \ \int_{B_4}(\sigmaone c_1(L_1) + c_1(F))^2 = \int_{B_4} c_1(K_{B_4})^2\, .
\end{align}

It is straightforward to check that the above analysis and \eqref{mformulaB4} are
 consistent with the formula for $c_1(F)$ for two-dimensional fixed point sets, \eqref{mformula}. For example, consider $B_4$ to be $\mathcal{O}(-p_2)\rightarrow \Sigma_g$. Then from \eqref{mformulaB4} one finds
\begin{equation}
\begin{split}
    \int_{\Sigma_g}c_1(F)&= \int_{\Sigma_g}\left[-\sigma^{(1)}c_1(L_1)\pm c_1(K_{B_4})\right]\\
    &=\sigmaone p_1 \mp p_2\mp \chi(\Sigma_g)\, ,
\end{split}
\end{equation}
where we have used the adjunction formula. One can then identify the signs here with the signs $\sigmatwo$ and $\sigmathree$ in 
\eqref{mformula}. 

Putting together the results of this section leads to the final formula \eqref{main}, which we present again for the convenience of the reader:
\begin{align}
     I = \ &   \bigg\{\sum_{\text{dim  }  0} \sigmaone\sigmatwo\sigmathree\frac{(\sigmaone\epsilon_1+\sigmatwo\epsilon_2+\sigmathree\epsilon_3)^3}{\epsilon_1\epsilon_2\epsilon_3} \\
+&\sum_{\text{dim  }  2} \sigmaone\sigmatwo\sigmathree\frac{
(\sigmaone\epsilon_1+\sigmatwo\epsilon_2)^2}{\epsilon_1\epsilon_2} \bigg[3\sigmathree \chi(\Sigma_g)\nonumber\\ & \qquad +
 \int_{\Sigma_g}2\big(\sigmaone c_1(L_1)+\sigmatwo c_1(L_2)\big)-\sigmaone \frac{\epsilon_1}{\epsilon_2}c_1(L_2)
 -\sigmatwo\frac{\epsilon_2}{\epsilon_1}c_1(L_1)\bigg]\nonumber\\
 +& \sum_{\text{dim  }  4} \bigg[  6 \chi(B_4)+9 \tau (B_4)+\int_{B_4}c_1(L_1)\wedge\big( c_1(L_1)\mp 3 \sigmaone c_1(K_{B_4})\big)\bigg]\bigg\} \frac{F_{S^5}}{27}\, .\nonumber
 \end{align}

Having derived the general formula for computing the renormalized on-shell action, in the following sections we will show how to apply it for a large class of examples. We begin by working with the more trivial examples before ramping up the difficulty as we proceed. We will show that our results recover known results whilst also giving new predictions for currently unknown solutions and dual field theory calculations. As a consistency check of our results, we show that by specializing the results for the more general topologies, i.e., by trivializing the normal bundles or turning off rotation parameters, we recover identically the earlier results. This is a non-trivial consistency check of our formulae since the fixed point sets often differ between the two setups.

 
\section{Examples: product spaces}\label{sec:examplesI}

The simplest examples to consider are product spaces,  where in particular the normal bundles to any fixed point sets are trivial; that is, we take $c_1 (L_i)=0$. 
The condition that $c_1 (L_i)=0$ gives a simplified form of the on-shell action in \eqref{main}, which then reads
\begin{align}\label{simplifiedaction}
I = &~    \bigg\{\sum_{\text{dim  }  0} \frac{\chi(\sigmaone\epsilon_1+\sigmatwo\epsilon_2+\sigmathree\epsilon_3)^3}{\epsilon_1\epsilon_2\epsilon_3} \\
&+3\sum_{\text{dim  }  2} \frac{\sigmaone\sigmatwo
(\sigmaone\epsilon_1+\sigmatwo\epsilon_2)^2}{\epsilon_1\epsilon_2}\chi(\Sigma_g)
 + 3\sum_{\text{dim  }  4} \big(2 \chi(B_4)+3 \tau(B_4)\big)\bigg\}\frac{F_{S^5}}{27}\, .\nonumber
\end{align}
Here recall that $\chi(X)$ denotes the Euler characteristic of the space $X$ and $\tau(X)$ is its signature. 

In this section, we will consider seven distinct examples, each having different contributions from the localization formula \eqref{simplifiedaction}, before comparing them to known field theory results. The first example we consider is the Euclidean hyperbolic black hole found in \cite{Alday:2014fsa} with topology $M_6=\mathbb{R}^2\times \mathbb{H}^4\cong \R^6$, which has a single fixed point at the origin. We then move on to consider solutions with topology $\mathbb{R}^4\times \Sigma$, where $\Sigma$ is a Riemann surface. This is split into two cases where we take the Riemann surface to be a two-dimensional fixed point set or in the case of a sphere we take the $S^2$ to be rotating. The latter means that the R-symmetry Killing vector $\xi$ has a component tangent to the two-sphere and $M_6$ has two distinct isolated fixed points.
Our final examples are of the form $\mathbb{R}^2 \times B_4$, and include the Chow black hole and the holographic duals of 5d SCFTs on compact four-manifolds mentioned in the introduction. 
 These examples, apart from being particularly simple, are also interesting as there are known solutions or field theory results with which we can compare our results. In all cases where field theory results exist, we find perfect agreement.

We note that some of the rotating black hole solutions we consider here and in the subsequent sections are complex, in particular they have complex rotation parameters. Strictly speaking, in deriving our formulae we have assumed everything is real. Nevertheless, applying the same formulae gives the correct result for the free energies. It is natural to conjecture that the reality conditions we assumed in section \ref{sec:SUGRA} are not necessary, and relaxing this assumption will not change the final form of the fixed point formula. This has also been noticed in 4d and 5d, see for example \cite{BenettiGenolini:2024lbj}.


\subsection{1/2 BPS black hole: \texorpdfstring{$M_6=\R^6$}{M6=R6} }\label{sec:hyperbolicBH}

Our first example is the hyperbolic black hole constructed in \cite{Alday:2014fsa}. There it was used to holographically compute the supersymmetric R\'enyi entropy of 5d USp$(2N)$ gauge theories. Globally the space is the direct product $\mathbb{R}^2\times \mathbb{H}^4$ where the $\mathbb{H}^4\cong\mathbb{R}^4$ has an $S^3$ slicing. The solution can be defined on $\mathbb{R}^6$ with the action of the maximal torus U$(1)^3\subset \mathrm{U}(1)\times \mathrm{SO}(1,4)$, making it naturally into $\mathbb{R}^6\cong \mathbb{R}^2\oplus \mathbb{R}^2_1\oplus \mathbb{R}^2_2$. We are interested in the boundary being the $n$-branched five-sphere which is conformal to $S^1\times \mathbb{H}^4$, where the Euclidean time circle has period $2\pi n$. Note that for $n=1$ this is nothing but Euclidean AdS$_6$.  

To proceed, we introduce $2\pi$ periodic coordinates on each of the $\mathbb{R}^2$ factors. Let $\partial_{\tau}$ be the Killing vector which generates the rotations of $\mathbb{R}^2$ 
and let $\partial_{\varphi_i}, i\in\{1,2\}$ denote the two Killing vectors generating the rotations of the $\mathbb{R}^2_i$ factors. In these coordinates the R-symmetry Killing vector is:\footnote{Since this background preserves more supersymmetry, quadruple of the required amount to use our localization formula, there are actually four naturally defined independent choices of the Killing vector field. In appendix \ref{app:half} we discuss this example in more detail, explaining the choices of the $\sigma$'s and why despite the different choices of vector field the final result is independent of the choice of $\xi$. }
\begin{equation}
\xi= \frac{1}{n} \partial_{\tau}+\partial_{\varphi_1}+\partial_{\varphi_2}\,.
\end{equation}
There is a single fixed point of the Killing vector at the centre of each of the $\mathbb{R}^2$ factors. In this case, the weights are easily read off from the Killing vector, and one finds 
\begin{equation}
(\epsilon_1,\epsilon_2,\epsilon_3)=\big(\tfrac{1}{n},1,1\big)\, .
\end{equation}
It remains to give  the signs $\sigma^{(i)}$ associated to the projection conditions. For our choice of Killing vector we have $\sigma^{(i)}=1$ for all $i$. In general, as we explain in appendix \ref{app:half}, the Killing vector field takes the universal form
\begin{equation}
\xi= \chi \Big(\frac{\sigmaone}{n} \partial_{\tau}+\sigmatwo\partial_{\varphi_1}+\sigmathree\partial_{\varphi_2}\Big)\,.
\end{equation}
There are four inequivalent distributions of signs, up to an overall orientation change;  nonetheless, for each of the four inequivalent choices one finds the on-shell action to be
\begin{equation}
    I = \frac{F_{S^5}}{27}\frac{\chi(\sigmaone\epsilon_1+\sigmatwo\epsilon_2+\sigmathree\epsilon_3)^3}{\epsilon_1\epsilon_2\epsilon_3}  = F_{S^5}\frac{(2n+1)^3}{27 n^2}\,.
    \end{equation}
This agrees with the free energy computed in  \cite{Alday:2014fsa} ({\it cf}. equation (2.37)). Notice also that, for $n=1$, one indeed finds the Euclidean AdS$_6$ result as expected.


\subsection{Topologically twisted 5d SCFTs on a Riemann surface: \texorpdfstring{$M_6=\R^4\times \Sigma_g$}{M6=R4xSg}}\label{sec:topRiemann}

We now turn our attention to solutions with topology $M_6=\R^4\times \Sigma_g$. Their conformal 
boundary is $M_5=\partial M_6 = S^3\times \Sigma_g$, where we view $\R^4$ as a ball with boundary 
three-sphere $S^3$. These solutions are holographically dual to the twisted compactification of the 5d Seiberg USp$(2N)$ SCFTs on the Riemann surface, and our on-shell action computes the free energy of the resultant 3d $\mathcal{N}=2$ SCFT on a squashed $S^3$.

Let us write $\R^4=\R^2_1\oplus \R^2_2$, and let $\varphi_i$ be $2\pi$-periodic coordinates such that $\partial_{\varphi_i}$ generates the U$(1)$ symmetry of $\R^2_i$. Using this basis, we write the R-symmetry Killing vector as $\xi = b_1\partial_{\varphi_1}+b_2\partial_{\varphi_2}$. For $b_1b_2\neq 0$, the fixed point set is $\mathscr{F}_2=\Sigma_g$, at the origin of the $\R^4$ factor, and therefore our localization formula \eqref{simplifiedaction} picks up a single contribution from this two-dimensional fixed point set. The weights of the action are simply the $b_i$'s and therefore the on-shell action is
\begin{equation}\label{Ib1b2}
\begin{split}
I= \ & \chi_{\mathbb{R}^4}\chi(\Sigma_g) \frac{( b_1+\chi_{\mathbb{R}^4} b_2)^2}{b_1 b_2}\frac{F_{S^5}}{9}\, ,
\end{split}
\end{equation}
where we have introduced the sign $\chi_{\mathbb{R}^4} \equiv \sigmaone \sigmatwo\in \{\pm 1\} $. The result is the large $N$ free energy for the Seiberg  USp$(2N)$ SCFTs compactified on a Riemann surface, with a topological twist and placed on a squashed U$(1)^2$-invariant three-sphere with the squashing parameters $b_i$. As expected, this 
free energy takes the same form as that in 
$D=4$ dimensions, {\it cf.} equation (23) in \cite{BenettiGenolini:2024xeo}, for example. 
Introducing the standard squashing variables 
\begin{align}\label{eq:squashingdef}
b\equiv \sqrt{\frac{b_1}{b_2}}\, , \qquad Q\equiv \tfrac{1}{2}\left(b+b^{-1}\right) \quad \Rightarrow \quad Q^2=\frac{(b_1+b_2)^2}{4b_1b_2}\, ,
\end{align}
and furthermore setting $\sigmaone=\sigmatwo$ 
so that $\chi_{\R^4}=1$, the result \eqref{Ib1b2} may be written as
\begin{align}\label{IR4Sg}
I = -\frac{8}{9}(g-1)Q^2  F_{S^5}\, .
\end{align}
This agrees precisely with the dual large $N$ field theory result, 
given in  (3.17) of \cite{Crichigno:2018adf}. Furthermore setting $b_1=b_2$, hence 
turning off the squashing parameter, $Q=1$, we find that the on-shell action \eqref{IR4Sg}
agrees precisely with equation (4.31) of \cite{Bobev:2017uzs}, 
which is a supergravity result. As far as we are aware, squashed 
supergravity solutions corresponding 
to the free energy in 
\eqref{IR4Sg} with $b_1\neq b_2$ are not known, but assuming they exist we have successfully matched to the dual field theory result.


\subsection{\texorpdfstring{$M_6=\R^4\times S^2$}{M6=R4xS2}, where \texorpdfstring{$S^2$}{S2} is rotating}\label{sec:R4S2unfib}

We may further extend the above example by restricting to the
case that the Riemann surface $\Sigma_{g=0}=S^2$ is a sphere, and  then turning on an equivariant parameter for rotation along the $S^2$. Geometrically this means that the R-symmetry Killing vector $\xi$ now has a component that is tangent to the sphere.
From the point of view of the associated black object, this is a chemical potential for the angular momentum. We write the R-symmetry Killing vector as
\begin{align}\label{eq:xiR4S2}
\xi =   b_1\partial_{\varphi_1}+b_2\partial_{\varphi_2} +\varepsilon \partial_{\varphi} \, ,
\end{align}
where $\varphi$ is the angular coordinate on $S^2$ with period $2\pi$ and $\varphi_i$ are as before. 
For $b_1b_2\varepsilon\neq 0$ there are now two isolated fixed points, which are located at the origin of $\R^4$ and at each one of the two poles of the two-sphere. We will call these the North (N) and South (S) pole contributions. When applying \eqref{simplifiedaction}, we find that the on-shell action takes the form 
\begin{equation}\label{eq:IR4S2initial}
    I = \frac{ F_{S^5}}{27} \left[\chi_N\frac{\big(\sigmaone_N \epsilon_{1}^N +\sigmatwo_N\epsilon_2^{N}+\sigmathree_N \epsilon^{N}_{3}\big)^3}{\epsilon^N_{1}\epsilon^{N}_{2} \epsilon^{N}_{3}} +\chi_S\frac{\big(\sigmaone_S \epsilon^{S}_{1} +\sigmatwo_S\epsilon^{S}_{2}+\sigmathree_S \epsilon^{S}_{3}\big)^3}{\epsilon^S_{1}\epsilon^{S}_{2} \epsilon^{S}_{3}}  \right ]
\end{equation}
where the weights and chiralities now carry an index for the associated fixed point. 

\begin{figure}[h!]
\begin{center}
\tdplotsetmaincoords{70}{50}
\begin{tikzpicture}
		[tdplot_main_coords,
			cube/.style={very thick,black},
			axisb/.style={->,blue,thick},
            axisr/.style={->,red,thick},
            axisg/.style={->,Green,thick},
			inf/.style={dashed,black}]

\shade[top color= Green!20,  bottom color =Green!50,fill opacity=0.2,shading angle=-25](0,0,0)-- (0,3,0) -- (0,3,3)--(0,0,3);
	 \shade[top color= white,  bottom color =red!50,fill opacity=1,shading angle=-25](0,3,0)-- (0,0,0) -- (3,0,0)--(3,3,0);
	\shade[top color= white,  bottom color =blue!50,fill opacity=0.7,shading angle=-25](0,0,3)-- (0,0,0) -- (3,0,0)--(3,0,3);
    \shade[left color= Green!0,  right color =Green!70,fill opacity=0.4,shading angle=-25](3,0,0)-- (3,3,0) -- (3,3,3)--(3,0,3);

	\draw[black](0,0,0)--(0,0,3);
    \draw[black](3,0,0)--(3,0,3);
    \draw[inf] (0,0,3) -- (0,3,3)--(0,3,0)--(3,3,0)--(3,3,3)--(3,0,3)--(0,0,3);
    \draw[inf](0,0,3)--(0,0,4);
    \draw[inf](3,0,3)--(3,0,4);
    \draw[inf](0,3,3)--(0,3,4);
    \draw[inf](3,3,3)--(3,3,4);
   \draw[inf] (3,3,0)--(3,4,0);
   \draw[inf](0,3,0)--(0,4,0);
\draw[inf] (0,3,3)--(3,3,3);
\draw[inf] (0,3,3)--(0,4,3);
\draw[inf] (3,3,3)--(3,4,3);
	 \node at (3,1.5,1.75) {\textcolor{Green}{{\scriptsize$\sigma^{(3)}_N$}}};
     \node at (1.5,0,1.2) {\textcolor{blue}{{\scriptsize $\sigma^{(2)}$}}};
     \node at (0,1.7,1.7) {\textcolor{Green}{{\scriptsize $\sigma^{(3)}_{S}$}}};
     \node at (1.5,1.9,0){\textcolor{red}{{\scriptsize $\sigma^{(1)}$}}};

 \draw[axisg] (3,1.5,1.5)--(4.5,1.5,1.5) node[anchor= south]{$v_3$};
     \draw[axisb] (1.5,0,1.5)--(1.5,-1.75,1.5) node[anchor=east]{$v_2$};
     
 \draw[axisr] (1.5,1.5,0)--(1.5,1.5,-1.5
) node[anchor=east]{$v_1$};
\draw[black] (0,3,0)-- (0,0,0) -- (3,0,0)--(3,3,0);
       
    \draw[axisg] (0,1.5,1.5)--(-1.5,1.5,1.5) node[anchor=south west]{$v_4$};

\end{tikzpicture}
\end{center}
\caption{The toric diagram for $\mathbb{R}^4\times S^2$. It is infinite in extent in two directions denoted by the dashed lines.}\label{fig:R4S2}
\end{figure}
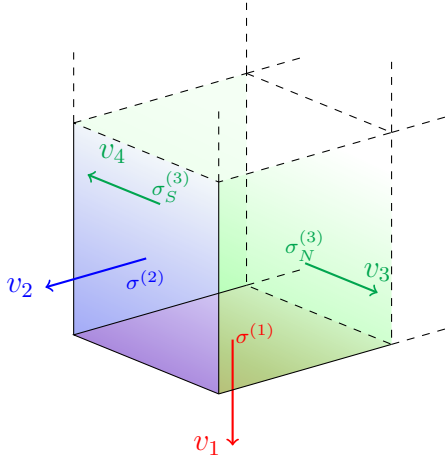

The space has a natural U$(1)^3$ toric action and it is useful to encode the information about the space in terms of a toric diagram specified by a set of vectors, see figure \ref{fig:R4S2}. From the toric vectors, the weights appearing in the localization formula can be extracted using the formulae in appendix \ref{app:weights}. The geometry in this example is simple enough that we can easily work out the weights without this technology; however, for completeness and a lighter introduction to the toric computations of later sections, we will spend a few lines presenting the toric computations here. The toric data is specified by four outward pointing normals, which we take to be 
\begin{equation}
  \begin{split}
    v_1 =(1, 0, 0)\, , \quad
       v_2 = (0, 1,0) \, ,\quad 
       v_3 = (0,0,1)\, ,\quad 
    v_4 = (0,0,-1)\, .
  \end{split}
\end{equation}
The diagram consists of four faces $D_i$, which correspond to the four distinct four-dimensional fixed point sets associated to the three rotation vectors. This is a cuboid of infinite extent in two directions, with two adjacent faces missing, see figure \ref{fig:R4S2}.

Each face, in addition to being labelled by the toric vector, is accompanied by a sign choice $\sigma^{(i)}$ which defines the projection condition of the spinor on the normal bundle of the associated face, as in section \ref{sec:charges}. At each vertex in the diagram, the three signs for the intersecting faces fix the chirality of the spinor at the fixed point. 
The proliferation of signs in \eqref{eq:IR4S2initial}, which in principle has $2^3=8$ distinct cases, can therefore be reduced. In particular, the signs for the two $\mathbb{R}^2$ factors are fixed to be the same at both fixed points, $\sigma_{N}^{(i)}=\sigma_{S}^{(i)}\equiv\sigma^{(i)}$ for $i=1,2$. We are therefore left with four independent choices which can be broken up into $\sigmaone\sigmatwo=\pm1$ and whether $\sigmathree_N=\pm\sigmathree_S$. 
Finally the weights, using the formulae in appendix \ref{app:weights}, are computed to be
\be
(\epsilon_1,\epsilon_2,\epsilon_3)=
\begin{cases}
(b_1, b_2,\varepsilon) & \text{at North pole},\\
(b_1, b_2,-\varepsilon)&\text{at South pole} \, .
\end{cases}
\ee
Putting everything together we find that the on-shell action is given by
\begin{equation}\label{eq:R4S2trivfib}
  I = \frac{ F_{S^5}}{27}\frac{\sigmaone\sigmatwo}{b_1b_2\varepsilon} \left[\sigma^{(3)}_{N}\big(\sigmaone b_1 +\sigmatwo b_2+\sigma^{(3)}_{N} \varepsilon\big)^3 -\sigmathree_{S} \big(\sigmaone b_1 +\sigmatwo b_2-\sigma^{(3)}_{S} \varepsilon\big)^3 \right ] \, .
\end{equation}
In addition to the on-shell action we can compute the magnetic flux threading through the two-sphere using \eqref{PhiF}, finding the simple result
\begin{equation}
m=-\sigmathree_N -\sigmathree_S\, .
\end{equation}

Before we conclude this section, let us study in a little more detail the distinct types of solution. To proceed, let us define $\sigmaone\sigmatwo=\chi_{\mathbb{R}^4}$ and study the two cases of $\sigmathree_N=\pm\sigmathree_S$ independently. For $\sigmathree_N=\sigmathree_S$, the spinor has the same chirality at the two poles of the $S^2$, and supersymmetry is preserved using a twist on the $S^2$. For the on-shell action and magnetic charge we find
\be \label{IR4S2rotating}
I=\frac{2 F_{S^5} \chi_{\mathbb{R}^4}}{27b_1 b_2 }\big(3(b_1+\chi_{\mathbb{R}^4}b_2)^2 +\varepsilon^2\big) \, ,\quad m=-2 \sigmathree_N\, ,
\ee
respectively. One can smoothly take the $\varepsilon\rightarrow 0$ limit, turning off the angular momentum chemical potential, and match the result with the one of the topologically trivial Riemann surface of the previous section, see equation \eqref{Ib1b2}, upon using $\chi(\Sigma_{g=0})=2$. Furthermore, the magnetic charge correctly gives the topological result derived in section \ref{sec:gaugefield}, as required. Thus we are describing a topological twist on the $S^2$.  

For the other choice of sign, the spinor on the $S^2$ at the two poles has a different chirality; this is known as an anti-twist \cite{Ferrero:2021etw}. The on-shell action and magnetic charge are
\be
\begin{split}\label{eq:IR4S2anti}
I=
\frac{2 F_{S^5} \chi_{\mathbb{R}^4}\sigmathree_N}{27b_1 b_2 \varepsilon}\big(\sigmaone b_1  +\sigmatwo b_2+ \sigmathree_N\varepsilon\big)^3\, ,\quad m=0\, ,
\end{split}
\ee
respectively. Note that, in this case, we cannot turn off the angular momentum chemical potential, and, in addition, the magnetic charge vanishes! It is easy to understand why we cannot turn off $\varepsilon$ by recalling that, on fixed point sets, the spinor has a definite chirality. In the $\varepsilon=0$ limit, the $S^2$ becomes a two-dimensional fixed point set which must have a definite chirality. This is inconsistent with having a different chirality at the two poles as the anti-twist requires. From figure \ref{fig:R4S2}, one sees this by sandwiching the two green faces together and requiring that they are compatible. This class of solution is reminiscent of the Kerr--Newman type solutions in 4d. 

The field theory duals of topologically twisted black holes in maximally supersymmetric AdS compactifications were studied in \cite{Hosseini:2021mnn}. The result is that the log of the partition function on a compact manifold $M$ is given by the sum of so-called \emph{gravitational blocks} \cite{Hosseini:2019iad} as
\begin{equation}
    \log Z_{M}(\Delta,\epsft)=-F_{M}=\sum_{\sigma}\mathbf{\mathcal{B}}(\Delta^{(\sigma)},\epsft^{(\sigma)})\, .
\end{equation}
The way the blocks are summed depends on the choice of gluing. For the theory on $S^3_b\times S^2_{\epsft}$, there are two relevant gluings, the A-gluing and the identity gluing, the former relevant for the topologically twisted solutions and the later for the Kerr--Newman solutions.

For the topologically twisted theory the free energy is given by \cite{Hosseini:2021mnn}
\begin{equation}
\begin{split}
    F_{S^3_b \times S^2_{\epsft}}(\Delta_i,\mathfrak{t}_i, \epsft |b)&= \frac{8}{27}\frac{Q^2}{\epsft}\left [F_{S^5} \left (\Delta_i + \frac{\epsft}{2}\mathfrak{t}_i \right )- F_{S^5} \left (\Delta_i - \frac{\epsft}{2}\mathfrak{t}_i \right )\right ]\,, 
\end{split}
\end{equation}
with 
\begin{equation}
    F_{S^5}(\Delta_i ) = \left (\Delta_1 \Delta_2\right )^{\frac{3}{2}}F_{S^5}\,,
\end{equation}
and $Q$ the squashing parameter introduced in \eqref{eq:squashingdef}. The $\Delta_i$'s are constrained chemical potentials and the $\mathfrak{t}_i$'s are the fluxes for the flavour symmetry. Since we are considering the theory without an additional vector multiplet we must set $\Delta_1=\Delta_2=1$ and similarly $\mathfrak{t}_1=\mathfrak{t}_2=1$ for the topologically twisted theory. The free energy is 
\begin{equation}
    F_{S^3_b\times S^2_{\epsft}}=\frac{2 F_{S^5} Q^2}{27}(12+\epsft^2)\, .
\end{equation}
To compare with our result in equation \eqref{IR4S2rotating}, we need to relate the gravity parameters with the field theory parameters. Introducing the squashing parameter variables in \eqref{eq:squashingdef} and defining 
\begin{equation}
\varepsilon=\frac{b_1+b_2}{2}\epsft\, ,
\end{equation}
we find that this gives precisely \eqref{IR4S2rotating} for $\chi_{\mathbb{R}^4}=1$.


\subsection{\texorpdfstring{$M_6=\mathbb{R}^2 \times S^2_{\varepsilon_1}\times S^2_{\varepsilon_2}$}{M6=R2xS2xS2}}

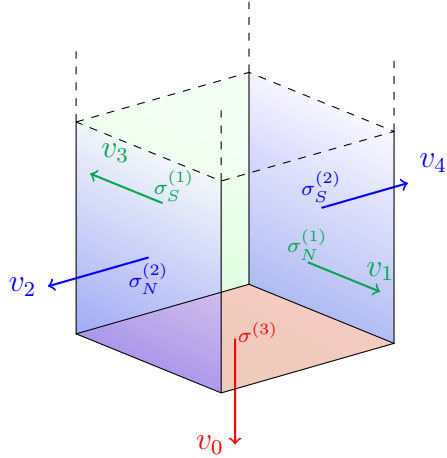
\begin{figure}[h!]
\begin{center}
\tdplotsetmaincoords{70}{50}
\tdplotsetmaincoords{70}{50}
\begin{tikzpicture}
		[tdplot_main_coords,
			cube/.style={very thick,black},
			axisb/.style={->,blue,thick},
            axisr/.style={->,red,thick},
            axisg/.style={->,Green,thick},
			inf/.style={dashed,black}]

    
\shade[top color= Green!20,  bottom color =Green!50,fill opacity=0.2,shading angle=-25](0,0,0)-- (0,3,0) -- (0,3,3)--(0,0,3);
	 \fill[red!50,fill opacity=0.5](0,3,0)-- (0,0,0) -- (3,0,0)--(3,3,0);
	\shade[top color= white,  bottom color =blue!50,fill opacity=0.7,shading angle=-25](0,0,3)-- (0,0,0) -- (3,0,0)--(3,0,3);
    \shade[top color= white,  bottom color =blue!50,fill opacity=0.7,shading angle=-25](0,3,3)-- (0,3,0) -- (3,3,0)--(3,3,3);
    \shade[left color= Green!0,  right color =Green!30,fill opacity=0.2,shading angle=-25](3,0,0)-- (3,3,0) -- (3,3,3)--(3,0,3);

    \draw[axisr] (1.5,1.5,0)--(1.5,1.5,-1.5) node[anchor=east]{$v_0$};
    \draw[axisg] (0,1.5,1.5)--(-1.5,1.5,1.5) node[anchor=south west]{$v_3$};
    \draw[axisb] (1.5,3,1.5)--(1.5,4.5,1.5) node[anchor=south west ]{$v_4$};

	\draw[black](0,3,0)-- (0,0,0) -- (3,0,0)--(3,3,0)--(0,3,0);
	\draw[black](0,0,0)--(0,0,3);
    \draw[black](3,0,0)--(3,0,3);
    \draw[black] (0,3,0)--(0,3,3);
    \draw[black](3,3,0)--(3,3,3);
    \draw[inf] (0,0,3) -- (0,3,3)--(3,3,3)--(3,0,3)--(0,0,3);
    
    \draw[inf](0,0,3)--(0,0,4);
    \draw[inf](3,0,3)--(3,0,4);
    \draw[inf](0,3,3)--(0,3,4);
    \draw[inf](3,3,3)--(3,3,4);

	 \node at (3,1.5,1.75) {\textcolor{Green}{{\scriptsize$\sigma^{(1)}_N$}}};
     \node at (1.5,0,1.2) {\textcolor{blue}{{\scriptsize $\sigma^{(2)}_N$}}};
     \node at (0,1.7,1.7) {\textcolor{Green}{{\scriptsize $\sigma^{(1)}_{S}$}}};
     \node at (1.5,1.9,0){\textcolor{red}{{\scriptsize $\sigma^{(3)}$}}};
\node at (1.5,3,1.8){\textcolor{blue}{{\scriptsize $\sigma^{(2)}_S$}}};

 \draw[axisg] (3,1.5,1.5)--(4.5,1.5,1.5) node[anchor= south]{$v_1$};
     \draw[axisb] (1.5,0,1.5)--(1.5,-1.75,1.5) node[anchor=east]{$v_2$};
	
\end{tikzpicture}

\end{center}
\caption{The toric diagram for $\mathbb{R}^2\times S^2_{\varepsilon_1}\times S^2_{\varepsilon_2}$. It is infinite in extent in one direction, following the dashed vertical lines. }\label{fig:R2S2S2}
\end{figure}

So far, the topologies we have considered have not contained a compact four-cycle over which we can define an electric charge. In this section we will consider an example which admits such a compact four-cycle. Similar to the previous example, we will consider the direct product of $\mathbb{R}^2$ with two two-spheres and turn on an equivariant parameter for rotations for each $S^2$. We take the R-symmetry Killing vector to be
\begin{equation}
    \xi= \varepsilon_1 \partial_{\varphi_1}+\varepsilon_2 \partial_{\varphi_2}+b\mskip1mu \partial_{\tau}\, ,
\end{equation}
where $\partial_\tau$ rotates $\mathbb{R}^2$ and $\partial_{\varphi_i}$ rotates one copy of the $S^2$. There are four isolated fixed points located at the centre of $\mathbb{R}^2$ and the poles of the two two-spheres. The toric data for the geometry is specified by the five outward pointing normals
\begin{equation}
\begin{split}
\begin{array}{lll}
    &v_0=(0,0,1)\,,\quad &v_1=(1,0,0)\, ,~~\quad v_2=(0,1,0)\,,\\
 &v_3=(-1,0,0)\, ,\quad &v_4=(0,-1,0)\, ,
 \end{array}
 \end{split}
\end{equation}
see figure \ref{fig:R2S2S2}. The vectors define five faces with $v_0$ defining the centre of $\mathbb{R}^2$, $v_1$ and $v_3$ the north and south poles of the first sphere respectively, and $v_2$ and $v_4$ the north and south poles of the second sphere. Observe that the fixed points are the four corners of the base. 

The weights at the fixed points are simple to compute and we find
\begin{equation}
    (\epsilon_1,\epsilon_2,\epsilon_3)=
    \begin{cases}
         (\varepsilon_1,\varepsilon_2,b) & \text{at NN}\, ,\\
          (-\varepsilon_1,\varepsilon_2,b) & \text{at SN}\, ,\\
           (\varepsilon_1,-\varepsilon_2,b) & \text{at NS}\, ,\\
            (-\varepsilon_1,-\varepsilon_2,b) & \text{at SS}\, .
    \end{cases}
\end{equation}
Inserting this into our  formula \eqref{main} we find that the on-shell action is 
\begin{align}
        I&=\frac{F_{S^5}\sigma^{(3)}}{27 b \varepsilon_1\varepsilon_2}\Big[\sigma^{(1)}_N\sigma^{(2)}_N\big(\sigmaone_N\varepsilon_1 +\sigmatwo_N\varepsilon_2+\sigmathree b\big)^3 -\sigma^{(1)}_S\sigma^{(2)}_N\big(-\sigmaone_S\varepsilon_1 +\sigmatwo_N\varepsilon_2+\sigmathree b\big)^3\nonumber\\
        &-\sigma^{(1)}_N\sigma^{(2)}_S\big(\sigmaone_N\varepsilon_1 -\sigmatwo_S\varepsilon_2+\sigmathree b\big)^3+\sigma^{(1)}_S\sigma^{(2)}_S\big(-\sigmaone_S\varepsilon_1 -\sigmatwo_S\varepsilon_2+\sigmathree b\big)^3\Big]\, .
    \end{align}
There are a plethora of signs; however, there are really only three independent cases to consider depending on the relative signs of $\sigma^{(i)}_N$ and $\sigma^{(i)}_{S}$. Before we consider the independent cases, let us compute the magnetic and electric charges and also the conserved charge associated to $* B$. 

For the magnetic charges we need to integrate $c_1(F)$ over compact two-cycles using the polyform \eqref{PhiF}. Clearly the two-cycles are the two two-spheres located at the centre of $\mathbb{R}^2$ and at either pole of the other sphere. Using \eqref{PhiF} we find that the magnetic charges are
\begin{equation}
m_1=-\sigmaone_N-\sigmaone_S\, ,\quad m_2=-\sigmatwo_N-\sigmatwo_S\, .
\end{equation}
Next consider the electric charge $q$, \eqref{eq:electricdef}. Using the BV--AB theorem, the electric charge threading through the compact four-cycle $S^2\times S^2$ is
\begin{align}\label{eq:qS2S2}
    q&=\frac{3 \sigmathree}{2\varepsilon_1\varepsilon_2}\Big[\sigmaone_N \sigmatwo_N\big(\sigmaone_N\varepsilon_1+\sigmatwo_N \varepsilon_2+\sigmathree b\big)^2 -\sigma^{(1)}_S\sigma^{(2)}_N\big(-\sigmaone_S\varepsilon_1 +\sigmatwo_N\varepsilon_2+\sigmathree b\big)^2\nonumber\\
        &-\sigma^{(1)}_N\sigma^{(2)}_S\big(\sigmaone_N\varepsilon_1 -\sigmatwo_S\varepsilon_2+\sigmathree b\big)^2+\sigma^{(1)}_S\sigma^{(2)}_S\big(-\sigmaone_S\varepsilon_1 -\sigmatwo_S\varepsilon_2+\sigmathree b\big)^2\Big]\, .
\end{align}
Finally we may compute the conserved charge associated to $* B$ using the polyform given in equation \eqref{eq:Phi*B}. We find the simple result
\begin{equation}
    \Bc=-\frac{9 \ii}{8}(\sigmaone_N+\sigmaone_S)(\sigmatwo_N+\sigmatwo_S)=-\frac{9 \ii}{4} m_1 m_2\, .
\end{equation}
The latter equality is obvious given that on a fixed point set $\Phi^{*B}\propto \Phi^F\wedge \Phi^F$ -- see section \ref{sec:fps}. Notice that, for $\sigma^{(i)}_N=\sigma^{(i)}_S$, we can send $\varepsilon_i\rightarrow0$ in all of our observables. For the anti-twist case, $\sigma^{(i)}_N=-\sigma^{(i)}_S$, the $\varepsilon_i\rightarrow 0$ limit is not smooth and the argument presented around equation \eqref{eq:IR4S2anti} about the obstruction to taking this limit equally applies. No field theory results nor explicit solutions are known in the general case with both equivariant parameters non-zero, however, by setting one of the equivariant parameters to vanish, without loss of generality, $\varepsilon_2$, we can make contact with field theory results in \cite{Hosseini:2021mnn}. Rather than taking the limit here we study the more general case where we have a single rotating two-sphere and the direct product with a generic Riemann surface in the next section.

\subsection{\texorpdfstring{$M_6 = \mathbb{R}^2 \times S^2_{\varepsilon} \times \Sigma_g $}{M6=R2xS2xSg}}\label{sec:R2S2Sg}

In this section we consider a geometry of the form $M_6 = \mathbb{R}^2 \times S^2_{\varepsilon} \times \Sigma_g $, where we turn on an equivariant parameter for rotations on the $S^2$. In the previous section we have computed the electric charge by using isolated fixed points. One of the novel features of this geometry is that the computation of the electric charge will involve integrating over two-dimensional fixed point sets. There are also field theory results within this class with which we may compare.

We take the R-symmetry vector field to be
\begin{equation}
    \xi= b \mskip1mu\partial_{\tau}+ \varepsilon \partial_{\varphi}\,, 
\end{equation}
where $\partial_{\tau}$ acts on $\mathbb{R}^2$ and $\partial_{\varphi}$ acts on the $S^2$ only. There are two connected two-dimensional fixed points sets, located at the centre of $\mathbb{R}^2$ and the poles of the two-sphere, with the Riemann surface fixed at both. 

The weights at the two fixed point sets are easy to determine using the results of appendix \ref{app:toricdata}, 
\begin{equation}
    (\epsilon_1,\epsilon_2)=\begin{cases}
        (b ,\varepsilon) & \text{at North pole},\\
        (b,-\varepsilon) & \text{at South pole},
    \end{cases}
\end{equation}
and from \eqref{main} we find that the on-shell action is given by
\begin{equation}\label{eq:IS2Sigmag}
    I=\frac{F_{S^5} \chi(\Sigma_g)\sigmaone}{9 b\varepsilon}\Big[(\sigmatwo_N-\sigmatwo_S)(b^2+\varepsilon^2)+4\sigmaone  b\mskip1mu \varepsilon\Big]\, ,
\end{equation}
where $\sigmatwo_{N}$ and $\sigmatwo_S$ are signs for the projection conditions for the spinor at the north and south pole of the two-sphere, and $\sigmaone$ is a sign for the projection condition for the spinor at the centre of $\mathbb{R}^2$.

We have two magnetic charges, and, interestingly in this case, they are determined in different ways from our localization formulae. For the magnetic charge threading through the Riemann surface the magnetic charge is computed from the topological constraint in section \ref{sec:gaugefield}, and we find
\begin{equation}
    m_{\Sigma_g}= -\sigmathree \chi(\Sigma_g)\, .
\end{equation}
For the magnetic charge threading through the two-sphere, we use \eqref{PhiF} and \eqref{XP} to find
\begin{equation}
    m_{S^2}=-\sigmatwo_N-\sigmatwo_S\, .
\end{equation}
We can also consider the two other conserved charges. For the electric charge we find
\begin{equation}\label{eq:qS2Sg}
    q=\frac{3 \sigmaone \chi(\Sigma_g)}{\varepsilon}\big(2\varepsilon+\sigmaone (\sigmatwo_N-\sigmatwo_S)b\big)-\frac{1}{2\pi\varepsilon}\left(\int_{\Sigma_g}C\big|_{N}-\int_{\Sigma_g}C\big|_{S}\right)\, .
\end{equation}
Note that our formulae do not allow us to determine the final integral. However if we specialise to the case that $\Sigma_g=S^2$ then we may take the $\varepsilon_2=0, \sigmatwo_N=\sigmatwo_S$ limit of equation \eqref{eq:qS2S2} and we find that the result is precisely the first term in equation \eqref{eq:qS2Sg}, i.e. the contributions from $C$ drop out. It would be interesting to see if this is a generic property or whether having a non-trivial contribution is possible.
Finally for the charge $\Bc$ we have
\begin{equation}
    \Bc= -\frac{9\ii}{4} m_{\Sigma_g}m_{S^2}\, .
\end{equation}

Explicit solutions of this form have been found in \cite{Hosseini:2020wag} by first performing a reduction on the Riemann surface to four-dimensional gauged supergravity. Rather than compare to these supergravity solutions we will compare with the field theory results in \cite{Hosseini:2021mnn}, since they are more general. There, the free energy has been computed for both the twist and anti-twist case, {\it cf.} equations (4.74) and (5.101) in \cite{Hosseini:2021mnn}. The results are written in terms of the free energy of the theory on $S^3\times \Sigma_g$:
\begin{equation}
     F_{S^3 \times \Sigma_g}(\Delta_i , \mathfrak{s}_i ) = \frac{8}{27\pi^2}F_{S^5}\sum_{i=1}^2 \mathfrak{s}_i \frac{\partial}{\del \Delta_i} (\Delta_1 \Delta _2 )^{\tfrac{3}{2}}\, ,
\end{equation}
where the $\mathfrak{s}_i$ are the magnetic charges on the Riemann surface and are constrained via $\sum_{i=1}^{2} \mathfrak{s}_i=\chi(\Sigma_g)$. Since we are working in Romans supergravity we must set $\Delta_1=\Delta_2\equiv \Delta$ and $\mathfrak{s}_1=\mathfrak{s}_2=\tfrac{\chi(\Sigma_g)}{2}$ in the following.

For the twist case ({\it cf.} equation (4.74) of \cite{Hosseini:2021mnn}) one has
\begin{equation}
    \log Z_{(S^2_{\epsilon }\times S^1 )\times \Sigma_g} = -\frac{\pi}{2\epsilon}\left [F_{S^3 \times \Sigma_g} \left (\Delta_i + \frac{\epsilon}{2}\mathfrak{t}_i, \mathfrak{s}_i \right )-F_{S^3 \times \Sigma_g} \left (\Delta_i - \frac{\epsilon}{2}\mathfrak{t}_i , \mathfrak{s}_i \right )\right ]\, ,
\end{equation}
where 
\begin{equation}
    \sum_{i=1}^{2}\Delta_i=2\pi\,,\qquad \sum_{i=1}^{2}\mathfrak{t}_i=2\, .
\end{equation}
For the Romans theory, we set $\Delta=\pi$ and $\mathfrak{t}_1=\mathfrak{t}_2=1$ and the final result is
\begin{equation}
    \log Z_{(S^2_{\epsilon }\times S^1 )\times \Sigma_g}=\frac{4 F_{S^5}\chi(\Sigma_g)}{9}\, , 
\end{equation}
which agrees precisely with \eqref{eq:IS2Sigmag} for the twist case $\sigmatwo_N=\sigmatwo_S$. 

For the anti-twist case, $\sigmatwo_N=-\sigmatwo_S$, \cite{Hosseini:2021mnn} find that the free energy is given by 
\begin{equation}
        \log Z_{(S^2_{\epsilon }\times S^1 )\times \Sigma_g} = -\frac{\pi}{2\epsilon}\left [F_{S^3 \times \Sigma_g} \left (\Delta_i + \frac{\epsilon}{2}\mathfrak{t}_i, \mathfrak{s}_i \right )+F_{S^3 \times \Sigma_g} \left (\Delta_i - \frac{\epsilon}{2}\mathfrak{t}_i , \mathfrak{s}_i \right )\right ]\, ,
\end{equation}
with 
\begin{equation}
    \sum_{i=1}^{2}\Delta_i=2\pi+\epsilon\, ,\quad \sum_{i=1}^{2}\mathfrak{t}_i=0\, .
\end{equation}
For the Romans theory, one finds that the free energy is 
\begin{equation}
     \log Z_{(S^2_{\epsilon }\times S^1 )\times \Sigma_g} =\frac{F_{S^5}\chi(\Sigma_g)(2\pi +\epsilon)^2}{9\pi \epsilon}\, .
\end{equation}
Identifying
\begin{equation}
    \epsilon= 2\pi \frac{\varepsilon}{b}\sigmatwo_N \sigmaone\, ,
\end{equation}
one finds that we have perfect agreement between the gravity result and the large $N$ field theory result.


\subsection{Chow black hole: \texorpdfstring{$M_6=\mathbb{R}^2\times S^4_{\varepsilon_1,\varepsilon_2}$}{M6=R2xS4}}

We now want to consider the Euclidean Chow black hole \cite{Chow:2008ip}. This is an electrically charged black hole with two angular momenta for the two orthogonal two-plane rotations on $S^4$ in global Euclidean AdS$_6$. We take $M_6=\mathbb{R}^2\times S^4_{\varepsilon_1,\varepsilon_2}$ where we turn on equivariant parameters for the two U$(1)$'s inside the $S^4$. The R-symmetry Killing vector is given by
\begin{equation}
    \xi=\varepsilon_1 \partial_{\varphi_1}+\varepsilon_2\partial_{\varphi_2}+b\mskip1mu\partial_{\tau} \, ,
\end{equation}
with $\partial_{\varphi_i}$ rotating the two orthogonal two-planes in the $S^4 \subset\mathbb{C}_1\oplus \mathbb{C}_2\oplus  \mathbb{R}$. For $b\mskip1mu\varepsilon_1\varepsilon_2\neq0$ there are two isolated fixed points at the poles of the $S^4$ and centre of $\mathbb{R}^2$. The topology of the $S^4$ fixes the weights and projection conditions uniquely. Following a similar argument to \cite{BenettiGenolini:2023kxp} we consider a linearly embedded $S^2$ inside $S^4$, $S_i^2\subset \mathbb{C}_i\oplus \mathbb{R}$ at the centre of $\mathbb{R}^2$. This can be chosen to be invariant under $\xi$ and is trivial in homology. Therefore the integral of $F$ over such a two-cycle must vanish. Since the weights are fixed such that $\varepsilon_1^N\varepsilon_2^N=-\varepsilon_1^S\varepsilon_2^S$, let us without loss of generality take $\varepsilon_1^N=\varepsilon^S_1$ and  $\varepsilon_2^N=-\varepsilon_2^S$. The (vanishing) magnetic charge over such two-cycles fixes
\begin{equation}
    \sigmaone_N=\sigmaone_S\, ,\qquad \sigmatwo_N=-\sigmatwo_S\,.
\end{equation}
Therefore we have that the products $\sigmaone \varepsilon_1$ and $\sigmatwo \varepsilon_2$ are necessarily the same at the two poles. It is then easy to show that the on-shell action is 
\begin{equation}
    I=\frac{2F_{S^5}\chi}{27 b\mskip1mu \varepsilon_1\varepsilon_2}(\sigmaone\varepsilon_1+\sigmatwo \varepsilon_2+\sigmathree b)^3\, .
\end{equation}
The electric charge, upon application of \eqref{eq:Phi*F}, is given by
\begin{equation}
    q=\frac{6 \chi }{\varepsilon_1\varepsilon_2}(\sigmaone\varepsilon_1+\sigmatwo \varepsilon_2+\sigmathree b)^2\, , 
\end{equation}
whilst the conserved charge associated to $* B$ vanishes. 

Upon defining
\begin{equation}
    \varepsilon_i=\sigma^{(i)}\sigmathree b \mskip1mu\omega_{i}  \, ,
\end{equation}
we match with the entropy function of \cite{Choi:2018fdc,Cassani:2019mms} which reads
\begin{equation}
    I=\frac{2 F_{S^5}}{27 \omega_1\omega_2}(1+\omega_1+\omega_2)^3\, .
\end{equation}


\subsection{\texorpdfstring{$M_6 = \mathbb{R}^2 \times B_4$}{M6=R2xB4}}\label{sec:R2xB4}

As a final example within the topologically trivial class, we will consider the case where $M_6 = \mathbb{R}^2 \times B_4$. This is the geometry for Euclidean static black saddles with horizon $B_4$. We take the R-symmetry vector field to be $\xi=b \mskip1mu\partial_{\varphi}$, where $\partial_\varphi$ generates rotations of $\mathbb{R}^2$. The fixed point set consists of a single connected four-dimensional submanifold, located at the origin of $\mathbb{R}^2$ leaving all of $B_4$ fixed. The on-shell action has a single contribution from the four-dimensional fixed point set,
\be\label{eq:IB4}
I= \frac{F_{S^5}}{9}\big[2 \chi(B_4)+3\eta \tau(B_4)\big]\, ,
\ee
where recall that $\eta$ satisfies $\chi=\sigmaone \eta $ (and generically we set $\eta=1$). 

This is the general result for any choice of $B_4$; and, to the best of our knowledge, such a formula has not been presented in the literature before. However, a special case is given by taking $B_4$ to be a complex surface, in which case the Euler characteristic and signature satisfy $\chi(B_4)=c_2(B_4)$, $\tau(B_4)=\frac{1}{3}\int_{B_4}[2c_2(B_4)-c_1^2(B_4)]$, where here the Chern classes $c_i(B_4)$ are those of the holomorphic tangent bundle of $B_4$. Thus, with $\eta=1$ we have $I=\frac{F_{S^5}}{9}\int_{B_4}c_1^2(B_4)$. This agrees with minus the Bekenstein--Hawking entropy of the black hole solutions constructed in \cite{Hosseini:2018usu}, where 
$B_4$ is K\"ahler--Einstein of negative curvature.\footnote{In the notation of \cite{Hosseini:2018usu} one should set $X^1=X^2=\pi$, $\mathfrak{s}_1=\mathfrak{s}_2=-1$, and $-I$ agrees with their (6.8) using $\mathrm{vol}(B_4)=\tfrac{(2\pi)^2}{2}\int_{B_4}c_1^2(B_4)$.}

We get a very similar result for the $B$-charge over $B_4$ by employing \eqref{BchargeB4}
\begin{equation}
   \mathfrak{b} = -\frac{9 \ii}{8 }\left [ 2\chi(B_4) + 3 \tau(B_4)\right ]=-\frac{81\ii}{8F_{S^5}}I\, .
\end{equation}


\paragraph{Suh black hole: $B_4 = \Sigma_{g_1}\times \Sigma_{g_2}$}
To compare with the literature, we take $B_4$ to be the direct product of two Riemann surfaces $B_4 = \Sigma_{g_1} \times \Sigma_{g_2}$. It follows that the Euler characteristic and signature become $\chi (B_4)= \chi(\Sigma_{g_1})\chi(\Sigma_{g_2})$ and $\tau(B_4)=0$. Inserting these into \eqref{eq:IB4}, we obtain the on-shell action
\be
\begin{split}
I&=\frac{2 F_{S^5}}{9}\eta\chi(\Sigma_1)\chi(\Sigma_2)\label{ISuhBH}\, .
\end{split}
\ee
This matches minus the entropy of the corresponding black hole as computed in \cite{Suh:2018tul} and the field theory result in \cite{Hosseini:2018uzp}. 
The $B$-charge is 
\begin{equation}
   \Bc = -\frac{9 \ii}{4 }\chi(\Sigma_1)\chi(\Sigma_2)\, .
\end{equation}


\section{Examples: fibred spaces}\label{sec:examplesII}

Having exhausted the topologically trivial examples, we will now allow for non-trivial normal bundles. We will consider a plethora of different examples, each exhibiting different types of interesting behaviour. For the solutions in section \ref{sec:S3overSigmag} and \ref{sec:S3overS2}, the conformal boundaries are squashed three-sphere bundles over the Riemann surface, while for the examples in sections \ref{sec:OpSg1Sg2}, \ref{sec:Op1p2S2Sigmag} and \ref{sec:O(-p1,-p2)xS2xS2}, the conformal boundaries are circle bundles over the product of Riemann surfaces.


\subsection{\texorpdfstring{$\mathcal{O}(-p_1)\oplus \mathcal{O}( -p_2 ) \rightarrow \Sigma_{g}$}{O+O->Sg}}\label{sec:S3overSigmag}

Our first example of a non-trivial normal bundle is $\R^4 \rightarrow \Sigma_g$, where now we take the geometry to be fibred. Concretely, we take the spacetime to be the bundle $M_6=\mathcal{O}(-p_1)\oplus \mathcal{O}( -p_2 ) \rightarrow \Sigma_{g}$. This is then the generalization of the example considered in section \ref{sec:topRiemann} to include a non-trivial normal bundle. The fixed point set is unaffected by the fibration and remains the same as in the unfibred case; it is simply the Riemann surface at the centre of $\mathbb{R}^4$.

We write the R-symmetry vector as $\xi=b_1\partial_{\varphi_1}+b_2\partial_{\varphi_2}$ with the $\varphi_i$ coordinates each rotating one of the copies of $\mathbb{C}$ inside $\mathbb{R}^4$. The weights at the fixed point locus are simply $\epsilon_i=b_i$ and, from \eqref{actionwithsigns} we find that the on-shell action is
\be\label{eq:IR4Sigmagps}
\begin{split}
I=&\frac{\chi F_{S^5}}{27 b_1^2b_2^2}\big(b_1 + \chi_{\mathbb{R}^4}b_2)^2\\
&\qquad\times \Big[3 \sigmathree b_1 b_2\chi(\Sigma_g)+\sigmaone p_2 b_1(b_1-2\sigma_{\mathbb{R}^4} b_2 )+\sigmatwo p_1 b_2 (b_2-2 \sigma_{\mathbb{R}^4} b_1)\Big]\, ,
\end{split}
\ee
where $\chi_{\R^4} = \sigmaone \sigmatwo$.
As a consistency check, upon setting $p_i=0$, we recover expression \eqref{Ib1b2} as expected.
As far as we are aware, there are no field theory results for this class, nor explicit supergravity solutions which could be used as additional checks. Our result is therefore a prediction for the on-shell action for this class of geometries. 


\subsection{\texorpdfstring{$\mathcal{O}(-p_1)\oplus \mathcal{O}(- p_2 ) \rightarrow S^2$}{O+O->S2} with rotation}\label{sec:S3overS2}

\begin{figure}[h!]
\begin{center}
\tdplotsetmaincoords{78}{-60}

\begin{tikzpicture}
		[tdplot_main_coords,
			cube/.style={very thick,black},
			axisb/.style={->,blue,thick},
            axisr/.style={->,red,thick},
            axisg/.style={->,Green,thick},
			inf/.style={dashed,black}]

     \draw[inf] (3,3,0)--(3,3,-5);
\draw[inf](3,0,0)--(4,0,0);
\draw[inf](0,3,0)--(0,4,0);
\draw[inf] (0,3,-4)--(0,4,-4.6);
\draw[inf] (3,0,-3)--(4.5,0,-3.5);

\shadedraw[dashed,bottom color=white, top color=red!40,fill opacity=0.9] (0,0,-2)--(3,0,-3)--(3,3,-5)--(0,3,-4)--(0,0,-2);
\shadedraw[dashed,top color=white, bottom color=red!50,fill opacity=0.5] (0,0,0)--(3,0,0)--(3,3,0)--(0,3,0)--(0,0,0);
\shadedraw[black,right color=blue!40, left color=white,fill opacity=0.8] (0,0,0)--(0,3,0)--(0,3,-4)--(0,0,-2)--(0,0,0);
\shadedraw[black,left color=Green!70, right color=Green!0,fill opacity=0.5] (0,0,0)--(3,0,0)--(3,0,-3)--(0,0,-2)--(0,0,0);

\draw[ thick,white] (0,3,0)--(0,3,-4);
\draw[ thick,white] (3,0,0)--(3,0,-3);
\draw[ dashed ,black] (0,3,0)--(0,3,-4);
\draw[ dashed ,black] (3,0,0)--(3,0,-3);

\draw[axisg] (1.5,0,-1.5)--(1.5,-1.5,-1.5);
\draw[axisb, xshift=-0.2cm] (0,1.5,-1.5)--(-1.5,1.5,-1.5);
\draw[axisr] (1.5,1.5,0)--(1.5,1.5,1);
\draw[axisr](1.5,1.5,-3.5)--(4.5,3,-5);

\node at (1.5,1.5,1.2) {\color{red}$v_1$};
\node at (1.7,1.2,0) {\scriptsize{\color{red}$\sigmathree_N$}};

\node at (1.5,-1.5,-1.8) {\color{Green}$v_2$};
\node at (1.5,0,-1.3) {\scriptsize{\color{Green}$\sigmaone$}};

\node at (-1.5,1.7,-1.8) {\color{blue}$v_3$};
\node at (-0.76,1.1,-1.1) {\scriptsize{\color{blue}$\sigmatwo$}};

\node at (4.5,2.9,-5.2) {\color{red}$v_4$};
\node at (1.5,1.5,-3.2 ){\scriptsize{\color{red}$\sigmathree_S$}};

\end{tikzpicture}
\end{center}
\caption{Toric diagram for $\mathcal{O}(-1)\oplus \mathcal{O}(-2)\rightarrow S^2$. The red faces are the north and south pole of the sphere, while the green and blue are the centres of the two copies of $\mathbb{C}$.}
\label{fig:toricR4S2ps}
\end{figure}
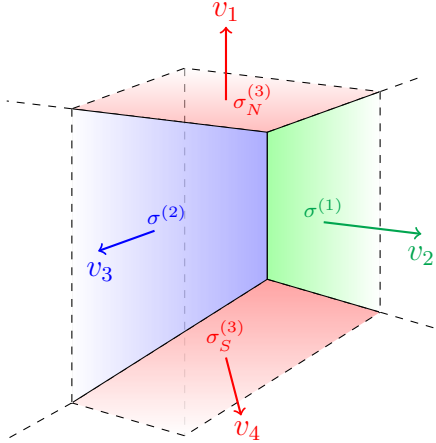

Analogously to the example in section \ref{sec:R4S2unfib}, we can allow for a chemical potential for the angular momentum when we restrict the Riemann surface to be a two-sphere. Turning on this chemical potential requires us to take the R-symmetry vector field to mix with the isometry of the two-sphere as in \eqref{eq:xiR4S2}. The setup is toric and as before this makes computing the weights particularly simple.  
The toric data consists of four outward pointing normals, which we take to be 
\be
v_1=(0,0,1)\,,\quad v_2=(1,0,0)\, ,\quad v_3=(0,1,0)\, ,\quad v_4=(p_1,p_2,-1)\, . 
\ee
For the interested reader, in appendix \ref{app:Op1Op2S2} we explain how one can compute, from first principles, the toric data for this geometry. The toric diagram is given in figure \ref{fig:toricR4S2ps} and should be contrasted with figure \ref{fig:R4S2}.  

We take the R-symmetry vector to be 
\begin{equation}
\xi=b_1\partial_{\varphi_1}+b_2\partial_{\varphi_2}+\varepsilon \partial_{\varphi}\, 
\end{equation}
which is the same as the one in section \ref{sec:R4S2unfib}.
The R-symmetry vector has two fixed points, located at the centre of $\mathbb{R}^4$ and at either of the poles of the two-sphere. In terms of toric geometry, the fixed points are determined by three vectors. The North pole contribution is determined by the three vectors $\{v_1,v_2,v_3\}$, while the South pole contribution is determined by $\{v_2,v_3,v_4\}$. 
Using the formulae in appendix \ref{app:weights}, the weights at the two fixed points are 
\be
(\epsilon_1,\epsilon_2,\epsilon_3)=
\begin{cases}
(b_1, b_2,\varepsilon) & \text{at North pole},\\
(b_1+p_1 \varepsilon, b_2+p_2\varepsilon,-\varepsilon)&\text{at South pole} \, .
\end{cases}
\ee
Substituting this into the on-shell action in equation \eqref{main}, we find
\be
I=\frac{\chi_N(\sigma^{(1)} b_1 +\sigma^{(2)} b_1+\sigma^{(3)}_N b_3)^3}{b_1 b_2 \varepsilon}-\frac{\chi_S(\sigma^{(1)}(b_1+p_1 \varepsilon) +\sigma^{(2)}(b_2+p_2 \varepsilon )-\sigma^{(3)}_S \varepsilon)^3}{(b_1+p_1 \varepsilon)(b_2+p_2 \varepsilon)\varepsilon}\, .
\ee
As before, we have the freedom to choose the relative sign between $\sigma^{(3)}_N$ and $\sigma^{(3)}_S$,  which corresponds to choosing either the twist or anti-twist. We can also compute the magnetic charge threading through the two-sphere using the polyform in \eqref{PhiF}. We find
\be
\begin{split}
m&=p_1 \sigmaone +p_2 \sigmatwo -\sigmathree_N -\sigmathree_S\, .
\end{split}
\ee

As a consistency check we proceed to compare this result with the previous results by taking various limits. 
To compare with the rotating and unfibred $S^2$ example of section \ref{sec:R4S2unfib}, we need to set $p_i=0$. It is not hard to see that the on-shell action is exactly \eqref{eq:R4S2trivfib} and the magnetic charge $m$ reduces to the expected topological term, \eqref{mformula}. The other limit we can take is to turn off the chemical potential for the angular momentum by setting $\varepsilon=0$. This works only for the twist case, $\sigmathree_N=\sigmathree_S$, wherein we obtain precisely equation \eqref{eq:IR4Sigmagps}. Note that this is another non-trivial consistency check of  the topological relation in \eqref{mformula}.


\subsection{\texorpdfstring{$\mathcal{O}(-p_1, -p_2 ) \rightarrow \Sigma_{g_1}\times \Sigma_{g_2}$}{O->Sg1xSg2}}\label{sec:OpSg1Sg2}

Let us now consider the fibred version of the four-dimensional fixed point sets of section \ref{sec:R2xB4}. We have a single four-dimensional fixed point set $B_4$ located at the centre of $\mathbb{R}^2$. For the moment we will consider $B_4$ to be the product of two Riemann surfaces, before considering $\mathbb{CP}^2$ in section \ref{sec:CP2}. Over each Riemann surface we can define a complex line bundle with Chern number $-p_i$ so that the full bundle is\footnote{Note that this definition of the $p_i$'s is different from the setup in section \ref{sec:gaugefield}: there we had one Riemann surface and two line bundles over it, while here it is two Riemann surfaces and one line bundle.} $\mathcal{O}(-p_1, -p_2 ) \rightarrow \Sigma_{g_1}\times \Sigma_{g_2}$. 

Recall from section \ref{sec:gaugefield} that the spinor is a section of the rank 2 bundle
\begin{equation}
\mathcal{E}\equiv\mathcal{S}^\eta_{B_4}\otimes L_1^{\sigmaone/2} \otimes \mathcal{L}^{1/2}\, ,
\end{equation}
where 
\begin{equation}
\mathcal{S}^+_{B_4} = K^{1/2}_{B_4}\oplus K^{-1/2}_{B_4}\, , \qquad \mathcal{S}^-_{B_4} = K^{1/2}_{B_4}\otimes \Lambda^{0,1}\, .
\end{equation}
Let $\KSone$ and $\KStwo$ be the canonical line bundles of $\Sigma_{g_1}$ and $\Sigma_{g_2}$ respectively. Since the manifold $B_4$ is a product of Riemann surfaces, its canonical bundle $K_{B_4}$ and the bundle of (0,1)-forms $\Lambda^{0,1} $ decompose as
\begin{align}\label{L01Sigmag1Sigmag2}
    K_{B_4} =& K_{\Sigma_{g_1}} \otimes K_{\Sigma_{g_2}} \, , \qquad
    \Lambda^{0,1}  = \KSone^{-1} \oplus \KStwo^{-1} \, . 
\end{align}
Recall that a necessary condition for the spinor to be well defined was that the bundle $\mathcal{E}$ splits as the sum of two bundles one of which is trivial. Since $\Lambda^{0,1}$ does split for the present case we can define spinors as sections of $\mathcal{E}$ with either chirality $\eta=\pm1$; this is in contrast to a generic $B_4$. Since we can further decompose the tangent planes on $B_4$ we may also correspondingly write $\eta=\sigmatwo\sigmathree$.

We can now use the results of section \ref{sec:gaugefield} to fix the magnetic charges. For the $\eta=1$ case we necessarily have $\sigmatwo=\sigmathree$ and then \eqref{mformulaB4} gives
\begin{equation}
    0=\sigmatwo c_1(K_{B_4})+ \sigmaone c_1(L_1)+c_1(F)\, ,
\end{equation}
and the magnetic charges threading through the two Riemann surfaces are
\begin{equation}
    m_1=\sigmaone p_1 -\sigmatwo \chi(\Sigma_{g_1})\, ,\qquad  m_2=\sigmaone p_2 -\sigmathree \chi(\Sigma_{g_2})\, .
\end{equation}
For $\eta=-1$ we once again have two cases, depending on which of $\sigmatwo$ or $\sigmathree$ is $-1$. The bundle $\mathcal{E}$ in this case is 
\begin{equation}
\mathcal{E}=K_{\Sigma_{g_1}}^{1/2}\otimes K_{\Sigma_{g_2}}^{-1/2} \otimes L_1^{\sigmaone/2} \otimes \mathcal{L}^{1/2} \oplus K_{\Sigma_{g_1}}^{-1/2}\otimes K_{\Sigma_{g_2}}^{1/2} \otimes L_1^{\sigmaone/2} \otimes \mathcal{L}^{1/2}\, ,
\end{equation}
and we now have a choice of which bundle to take to be trivial. In general we have
\begin{equation}
    0=\sigmatwo c_1(K_{\Sigma_{g_1}})+\sigmathree c_1(K_{\Sigma_{g_2}})+\sigmaone c_1(L_1)+c_1(F)\, .
\end{equation}
The magnetic charges are 
\begin{equation}
     m_1=\sigmaone p_1 -\sigmatwo \chi(\Sigma_{g_1})\, ,\qquad  m_2=\sigmaone p_2 -\sigmathree \chi(\Sigma_{g_2})\, ,
\end{equation}
note that they take exactly the same form as in the $\eta=1$ case.

To compute the on-shell action we now need to substitute our results into the general formula. Since our master formula in equation \eqref{main} sets $\eta=1$ we will use equation \eqref{actionwithsigns} in the following, which gives
\begin{equation}
    I=\frac{F_{S^5}}{27}\chi \sigmaone \left[6 m_1 m_2 -3 \sigmaone (m_1 p_2+m_2 p_1)+2 p_1 p_2\right]\, .
\end{equation}
Using our results for the magnetic charges, we find that the on-shell action is
\begin{equation}
    I=\frac{F_{S^5}}{27}\eta \Big[2 p_1 p_2+6 \eta \chi(\Sigma_{g_1})\chi(\Sigma_{g_2})-3 \sigmaone \big(\sigmatwo p_2 \chi(\Sigma_{g_1})+\sigmathree p_1 \chi(\Sigma_{g_2})\big)\Big]\, .
\end{equation}
Note that this correctly reduces to the unfibred case \eqref{ISuhBH} upon setting $p_1=p_2=0$.
Finally we find that the conserved charge $\Bc$ is
\begin{equation}
    \mathfrak{b}=-\frac{9\ii}{4} m_1 m_2\, .
\end{equation}

\subsection{\texorpdfstring{$\mathcal{O}(-p_1, -p_2 ) \rightarrow S^2_{\varepsilon}\times \Sigma_g$}{O(-p1,-p2)->S2xSg}}\label{sec:Op1p2S2Sigmag}

Consider now the case where one of the Riemann surfaces is a sphere with an equivariant parameter for rotations turned on. As in the example studied in section \ref{sec:R2S2Sg}, there are two connected two-dimensional fixed points sets, located at the centre of $\mathbb{R}^2$ and the poles of the two-sphere, with a copy of the Riemann surface fixed at both. To determine the weights at the fixed point set, we will use toric geometry, the data for which has been derived in appendix \ref{sec:ToricO(p)S2}. The toric data for $\mathcal{O}(-p_1, -p_2 ) \rightarrow S^2_{\varepsilon}\times \Sigma_g$ at a fixed point on the Riemann surface is 
\begin{equation}
    v_0=(1,0)\, ,\quad v_1=(0,1)\, ,\quad v_2=(p_1,-1)\, ,
\end{equation}
where the R-symmetry vector field in this basis is
\begin{equation}
    \xi= b\mskip1mu \partial_{\tau}+\varepsilon \partial_{\varphi}\, .
\end{equation}
The weights at the two fixed points are
\begin{equation}
(\epsilon_1,\epsilon_2)=\begin{cases}
        (b,\varepsilon) &\text{at North pole},\\
        (b+p_1\varepsilon, -\varepsilon)& \text{at South pole}\, .
    \end{cases}
\end{equation}
Inserting these weights into our master formula, \eqref{main}, we find the on-shell action 
\begin{align}
    I&=\frac{F_{S^5}\sigmaone\sigmathree}{27\varepsilon}\Bigg[\frac{\sigmatwo_N\big(\sigmaone b+\sigmatwo_N \varepsilon\big)^2\big(3 \sigmathree b \chi(\Sigma_g)+\sigmatwo_N p_2 \varepsilon  -2\sigmaone  p_2 b  \big) }{b^2}\\
    &-\frac{\sigmatwo_S \big(\sigmaone (b+p_1\varepsilon )-\sigmatwo_S \varepsilon\big)^2\big(3\sigmathree(b+ p_1\varepsilon)\chi(\Sigma_g)-\sigmatwo_S p_2\varepsilon  -2 \sigmaone p_2(b+p_1\varepsilon )\big)}{(b+ p_1\varepsilon)^2}\Bigg]\nonumber\, .
\end{align}
There are two distinct cases to consider depending on the twist or anti-twist. We will not further analyse these different cases since there are no field theory nor gravity results with which to compare. Thus, this stands as a prediction. Note that the result is consistent with the unfibred case in section \ref{sec:R2S2Sg}.


\subsection{\texorpdfstring{$\mathcal{O}(-p_1, -p_2 ) \rightarrow S^2_{\varepsilon_1}\times S^2_{\varepsilon_2}$}{O->S2xS2}}\label{sec:O(-p1,-p2)xS2xS2}

Our final example in this section is to consider fibering $\mathbb{R}^2$ over two rotating two-spheres. The R-symmetry vector field is $\xi=b\mskip1mu \partial_{\tau}+\varepsilon_1\partial_{\varphi_1}+\varepsilon_2\partial_{\varphi_2}$, where $\tau$ is the angular coordinate on $\mathbb{R}^2$, while $\varphi_i$ is the azimuthal angle on $S^2_{\varepsilon_i}$.  The full six-dimensional manifold is toric and is characterized by the toric vectors
\begin{align}
    v_0 & =  (0,0, 1)\, ,\quad 
    v_1 = ( 1,0, 0)\, ,\quad 
    v_2 = (0,  1,0) \, ,\quad
    v_3 = (-1,  0,p_1)\, , \nonumber\\
    v_4 & = (0, -1,p_2)\, ,
\end{align}
which is derived in appendix \ref{app:O(p1p2)S2S2}, and forms the polytope in figure \ref{fig:ToricO(1,2)S2S2}. We have labelled the polytope with the signs for the projection conditions.

\begin{figure}[h!]
\begin{center}
\tdplotsetmaincoords{70}{-60}

\begin{tikzpicture}
		[tdplot_main_coords,
			cube/.style={very thick,black},
			axisb/.style={->,blue,thick},
            axisr/.style={->,red,thick},
            axisg/.style={->,Green,thick},
			inf/.style={dashed,black}]

\shade[top color=Green!50,bottom color=white, fill opacity=0.5] (0,0,0)--(0,0,-3)--(-6,0,-3)--(-3,0,0)--(0,0,0);
\shade[top color=blue!50, bottom color=white, fill opacity =0.7 ](0,0,0)--(0,0,-3)--(0,-9,-3)--(0,-3,0)--(0,0,0);
\fill[black,red!40,fill opacity=0.8] (0,0,0)--(-3,0,0)--(-3,-3,0)--(0,-3,0)--(0,0,0);
\draw[black] (0,0,0)--(-3,0,0)--(-3,-3,0)--(0,-3,0)--(0,0,0);

\shade[top color=Green!50,bottom color=white, fill opacity=0.3](0,-3,0)--(0,-9,-3)--(-6,-9,-3)--(-3,-3,0)--(0,-3,0);
\shade[top color=blue!50, bottom color=white, fill opacity =0.2 ](-3,0,0)--(-6,0,-3)--(-6,-9,-3)--(-3,-3,0)--(-3,0,0);

\draw[axisg](-1.5,0,-1.5)--(-1.5,1.5,-1.5);
 \node at (-1.5,1.75,-1.5) {\color{Green}$v_2$};
 \node at (-1.3,0,-1.3) {\scriptsize{\color{Green}$\sigmatwo_N$}};

\draw[black](-3,-3,0)--(-6,-9,-3);
\draw[black](-6,0,-3)--(-3,0,0);
\draw[black](0,-3,0)--(0,-9,-3);
\draw[dashed](0,0,0)--(0,0,-3.5);
\draw[dashed](0,0,-3)--(0,-9,-3);
\draw[dashed](0,0,-3)--(-6,0,-3);
\draw[dashed](-6,0,-3)--(-6,-9,-3);
\draw[dashed](-6,-9,-3)--(0,-9,-3);

\draw[dashed](-6,-9,-3)--(-6.5,-10,-3.5);
\draw[dashed](0,-9,-3)--(0,-10,-3.5);
\draw[dashed](-6,0,-3)--(-6.5,0,-3.5);

\draw[axisr](-1.5,-1.5,0)--(-1.5,-1.5,1.5);
 \node at (-1.5,-1.5,1.7) {\color{red}$v_0$};
 \node at (-1.5,-1.8,0) {\scriptsize{\color{red}$\sigmathree$}};

\draw[axisg](-1.5,-6,-1.5)--(-1.5,-7,0.5);
 \node at (-1.5,-7.2,0.7) {\color{Green}$v_4$};
 \node at (-1.5,-6.4,-1.7) {\scriptsize{\color{Green}$\sigmatwo_S$}};

\draw[axisb](-5,-3.5,-2)--(-6,-3.5,-1.3);
 \node at (-6.3,-3.5,-1) {\color{blue}$v_3$};
 \node at (-4.7,-3.5,-2.3) {\scriptsize{\color{blue}$\sigmaone_S$}};

\draw[axisb](0,-3,-2)--(1.5,-3,-2);
\node at (1.8,-3,-2) {\color{blue}$v_1$};
 \node at (-0.2,-3.3,-2.2) {\scriptsize{\color{blue}$\sigmaone_N$}};
\end{tikzpicture}
\end{center}
\caption{Toric diagram for $\mathcal{O}(-1,-2) \rightarrow S^2 \times S^2$. }
 \label{fig:ToricO(1,2)S2S2}
\end{figure}
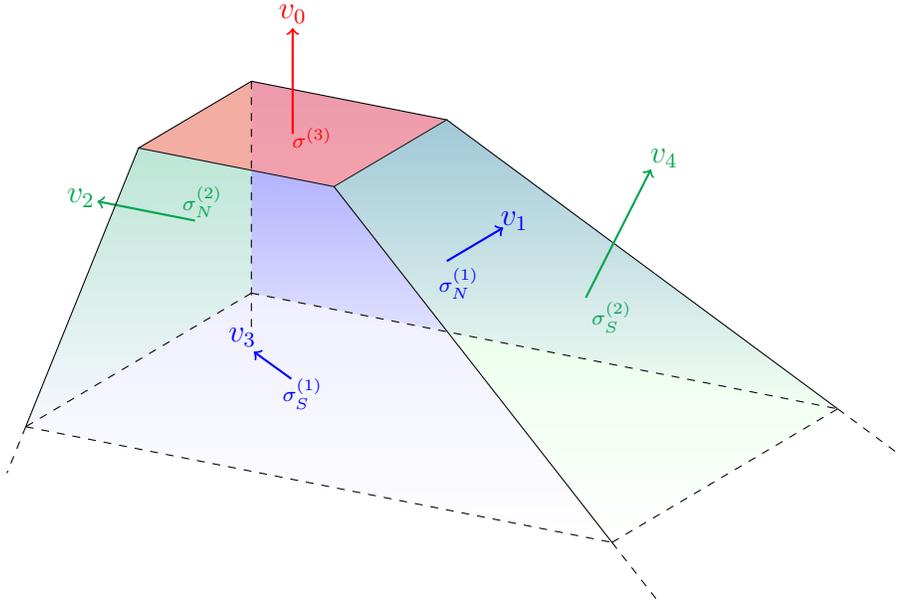

We have four isolated fixed points which we denote by the pole type on the two $S^2$'s. The weights at the fixed points, computed using the formulae in appendix \ref{app:toricdata}, are
\begin{equation}
(\epsilon_1,\epsilon_2,\epsilon_3)=\begin{cases}
(b,\varepsilon_1,\varepsilon_2) & NN\, ,\\
(b+p_1\varepsilon_1 +p_2\varepsilon_2 ,-\varepsilon_1,-\varepsilon_2) & SS\, ,\\
(b+p_1\varepsilon_1 ,-\varepsilon_1,\varepsilon_2) & SN\, ,\\
(b+p_2\varepsilon_2 ,\varepsilon_1,-\varepsilon_2,) & NS\, .
\end{cases}
\end{equation} 
From here, it is just a matter of inserting the weights and the signs from the projection conditions into our master formula \eqref{main} in order to compute the on-shell action. The unwieldy result is
\begin{equation}\label{R2fibS2S2Odim}
\begin{split}
I&=\frac{F_{S^5}\sigmathree}{27\varepsilon_1 \varepsilon_2}\Bigg[\frac{\sigma^{(1)}_{N}\sigma^{(2)}_{N}\big( \sigmathree b+\sigma^{(1)}_{N}\varepsilon_1+\sigma^{(2)}_{N}\varepsilon_2\big)^3}{b}\\
&-\frac{\sigma^{(1)}_{S}\sigma^{(2)}_{N}\big(\sigmathree(b +p_1\varepsilon_1 )-\sigma^{(1)}_{S}\varepsilon_1+\sigma^{(2)}_{N}\varepsilon_2\big)^3}{b+p_1\varepsilon_1 }\\
&-\frac{\sigma^{(1)}_{N}\sigma^{(2)}_{S}\big ( \sigmathree(b +p_2 \varepsilon_2)+\sigma^{(1)}_{N}\varepsilon_1-\sigma^{(2)}_{S}\varepsilon_2\big)^3}{b+p_2\varepsilon_2 }\\
&+\frac{\sigma^{(1)}_{S}\sigma^{(2)}_{S}\big(\sigmathree(b +p_1 \varepsilon_1+p_2\varepsilon_2)-\sigmaone_S\varepsilon_1-\sigma^{(2)}_{S}\varepsilon_2\big)^3}{b+p_1\varepsilon_1 +p_2 \varepsilon_2}\Bigg]\, .
\end{split}
\end{equation}
The magnetic charges threading through the two two-spheres are easy to compute. We fix the two cycles to be evaluated at either of the poles of the other two-sphere; these cycles are homologous and we find the same result for both poles:
\be
m_1=-\sigma^{(1)}_{N}-\sigma^{(1)}_{S}+\sigma^{(3)}p_1\, ,\quad m_2=-\sigma^{(2)}_{N}-\sigma^{(2)}_{S}+\sigma^{(3)}p_2\, .
\ee
The other charges can be computed similarly; however, they are particularly unwieldy and not very illuminating and so we do not give them here. 

One could now consider the various limits of turning off the $p_i$'s and $\varepsilon_i$'s. The same comments from the earlier sections follow almost verbatim and we find that the results are fully consistent with all of the previous results. As in the previous section, there are no known field theory results, nor explicit supergravity solutions and it would be interesting to test this prediction through either method.


\section{Examples: 4d horizons}\label{sec:examplesIII}

In this final section we will consider four examples each with a four-dimensional space which is naturally interpreted as the horizon of a black hole. The four-manifold for the first three examples will be $\mathbb{CP}^2$ while our final example will consider a general four-dimensional toric orbifold.


\subsection{Complex projective plane}\label{sec:CP2}

We begin by considering $M_6$ to be a non-trivial complex line bundle over $\mathbb{CP}^2$, that is $M_6=\mathcal{O}(-p)\rightarrow \mathbb{CP}^2$. Although $\mathbb{CP}^2$ does not admit a spin structure, it does admit a spin$^c$ structure which is all that we require. 
It is well-known that $\mathbb{CP}^2$ has 
$H^2(\mathbb{CP}^2,\Z)\cong\Z$. 
A generator  of this class is known as the \emph{hyperplane class}, and we denote it by $H$ in the following. We normalize such that $\int_{\mathbb{CP}^2}H\wedge H=1$. Since 
any two-form cohomology class
is then a multiple of $H$, 
 it follows that we may write
\begin{equation}
    c_1(F)= m H\, ,\qquad c_1(\mathcal{O}(-p))=-p H\,.
\end{equation}
Moreover, it is known that for $\mathbb{CP}^2$ one  has  $c_1(K^{-1}_{\mathbb{CP}^2})=3 H$, $\chi(\mathbb{CP}^2)=3$ and $\tau(\mathbb{CP}^2)=1$.
Since the bundle of holomorphic one-forms $\Lambda^{0,1}(\mathbb{CP}^2)$ does not decompose into the sum of two line bundles on $\mathbb{CP}^2$, from the discussion in section \ref{sec:gaugefield}, it follows that the spinor is necessarily of positive chirality. Therefore, the magnetic charge is fixed using \eqref{mformulaB4} to be
\begin{equation}\label{eq:CP2m}
m=\sigmaone p\pm 3\, .    
\end{equation}
One can also see why $\eta=-1$ is not allowed by explicitly computing the topological restriction \eqref{identity}, which reads 
\begin{equation}
   6 \eta+3\equiv 2\eta \chi(\mathbb{CP}^2)+3 \tau(\mathbb{CP}^2)=(m-\sigmaone p)^2\, .
\end{equation}
Solving for $m$ one finds $m=\sigmaone p \pm \sqrt{3+6\eta}$, which is obviously complex for $\eta=-1$ and reduces to \eqref{eq:CP2m} for $\eta=1$ as it should.

The on-shell action is obtained by substituting the topological data above into our main formula, \eqref{main} and reads
\begin{equation}\label{ActionCP2}
    I = \frac{F_{S^5}}{27} \left(27+p^2 \pm 9 p \sigmaone\right).
\end{equation}
Curiously for $p=0$ one obtains the same result for the on-shell action as for the Euclidean AdS$_6$ vacuum solution.
Using \eqref{BchargeB4}, the charge $\Bc$ is  
\begin{equation}
    \Bc = -\frac{9\ii}{8}m^2 \, .
\end{equation}


\subsection{Complex projective plane with rotation}\label{sec:CP2rot}

Having studied the fibred $\mathbb{CP}^2$ example, let us now turn on equivariant parameters for the two U$(1)$ symmetries of $\mathbb{CP}^2$, i.e. we consider the $\mathbb{CP}^2$ to be ``rotating''. We take the R-symmetry vector field to be
\begin{equation}
    \xi=\varepsilon_1 \partial_{\varphi_1}+\varepsilon_2\partial_{\varphi_2}+ b \mskip1mu\partial_{\tau}\, ,
\end{equation}
where $\partial_{\tau}$ rotates $\mathbb{R}^2$ and $\partial_{\varphi_i}$ is a basis for the $\mathbb{T}^2$ toric action on $\mathbb{CP}^2$. In appendix~\ref{app:CP2}, we derive from first principles the toric data for $\mathcal{O}(-p)\rightarrow \mathbb{CP}^2$ and work in the basis of $\mathbb{T}^2=\mathrm{U}(1)^2$ defined there. The toric data we use in the following is
\begin{equation}
  v_0 =  (0,0,1)\, ,\quad
    v_1 =  (1, 0, 0)\, ,\quad 
    v_2 = (0, 1, 0)\, ,\quad 
    v_3 =  (-1, -1,p)\, .
\end{equation}
Note that the order of the basis vector fields used to give $\xi$ is the order of the basis in the toric data. To each facet of the polytope we associate a sign $\sigma$ for the projection condition of the spinor on that facet; in total we have four such signs.

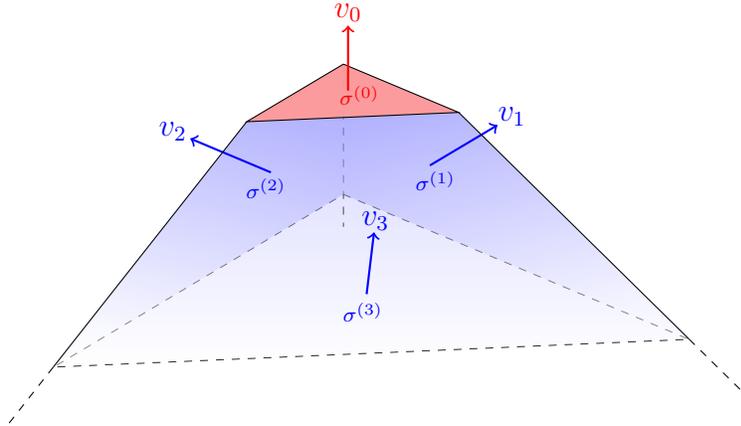
\begin{figure}[h!]
\begin{center}
\tdplotsetmaincoords{60}{-50}

\begin{tikzpicture}
		[tdplot_main_coords,
			cube/.style={very thick,black},
			axisb/.style={->,blue,thick},
            axisr/.style={->,red,thick},
            axisg/.style={->,Green,thick},
			inf/.style={dashed,black}]

\draw[inf](-6,0,-2)--(-7,0,-2.5);
\draw[inf](0,-6,-2)--(0,-7,-2.5);
\draw[inf](0,0,0)--(0,0,-2.5);
\draw[inf](0,0,-2)--(0,-6,-2)--(-6,0,-2)--(0,0,-2);

\shade[right color=white, left color=blue!40,fill opacity=0.6,shading angle=-30](0,0,0)--(0,0,-2)--(-6,0,-2)--(-2,0,0)--(0,0,0);
\shade[right color=white, left color=blue!40,fill opacity=0.6,shading angle=30](0,0,0)--(0,0,-2)--(0,-6,-2)--(0,-2,0)--(0,0,0);
\shade[bottom color=white, top color=blue!40,fill opacity=0.4,shading angle=0](0,-6,-2)--(0,-2,0)--(-2,0,0)--(-6,0,-2)--(0,-6,-2);
\fill[red!40,fill opacity=0.95](0,0,0)--(-2,0,0)--(0,-2,0)--(0,0,0);

\draw[axisb] (0,-1.5,-1)--(1.4,-1.5,-1);
  \node at (1.7,-1.5,-1){\color{blue}$v_1$};
 \node at (0,-1.6,-1.2) {\scriptsize{\color{blue}$\sigma^{(1)}$}};

\draw[axisb] (-1.5,0,-1)--(-1.5,1.4,-1);
  \node at (-1.5,1.7,-1){\color{blue}$v_2$};
 \node at (-1.6,0,-1.2) {\scriptsize{\color{blue}$\sigma^{(2)}$}};
 
\draw[axisb] (-2.5,-2.5,-1.5)--(-3.3,-3.3,0.1);
 \node at (-3.45,-3.45,0.4){\color{blue}$v_3$};
 \node at (-2.1,-2.1,-2.1) {\scriptsize{\color{blue}$\sigma^{(3)}$}};

\draw[black](0,0,0)--(-2,0,0)--(0,-2,0)--(0,0,0);

\draw[black](-2,0,0)--(-6,0,-2);

\draw[black](0,-2,0)--(0,-6,-2);

 \draw[axisr] (-0.5,-0.5,0)--( -0.5,-0.5,1);
\node at (-0.5,-0.5,1.2) {\color{red}$v_0$};
 \node at (-0.5,-0.7,0) {\scriptsize{\color{red}$\sigma^{(0)}$}};

\end{tikzpicture}
\end{center}
\caption{Toric diagram for $\mathcal{O}(-2)\rightarrow \mathbb{CP}^2$.}
\label{fig:toricCP2}
\end{figure}

For $b\mskip1mu\varepsilon_1\varepsilon_2\neq0$ there are three fixed points consisting of the centre of $\mathbb{R}^2$ and the three vertices of the polytope describing $\mathbb{CP}^2$ as a toric variety. In figure \ref{fig:toricCP2} the fixed points are the three vertices. It is simple to compute the weights of $\xi$ at these three vertices, which we label by the intersecting facets, using the results of appendix \ref{app:toricdata}
\begin{equation}
(\epsilon_1,\epsilon_2,\epsilon_3)=\begin{cases}
(b,\varepsilon_1,\varepsilon_2) & \text{at vertex }012\, ,\\
(b+p\mskip1mu \varepsilon_2,\varepsilon_1-\varepsilon_2,-\varepsilon_2) & \text{at vertex }013\,,\\
(b+p\mskip1mu \varepsilon_1, \varepsilon_2-\varepsilon_1,-\varepsilon_1) & \text{at vertex }023\, .
    \end{cases}
\end{equation}
Observe that the labelling of the vertex corresponds to the degenerating vector and therefore unambiguously fixes the sign of the projection conditions we need to use in our formulae.

The action, upon using \eqref{main}, is readily found to be
\begin{equation}
\begin{split}
I&=\frac{F_{S^5} \sigma^{(0)}}{27}\bigg[\frac{\sigmaone\sigmatwo \big(\sigma^{(0)} b+\sigmaone \varepsilon_1+\sigmatwo \varepsilon_2\big)^3}{b \varepsilon_1\varepsilon_2}\\
&+\frac{\sigmatwo \sigmathree \big(\sigma^{(0)}(b+p \varepsilon_1)+\sigmatwo(\varepsilon_2-\varepsilon_1)-\sigmathree \varepsilon_1\big)^3}{(b+p \varepsilon_1)\varepsilon_1(\varepsilon_1-\varepsilon_2)}\\
&+\frac{\sigmaone \sigmathree\big(\sigma^{(0)}(b+p \varepsilon_2)+\sigmaone(\varepsilon_1-\varepsilon_2)-\sigmathree\varepsilon_2\big)^3}{(b+p\varepsilon_2)(\varepsilon_2-\varepsilon_1)\varepsilon_2}
\bigg]\, .
\end{split}
\end{equation}
We can also compute the magnetic charge over the Poincar\'e dual of the hyperplane class of $\mathbb{CP}^2$ by using equation \eqref{mformulatoric}. The edges connecting two vertices in figure \ref{fig:toricCP2} give rise to compact two-cycles. The cycles are not independent, since there is only one non-trivial two-cycle, and all give the dual two-cycle to the hyperplane class. The magnetic charge is
\begin{equation}
    m = -(\sigmaone + \sigmatwo + \sigmathree ) +  \sigma^{(0)} p \, .
\end{equation}
Observe that if we turn off the rotation we must set $\sigmaone=\sigmatwo=\sigmathree$, and this result reduces to the one in the previous section, \eqref{eq:CP2m}.

The evaluation of the electric and the $B$-charge follow similarly to our previous examples. The only relevant compact four-cycle is $\mathbb{CP}^2$ at the centre of $\mathbb{R}^2$ and we can directly apply \eqref{electriccharge} and \eqref{Bcharge} to obtain the electric charge
\begin{equation}\label{qCP2toric}
\begin{split}
    q &= \frac{3 \sigma^{(0)}}{2 \varepsilon_1 \varepsilon_2 (\varepsilon_1-\varepsilon_2)}\bigg[ \sigmaone\sigmatwo (\varepsilon_1-\varepsilon_2)(\sigma^{(0)}b +\sigmaone \varepsilon_1+\sigmatwo \varepsilon_2)^2 \\
    &+\sigmatwo\sigmathree \varepsilon_2(\sigma^{(0)}(b+p\varepsilon_2)+\sigmatwo(\varepsilon_2-\varepsilon_1)-\sigmathree \varepsilon_1)^2\\
    &-\sigmaone\sigmathree \varepsilon_1( \sigma^{(0)}(b +p\varepsilon_2)+\sigmaone(\varepsilon_1-\varepsilon_2)-\sigmathree \varepsilon_2)^2\bigg] \, ,
\end{split}
\end{equation}
and the $B$-charge
\begin{equation}
    \mathfrak{b} = -\frac{9 \ii}{4} m^2 \, .
\end{equation}
One can also take the limit of vanishing equivariant parameters to compare with the previous section, one finds that the electric charge is
$q=\tfrac{3}{2} \sigma^{(0)}(p \sigma^{(0)}-3 \sigmaone)^2$, where we have set $\sigmaone  =  \sigmatwo =  \sigmathree$.

\subsection{Blow up of \texorpdfstring{$\mathbb{C}^3$}{C3}}\label{sec:Blowup}

\begin{figure}[h!]
\begin{center}
\tdplotsetmaincoords{60}{110}

\begin{tikzpicture}
		[tdplot_main_coords,
			axisb/.style={->,blue,thick},
            axisr/.style={->,red,thick},
            axisg/.style={->,Green,thick},
			inf/.style={dashed,black}]
            
\fill[red!50,opacity=0.6] (0,0,2)--(0,2,0)--(2,0,0)--(0,0,2);
\draw[black] (0,0,2)--(0,2,0)--(2,0,0)--(0,0,2);

\draw[axisb] (2,2,0)--(2,2,-1.5);
\draw[axisb] (0,2,2)--(-1.5,2,2);
\draw[axisb] (2,0,2)--(2,-1.5,2);
\draw[axisr] (2/3,2/3,2/3)--(-1/3,-1/3,-1/3);

\node at (-1.8,2,2){\color{blue}$v_3$};

 \shade[right color=white, left color=blue!40,fill opacity=0.6,shading angle=135](0,0,4)--(0,4,4)--(0,4,0)--(0,2,0)--(0,0,2);
 \shade[left color=white, right color=blue!30,fill opacity=0.6,shading angle=90](0,0,4)--(4,0,4)--(4,0,0)--(2,0,0)--(0,0,2);
 \shade[bottom color=white, top color=blue!30,fill opacity=0.6,shading angle=0](4,0,0)--(4,4,0)--(0,4,0)--(0,2,0)--(2,0,0);

\draw[inf] (0,0,4)--(0,4,4)--(0,4,0);
\draw[inf] (0,0,4)--(4,0,4)--(4,0,0);
\draw[inf] (4,0,0)--(4,4,0)--(0,4,0);

\draw[inf] (0,0,4)--(0,0,5);
\draw[inf] (0,4,0)--(0,5,0);
\draw[inf] (4,0,0)--(5,0,0);

\draw[black](0,0,2)--(0,0,4);
\draw[black](0,2,0)--(0,4,0);
\draw[black](2,0,0)--(4,0,0);
  \node at (2,2,-1.7){\color{blue}$v_1$};
 \node at (1.75,1.9,0) {\scriptsize{\color{blue}$\sigma^{(1)}$}};
  \node at (2,-1.7,2){\color{blue}$v_2$};
 \node at (1.75,0,2.2) {\scriptsize{\color{blue}$\sigma^{(2)}$}};
 \node at (0,1.9,1.8) {\scriptsize{\color{blue}$\sigma^{(3)}$}};
 \node at (0.7,0,0){\color{red}$v_0$};
 \node at (1.2,1.2,1.2) {\scriptsize{\color{red}$\sigma^{(0)}$}};
 
\end{tikzpicture}
\end{center}
\caption{Toric diagram for the blow-up of the centre of $\mathbb{C}^3$ with a $\mathbb{CP}^2$.}
\label{fig:Blowup}
\end{figure}
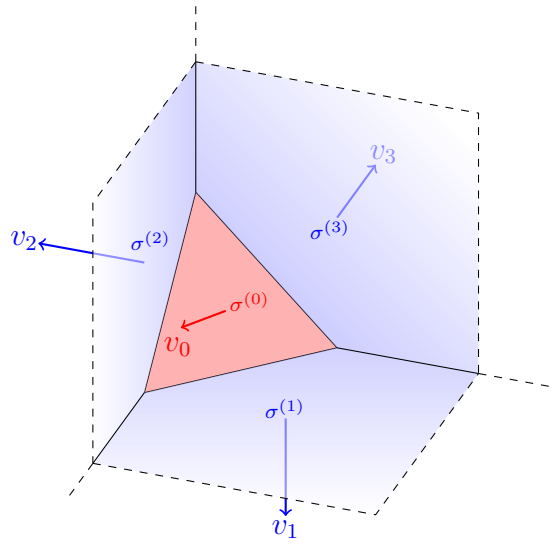

An interesting extension of the Euclidean AdS$_6$ solution is to consider blowing up the centre of $\mathbb{C}^3\cong\R^6$. We take the toric data of $\mathbb{C}^3$ to be 
\begin{equation}
    v_1=(1,0,0)\, ,\quad v_2=(0,1,0)\, ,\quad v_3=(0,0,1)\, ,
\end{equation}
and to blow up the centre of $\mathbb{C}^3$ we add the additional toric vector
\begin{equation}
    v_0=v_1+v_2+v_3=(1,1,1)\, .
\end{equation}
The resultant polytope is given in figure \ref{fig:Blowup}. As usual to each facet we associate a sign for the projection condition of the spinor at that facet. 

We take the R-symmetry vector field to be 
$\xi=\sum_{i=1}^{3}\varepsilon_i \partial_{\varphi_i}$ where each of the $\varphi_i$ are azimuthal coordinates on a copy of $\mathbb{C}\subset \mathbb{C}^3$. 
The weights are easily computed using appendix \ref{app:weights}, 
\begin{equation}
(\epsilon_1,\epsilon_2,\epsilon_3)=\begin{cases}
(\varepsilon_3, \varepsilon_1-\varepsilon_3,\varepsilon_2-\varepsilon_3) & \text{at vertex 012}\, ,\\
(\varepsilon_2, \varepsilon_1-\varepsilon_2,\varepsilon_3-\varepsilon_2) & \text{at vertex 013}\, ,\\
(\varepsilon_1,\varepsilon_2-\varepsilon_1,\varepsilon_3-\varepsilon_1) & \text{at vertex 023}\, .
\end{cases}
\end{equation} 
The on-shell action for the blow-up is given by
\begin{equation}\label{IBlowUp}
\begin{split}
I&=\frac{F_{S^5}\sigma^{(0)}}{27}\bigg[\frac{\sigmaone\sigmatwo\big(\sigmaone\varepsilon_1+\sigmatwo \varepsilon_2+(\sigma^{(0)}-\sigmaone-\sigmatwo)\varepsilon_{3}\big)^{3}}{(\varepsilon_1-\varepsilon_3)(\varepsilon_2-\varepsilon_3)\varepsilon_3}\\
&+\frac{\sigmaone \sigmathree \big(\sigmaone \varepsilon_3+(\sigma^{(0)}-\sigmaone-\sigmathree)\varepsilon_2+\sigmathree \varepsilon_3\big)^3}{(\varepsilon_1-\varepsilon_2)\varepsilon_2(\varepsilon_3-\varepsilon_2)}\\
&+\frac{\sigmatwo \sigmathree\big((\sigma^{(0)}-\sigmatwo-\sigmathree)\varepsilon_1+\sigmatwo \varepsilon_2+\sigmathree\varepsilon_3\big)^3}{\varepsilon_1 (\varepsilon_2-\varepsilon_1)(\varepsilon_3-\varepsilon_1)}
\bigg]\, .
\end{split}
\end{equation}
The magnetic flux is 
\begin{equation}
    m=\sigma^{(0)}-\sigmaone-\sigmatwo-\sigmathree\, ,
\end{equation}
and $\Bc=-\tfrac{9\ii}{4}m^2$ while the electric charge is more unwieldy and we suppress presenting it. 

A curiosity of the blow-up on-shell action is its relation to the on-shell action of the omega-deformed theory. For the latter we compute the on-shell action on $\mathbb{C}^3$ with three equivariant parameters turned on for the rotations in the three planes. The result is
\begin{equation}\label{IC3}
I_{\mathbb{C}^3_{\varepsilon_i}}=\frac{F_{S^5}}{27 \varepsilon_1\varepsilon_2\varepsilon_3}(\varepsilon_1+\varepsilon_2+\varepsilon_3)^3\, .
\end{equation}
To compare with the blow-up, we set all the $\sigma$'s to be equal. One can understand this requirement by thinking about taking the smooth limit in which the blow-up is removed. Then one finds the curious relation
\begin{equation}
   I_{\mathbb{C}^3_{\varepsilon_i}}-I_{\text{Blow-up}_{\varepsilon_i}} =\frac{8 F_{S^5}}{27}\,,
\end{equation}
and, in particular, the  difference between the two on-shell actions is independent of the equivariant parameters! The same phenomenon occurs for the other conserved charges, in particular taking all the $\sigma$'s equal for the blow-up one has
\begin{equation}
    m=-2\sigma^{(0)}\, ,\quad q=6 \sigma^{(0)}\, ,\quad \Bc =-\frac{9 \ii}{2}\, ,
\end{equation}
while for $\mathbb{C}^3$ they all vanish due to the lack of compact cycles. This observation extends the analogous result in four dimensions \cite{BenettiGenolini:2019jdz}. It would be interesting to understand this from a dual field theory perspective. The conformal boundary of the blown up geometry is also $S^5$, suggesting there may be additional 
saddle point contributions in the path integral.
This is known to 
happen in large $N$ 3d field theories on $S^3$,
holographically dual to minimal $D=4$ gauged supergravity on $\C^2$ blown up at the origin \cite{Martelli:2012sz, Toldo:2017qsh} (the so-called supersymmetric Taub-Bolt AdS solutions). We anticipate an entirely analogous story also in $D=6$ Romans supergravity 
and the dual large $N$ 5d Seiberg theories on $S^5$.


\subsection{Four-dimensional orbifolds}

As a final application of our results, we will study the on-shell action for black holes with four-dimensional orbifold horizons. Using our results we derive the previously conjectured entropy functions for the AdS$_2\times M_4$ orbifold solutions of \cite{Faedo:2022rqx}. 
Globally the topology of the Euclidean solutions is
\begin{equation}
M_6=\mathbb{R}^2\times M_4\, ,
\end{equation}
where we will restrict $M_4$ to be a four-dimensional toric orbifold. 
We will briefly comment at the end of the section how the results may be straightforwardly extended to the case where $\mathbb{R}^2$ is non-trivially fibred over $M_4$, as in section \ref{sec:CP2rot}.

The solutions will have fixed points located at the centre of $\mathbb{R}^2$ and at the corners of the polytope describing $M_4$. It is simple to give toric data for the setup. Let $\vv_a\in \mathbb{Z}^2$, with $a=1,...,d$, denote the toric vectors of $M_4$ with the correct ordering. We remove the usual restriction that the components of the $\vv_a$ need to be relatively prime, which implies that the normal space to such a vector is then an orbifold. The corners of the polytope are smooth if, for neighbouring toric vectors, we have $\det(\vv_a,\vv_{a+1})=\pm 1$. When this is not unity, in this case we have that the space is locally $\mathbb{R}^4/\Gamma_a$, with $\Gamma_a$ a finite group of rank $d_a\equiv|\Gamma_a|=|\det(\vv_a,\vv_{a+1})|$.

We now want to lift the four-dimensional toric vectors to toric vectors of the six-dimensional Euclidean spacetime. In order to do this, we make them 3-dimensional vectors by appending 0 in the first component. We also need to add in the toric vector corresponding to the degeneration of the Killing vector in $\mathbb{R}^2$, in our adapted basis this is $v_0=(1,0,0)$. Therefore the toric vectors $v_{\alpha}$, $\alpha=0,...,d$ are
\begin{equation}
v_0= (1,0,0)\, ,\quad v_a=(0,\vv_a)\, ,
\end{equation}
and to each vector (and therefore facet) we associate a sign for the projection condition $\sigma^{(\alpha)}$.

The R-symmetry vector field is a linear combination of the three independent U$(1)$'s, 
\begin{equation}\label{eq:weights4d}
\xi=b_0 \partial_{\tau}+b_1 \partial_{\varphi_1}+b_2 \partial_{\varphi_2}\, ,
\end{equation}
where $\partial_{\tau}$ rotates $\mathbb{R}^2$, and $\partial_{\varphi_i}$ form a basis of the toric action on the orbifold adapted to the toric data above. For $b_0b_1b_2\neq 0$, as we will assume, the fixed point set consists only of isolated fixed points located at the centre of $\mathbb{R}^2$ and at the corners of the toric diagram describing $M_4$. The 3d toric polytope is the semi-infinite cylinder with base the 2d toric polytope for $M_4$, see for example figure \ref{fig:R2S2S2}, where the base is $S^2\times S^2$. The fixed points are the vertices in the base, described by the neighbouring vectors $v_a, v_{a+1}$ and $v_0$. Using the formulae in appendix \ref{app:toricdata}, we can compute the weights of the corresponding fixed points. Let the vector field $\xi$ be given by $\vec{b}=(b_0,b_1,b_2)$ in the basis of $\mathbb{T}^3$ defined above. At the fixed point given by the intersection of the facets normal to $v_0, v_a, v_{a+1}$ the weights are
\begin{equation}\label{OrbifoldWeights}
\epsilon_1^{(a)}=\frac{\det(\vec{b}, v_a,v_{a+1})}{d_a}\, ,\quad \epsilon_2^{(a)}=\frac{\text{det}(\vec{b},v_{a+1},v_0)}{d_a}\, ,\quad \epsilon_3^{(a)}=\frac{\text{det}(\vec{b},v_{0},v_a)}{d_a} \, ,
\end{equation}
where
\begin{equation}
d_a=\det(v_0,v_a,v_{a+1})\, .
\end{equation}
Notice that given the above toric vectors we have
\begin{equation}
d_a=\det(\vv_a,\vv_{a+1})\, ,\quad \epsilon_1= b_0\, ,\quad \det(\vec{b},v_{0},v_a)=-\det{(\vec{\mathrm{b}},\vv_a)}\, ,
\end{equation}
where $\vec{\mathrm{b}}$ picks out the components of the R-symmetry vector along $M_4$.
To compare with \cite{Faedo:2022rqx}, we define $\varepsilon_1= -2 \sigma^{(0)}b_1/b_0$ and $\varepsilon_1= -2 \sigma^{(0)}b_2/b_0$, where $\varepsilon_i$ are the equivariant parameters of \cite{Faedo:2022rqx}.

Then, substituting these results into the localization formula \eqref{simplifiedaction}, we have
\begin{equation}
\label{Iorbs}
I=\frac{F_{S^5}}{27}\sum_{a} \chi\frac{(\sigma^{(0)}\epsilon_1^{(a)}+\sigma^{(a)}\epsilon_2^{(a)}+\sigma^{(a+1)}\epsilon_3^{(a)} )^3}{d_{a}\epsilon_1^{(a)}\epsilon_2^{(a)}\epsilon_3^{(a)}}\, .
\end{equation}
We can now compare this with equation (5.7) of \cite{Faedo:2022rqx}. We note that we are working in the minimal theory and therefore we need to set $\Phi_1=\Phi_2$ and $\mathfrak{p}_1=\mathfrak{p}_2$ in their formulae. We identify $\chi|_{\mathcal{F}_a}=\sigma^{(0)}\eta_a$ and without loss of generality we can set $\vec{W}=0$ by only summing over the independent line bundles.  One then needs to set the $\varphi_i$ parameters in \cite{Faedo:2022rqx} equal, $\varphi_1=\varphi_2=1$, and identify $\mathfrak{p}_1^{(a)}=\sigma^{(a)}/2$, where the $a$ label indicates the facet of the polytope. With this identification we find perfect agreement. The apparent anomalous factors of $m_l$ are related to the choice of vectors or ``long" vectors as we have chosen. We have therefore proven the form of the entropy functional that was conjectured in \cite{Faedo:2022rqx}! 

We can also compute the magnetic charges of the solution. To each edge of the toric diagram there is an associated compact divisor $D_a$. Computing the magnetic charge threading through this divisor we find
\begin{equation}\label{maorb}\begin{split}
    m_a&=\int_{D_a}c_1(F)\\
    &=\frac{\sigma^{(0)} b_0+\sigma^{(a-1)}\epsilon_2^{(a-1)}+\sigma^{(a)}\epsilon_{3}^{(a-1)}}{d_{a-1}\epsilon_{2}^{(a-1)}}+\frac{\sigma^{(0)}b_0+\sigma^{(a)}\epsilon_2^{(a)}+\sigma^{(a+1)}\epsilon_3^{(a)}}{d_{a}\epsilon^{(a)}_{3}}\\
    &=\frac{\sigma^{(a-1)}}{d_{a-1}}+\frac{\sigma^{(a+1)}}{d_{a}}+D_{aa}\sigma^{(a)}\, ,
    \end{split}
\end{equation}
where $D_{ab}$ is the intersection matrix of the 4d orbifold $M_4$ which is given by
\begin{equation}
    D_{ab}=\begin{cases}
        1/d_{a-1} & b=a-1\, ,\\
        -\det(\mathrm{v}_{a-1},\mathrm{v}_{a+1})/(d_{a-1}d_{a}) & b=a\, ,\\
        1/d_a & b=a+1\, ,\\
        0 &\text{otherwise}\,.
    \end{cases}
\end{equation}
Notice that the $b_0$ cancels because $d_{a-1}\epsilon_2^{(a-1)}=-d_a\epsilon_{3}^{(a)}$, as can be seen from \eqref{OrbifoldWeights}.
Our result for the magnetic charges agrees with equation (5.14) of \cite{Faedo:2022rqx}.

We can also compute the electric charge threading through the orbifold, finding
\begin{equation}
    q=\frac{3 \sigma^{(0)}}{2}\sum_{a=1}^{d}\frac{\sigma^{(a)}\sigma^{(a+1)}}{d_a\epsilon_2^{(a)}\epsilon_3^{(a)}}\big(\sigma^{(0)}b_0+\sigma^{(a)}\epsilon_2^{(a)}+\sigma^{a+1}\epsilon^{(a)}_{3}\big)^2\, ,
\end{equation}
where the weights are as in \eqref{OrbifoldWeights}. It is interesting to note that, if one takes all the $\sigma^{(a)}$'s to be equal, the electric charge is independent of the choice of the R-symmetry vector field. This is analogous to the $\mathbb{CP}^2$ example studied in the previous section.

A natural extension is to consider a non-trivial bundle over the four-dimensional orbifold giving the six-dimensional version of the toric gravitational instantons of \cite{BenettiGenolini:2024hyd}. Here we briefly present how the above formulae extend, following \cite{BenettiGenolini:2023yfe} where essentially the same geometry also appears in a different context. Fibring the $\mathbb{R}^2\cong\mathbb{C}$ over $M_4$, and choosing a lifting of the U$(1)^2$ action on $M_4$ to this complex line bundle, is the same as specifying an \emph{equivariant line bundle} $\mathcal{L}$ over $M_4$. On the other hand, 
a basis of such line bundles $L_a$ is provided by the toric divisors $D_a$, where 
the various U$(1)$ subgroups degenerate. The corresponding equivariant
first Chern class $c^\xi_1(L_a)$, when restricted to the fixed point
labelled by $b$ in $M_4$, is given by the formula
\begin{align}\label{equivc1}
    c^\xi_1(L_a)\mid_b = 
    \delta_{a,b}\epsilon_2^{(a)} + \delta_{a,b+1}\epsilon_3^{(b)}\, ,
\end{align}
where as in the unfibred case the weights are given by 
\begin{align}
\epsilon_2^{(a)}=\frac{\det(\vec{\mathrm{b}},{\mathrm{v}}_{a+1})}{d_a}\, , \qquad \epsilon_3^{(a)}=-\frac{\det(\vec{\mathrm{b}},{\mathrm{v}}_{a})}{d_a}\, .
\end{align}
We may then write $\mathcal{L}=-\sum_{a=1}^dp_a L_a$, with the parameters $p_a\in\Z$ specifying both the topology of this line bundle, but also the lifting of the U$(1)^2$ action. The weight 
$\epsilon_1^{(a)}$ of the 
R-symmetry Killing vector 
on the complex line fibre 
over $M_4$ is (from \eqref{equivc1}) given by 
\begin{align}
    \epsilon_1^{(a)}=b_0-\sum_{b=1}^d p_b  \, c^\xi_1(L_b)\mid_a = b_0 
    - p_a\epsilon_2^{(a)}-p_{a+1}\epsilon_3^{(a)}\, .
\end{align}
With these geometric weights to hand, the on-shell action 
is again given by the same formula \eqref{Iorbs}.

The magnetic charges 
may be computed as in 
\eqref{maorb}, and one finds
\begin{align}
    m_a  = \ &\frac{\sigma^{(a-1)}}{d_{a-1}}+\frac{\sigma^{(a+1)}}{d_{a}}+D_{aa}(\sigma^{(a)}-p_a \sigma^{(0)})-\frac{\sigma^{(0)}}{ d_{a-1}d_a}\big[p_{a+1}d_{a-1}+p_{a-1}d_{a} \big]\, .
\end{align}
Notice that the final result does not depend on the continuous choice of R-symmetry Killing vector $b_i$, as must be the case.


\section{Conclusion}\label{sec:discussion}

In this paper we have derived the general fixed point formula \eqref{main}
 for the holographically renormalized Euclidean on-shell action of 
 supersymmetric solutions to Romans $F(4)$ gauged supergravity theory. 
 As well as recovering results for known solutions, proving some conjectured 
 formulae in the literature, and matching to various dual large $N$ field theory computations, we have also presented large classes of
new results. This opens up various questions and directions for future research.

Firstly, while \eqref{main} is very general, we have made some simplifying 
assumptions in deriving this formula. As in \cite{Alday:2015jsa}, we have studied a ``real'' class of Euclidean solutions in which the non-Abelian R-symmetry gauge field has been truncated to an Abelian sector. Both assumptions exclude some interesting known solutions, 
for example, the former (formally) excludes the complex Euclidean rotating black holes in \cite{Cassani:2019mms}, while the latter excludes 
some of the (Cayley four-cycle) solutions in \cite{Suh:2018tul}.
 It would be interesting to relax these assumptions so as to cover such solutions. 
More interesting would be to couple the minimal Romans 
theory to matter \cite{DAuria:2000afl, Andrianopoli:2001rs}, which would, for example, cover the more general supergravity solutions constructed in \cite{Faedo:2022rqx, Couzens:2022lvg}, 
and the field theory results of \cite{Hosseini:2021mnn}. 
Furthermore, we expect that the Euclidean Romans theory coupled to an arbitrary number of vector multiplets will be equipped with a set of equivariantly closed forms and localization properties, generalizing the recent results of \cite{BenettiGenolini:2024xeo, BenettiGenolini:2024lbj} in $D=4$ to $D=6$. 

Finally, we have presented new results for on-shell actions for various topologies for $M_6$, 
where the conformal boundaries $M_5=\partial M_6$ are 
likewise various interesting five-manifolds. 
The examples in section \ref{sec:examplesII} 
include non-trivial $S^3$ bundles over Riemann surfaces 
and non-trivial $S^1$ bundles over four-manifolds, where the R-symmetry Killing vector has a component along the fibres, but may also mix with the base. As far as we are aware, there are no general field theory computations available to compare to. 
However, 
\cite{Alday:2015lta} analysed the most general 
class of such supersymmetric five-manifold backgrounds, 
and it would clearly be interesting 
to compute the large $N$ limits of 
five-dimensional supersymmetric partition functions
on these backgrounds, generalizing 
\cite{Crichigno:2018adf, Hosseini:2018uzp, Hosseini:2021mnn}, and compare these to our 
Romans supergravity predictions.

\section*{Acknowledgements}
This work was supported in part by STFC grant 
ST/X000761/1 and by CNPq - Brasil (130885/2023-1). TS is supported by Hans und Ria Messer Stiftung.
We would like to thank Alice L\"uscher for useful discussions and Tikz help. 


\appendix

\section{Charge of the Killing spinor}\label{app:charge}

The key result we needed to derive equation \eqref{XP} is that the Killing spinor has zero charge under $\xi$ when we take the supersymmetric gauge for the gauge field:
\begin{equation}\label{eq:KScharge}
    \mathcal{L}_{\xi}\epsilon=0\quad \Leftrightarrow \quad \xi \hook A = \sqrt{2}XP\, ,
\end{equation}
a claim we prove in this appendix. We will give many details since this is representative of some of the computations for computing the polyforms in the main text. 

Computationally, it is easier to prove \eqref{eq:KScharge} by instead showing the equivalent statement that $\bar{\epsilon}\mathcal{L}_{\xi}\epsilon=0$ in the supersymmetric gauge. Recall that the spinorial Lie derivative is
\begin{equation}\label{marlon}
    \mathcal{L}_{\xi}\epsilon\equiv \xi^{\mu}\nabla_{\mu}\epsilon+\frac{1}{8}\dd\xi_{\mu\nu}^{\flat}\gamma^{\mu\nu}\epsilon\, .
\end{equation}
Moreover, provided $\xi$ generates a symmetry of the full solution, the Lie derivative $\mathcal{L}_\xi\epsilon$ of the spinor will itself solve the Killing spinor equation, and without loss of generality we may then consider a spinor satisfying 
$\mathcal{L}_\xi\epsilon=\ii c \mskip2mu \epsilon$, for some constant $c$. 
From \eqref{marlon} we have
\begin{equation}
\bar{\epsilon}\mathcal{L}_{\xi}\epsilon=\xi^{\mu}\bar{\epsilon}(D_{\mu}-\tfrac{\ii}{2}A_{\mu})\epsilon+\frac{1}{8}(\dd \xi)_{\mu\nu}\bar{\epsilon}\gamma^{\mu\nu}\epsilon\, ,
\end{equation}
where $D_{\mu}$ is the gauge covariant derivative defined below \eqref{dilatino}. We may eliminate the $D_{\mu}\epsilon$ term by using \eqref{KSE}, use \eqref{bitK} to rewrite the $\dd \xi^\flat$ term and use the supersymmetric gauge condition to find the unwieldy result
\begin{equation}
\begin{split}\label{Hequation}
 \bar{\epsilon}\mathcal{L}_{\xi}\epsilon =& -\tfrac{\ii}{\sqrt{2}} (XP) S - \tfrac{1}{24\sqrt{2}}X^{-1} B_{\nu \rho}(\xi \hook V)^{\nu \rho} + \tfrac{3 \ii X^{-1}}{8\sqrt{2}} (\xi \hook F) \hook \tilde{K}\\
  &+ \tfrac{\ii }{8 X^2}\left( \tfrac{2\sqrt{2}}{3} X^{-1} \tilde{Y} + \ii  X^4 \xi\hook *H + 
  \sqrt{2} X (P{F} -\tfrac{2}{3} \ii S B)\right)_{\mu\nu} Y^{\mu \nu} \epsilon \\
  &-  \tfrac{1}{4X} (\dd X \wedge \xi^{\flat})_{\mu \nu} \bar\epsilon \gamma^{\mu \nu} \epsilon
  - \tfrac{X^2 \xi^{\mu}}{48}( 3 (*H)_\mu{}^{\tau \sigma} \ii \bar \epsilon \gamma_{\tau \sigma} \epsilon + 3 \bar \epsilon H_\mu{}^{\rho \sigma} \gamma_{\rho \sigma}\gamma_7\epsilon)\, .
\end{split}
\end{equation}
To proceed, one needs to use an assortment of algebraic conditions following from \eqref{dilatino}, and in particular, from Appendix A of \cite{Alday:2015jsa}, one finds
\begin{equation}\label{eq:dilalg}
    \begin{split}
        \partial^{\mu}X \bar{\epsilon}[\mathbb{A},\gamma_{\mu}]_{\mp}\epsilon&=-\frac{\ii}{2\sqrt{2}}(X^2-X^{-2})\bar{\epsilon}[\mathbb{A},\gamma_7]_{\pm}\epsilon+\frac{1}{24}X^3 H^{\mu\nu \rho}\bar{\epsilon}[\mathbb{A},\gamma_{\mu\nu\rho}\gamma_7]_{\pm}\epsilon\\
        &+\frac{\ii}{12\sqrt{2}}B^{\mu\nu}\bar{\epsilon}[\mathbb{A},\gamma_{\mu\nu}]_{\pm}\epsilon-\frac{1}{8\sqrt{2}}F^{\mu\nu}\bar{\epsilon}[\mathbb{A},\gamma_{\mu\nu}\gamma_7]_{\pm}\epsilon\, ,
    \end{split}
\end{equation}
with $\mathbb{A}$ any combination of gamma matrices. One needs to use this identity three times. Taking $\mathbb{A}=1$ and the upper sign in \eqref{eq:dilalg} allows one to eliminate the $B_{\mu\nu}Y^{\mu\nu}$ terms. Using $\mathbb{A}=\gamma_{\alpha}$ with the upper sign again we eliminate the $B_{\nu\rho}(\xi\hook V)^{\nu\rho} $ term and finally for $\mathbb{A}=\gamma_{\alpha}$ this time with the lower sign, one eliminates the remaining $(\xi \hook H)^{\rho \sigma} \tilde{Y}_{\rho \sigma}$ term. Once the dust settles one is left with the result  
\begin{equation}
  \bar\epsilon \mathcal{L}_{\xi} \epsilon =   0\, ,
\end{equation}
as claimed.


\section{Toric geometry from topology}\label{app:toricdata}

For some of our examples it is useful to define the spaces as toric manifolds (orbifolds) in order to compute the weights at fixed points. In this appendix we will first explain how, given the toric data, one can extract out the weights and secondly, how, given a particular topology, one can work out the toric data. 

There are a number of different definitions of, and approaches to, ``toric geometry'', which are largely equivalent, and 
certainly equivalent for our purposes. Here one can either start with a symplectic manifold (orbifold) or a complex 
manifold (orbifold), $X_{2n}$, which admits an effective $n$-dimensional compact torus action preserving the geometric structure. Our supergravity solutions of interest do not necessarily come canonically equipped with a symplectic or complex structure; rather we are simply using the existence of such structures on the underlying manifolds (orbifolds), in the examples we study in the main text, in order to describe global topological data of these spaces using simple combinatorics.
In the symplectic viewpoint, a moment map for the torus action exhibits $X_{2n}$ as a $\mathbb{T}^n$ fibration over a rational simple polytope $\mathcal{P}\subset\mathbb{R}^n$. The facets of the polytope are defined by a set of $d$ outward pointing vectors which define the degeneration of a $S^1\subset \mathbb{T}^n$ on the facet. The polytope is rational if the vectors defining the facets have integer entries whilst the polytope is \emph{simple} if each vertex lies at the intersection of precisely $n$ facets, with the corresponding normal vectors forming a basis for $\Z^n$. In the orbifold case, each facet comes equipped with a positive integer $n_\alpha$ such that the structure group of every point in the inverse image of the facet is $\mathbb{Z}_{n_\alpha}$. Moreover, at the intersection of $n$ facets, defined by the vectors $\{v_1,...,v_n\}$ the singularity has order $|\det(v_1,....,v_n)|$. For our setup to each face we also associate a sign $\sigma$, according to the projection condition \eqref{sigmai}. At each vertex, given that the R-symmetry vector field is a non-trivial linear combination of the $n$ U$(1)$'s, the chirality of the spinor is precisely the product of the $n$ $\sigma$'s of the intersecting faces.


\subsection{Weights from toric geometry}\label{app:weights}

One of the benefits of employing toric geometry is the ability to read off the weights at the fixed points of $\xi$ and to keep track of the independent projection condition signs $\sigma^{(i)}$. Let $\xi$ be a full (i.e. generic coefficients) linear combination of the $n$ U$(1)$'s inside $\mathbb{T}^n$. The fixed point set of $\xi$ are then isolated fixed points where the full torus $\mathbb{T}^n$ degenerates due to our simple assumption. These isolated points are located at the vertices of the polytope, i.e. the intersection of $n$ facets.

We may choose local coordinates $\varphi_{\alpha}$ such that on each of the $n$ facets a single Killing vector degenerates $\partial_{\varphi_{\alpha}}|_{\mathcal{F}_{\alpha}}=0$. With this basis, the vectors $v_i$ defining the vertex are the basis vectors in $\mathbb{R}^n$. In these coordinates the R-symmetry vector field is written as
\begin{equation}
    \xi= \sum_{\alpha=1}^{n}\epsilon_\alpha \partial_{\varphi_{\alpha}}\, ,
\end{equation}
and we can simply read off the weights at the isolated fixed point -- they are just the coefficients $\epsilon_\alpha$. This, however, is not generic. At another vertex these coordinates will not be correctly adapted to that facet and consequently we cannot simply just read off the weights. 

\begin{figure}[h!]
\begin{center}
\tdplotsetmaincoords{60}{110}
\begin{tikzpicture}
		[tdplot_main_coords,
			cube/.style={very thick,black},
			axis/.style={->,blue,thick},
            axisu/.style={->,red,thick},
			inf/.style={dashed,black}]

    \draw[black] (0,0,0) -- (3,0,0) -- (3,3,0) -- (0,3,0) -- (0,0,0);
    \draw[black] (0,0,0) -- (0,0,-2);
    \draw[black] (3,0,0) -- (3,0,-2);
    \draw[black] (0,3,0) -- (0,3,-2);
    \draw[black] (3,3,0) -- (3,3,-2);
	
    \draw[inf] (0,0,-2)--(0,0,-3);
    \draw[inf] (3,0,-2) -- (3,0,-3);
    \draw[inf] (0,3,-2) -- (0,3,-3) ;
	\draw[inf](3,3,-2)--(3,3,-3);
	
\draw[axis] (1.5,3,-1.5)--(1.5,4.5,-1.5) ;
\draw[axis] (0,1.5,-1)--(-1.5,1.5,-1) node[anchor=south]{$v_4$};
\draw[axis] (1.5,1.5,0)--(1.5,1.5,1.5) ;
\draw[axis] (1.5,0,-2)--(1.5,-1.5,-2) node[anchor=east]{$v_3$};
\draw[axis] (3,1.5,-1)--(4.5,1.5,-1) ;

\node at (1.7,4.7,-1.7) {\color{blue} $v_2$};
\node at (1.5,1.5,1.7) {\color{blue} $v_0$};
\node at (4.7,1.7,-1.2) {\color{blue} $v_1$};

\draw[axisu] (3,3,0) -- ( 3,4,0);
\node at (3,3.8,0.4) {\color{red} $u_2^{(1)}$};
\draw[axisu] (3,3,0) -- ( 4,3,0);
\node at (4.2,2.8,0.2) {\color{red} $u_1^{(1)}$};
\draw[axisu] (3,3,0) -- ( 3,3,1);
\node at (3,3.4,1.2) {\color{red} $u_0^{(1)}$};

\draw[axisu] (3,0,0) -- ( 3,-1,0);
\draw[axisu] (3,0,0) -- ( 4,0,0);
\draw[axisu] (3,0,0) -- ( 3,0,1);

\node at (3,-1.1,0.35) {\color{red} $u_3^{(4)}$};
\node at (4.3,-0.1,0.3) {\color{red} $u_1^{(4)}$};
\node at (3,-0.2,1) {\color{red} $u_0^{(4)}$};

\draw[axisu] (0,3,0) -- ( -1,3,0);
\draw[axisu] (0,3,0) -- ( 0,4,0);
\draw[axisu] (0,3,0) -- ( 0,3,1);
\node at (-1.2,3.3,0) {\color{red} $u_4^{(2)}$};
\node at (0.8,4.4,0.25) {\color{red} $u_2^{(2)}$};
\node at (0,3.2,1.35) {\color{red} $u_0^{(2)}$};

\draw[axisu] (0,0,0) -- ( -1,0,0);
\draw[axisu] (0,0,0) -- ( 0,-1,0);
\draw[axisu] (0,0,0) -- ( 0,0,1);
\node at (-1.5,0,0) {\color{red} $u_4^{(3)}$};
\node at (-0.2,-1.2,0) {\color{red} $u_3^{(3)}$};
\node at (0,-0.25,1.25) {\color{red} $u_0^{(3)}$};

\end{tikzpicture}
\end{center}
\caption{Toric diagram for $\R^2 \times S^2 \times S^2$, complete with both the vectors $v_i$ and dual vectors $u_i$. }
\label{fig:ToricR2S2S2}
\end{figure}
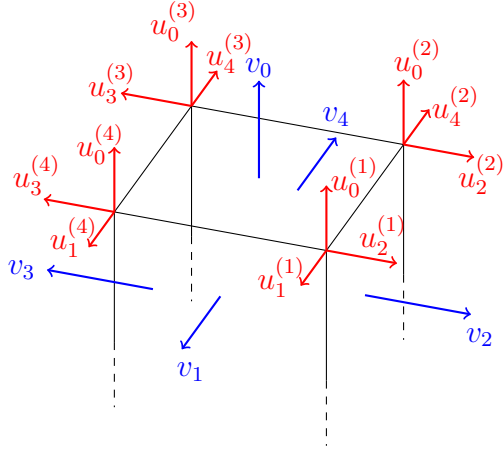

Fortunately toric geometry allows us to understand how we can change to adapted coordinates at each vertex. Let us work in a basis where the $\mathbb{T}^n$ is generated by $\partial_{\varphi_{\alpha}}$, and consequently the outward pointing normals are defined in this basis. Define a vertex by $n$ outward pointing normals $v_i$, $i=\{1,...,n\}$, which are arranged in a clockwise manner around the vertex. The Killing vector which degenerates on each facet is given by
\begin{equation}
\partial_{\psi_i}=\sum_{\alpha=1}^{n}v_i^{\alpha}\partial_{\varphi_{\alpha}}\, .
\end{equation}
If we write the Killing vector $\xi$ in terms of these coordinates the weights can be read off as above. We therefore want to find a way to express $\xi$ in terms of this basis. It is convenient to introduce the dual vectors $u_i$ such that $u_i\cdot v_j=\delta_{ij}$, see figure \ref{fig:ToricR2S2S2}. These are outward pointing vectors along each facet. Computationally, one can read these vectors off by doing
\begin{equation}\label{eq:ufromv}
    (u_1,...,u_n)^{T}=(v_1,...,v_n)^{-1}\, ,
\end{equation}
where we view the vectors as the rows of the matrices. To each vertex we then associate $n$ $v$'s and $n$ $u$'s, where the latter are the dual basis of the $v$'s. Writing the Killing vector $\xi$ as the vector $\xi=(b_1,..,b_n)$, the weights can be obtained by computing 
\begin{equation}\label{eq:weightsu}
    (\epsilon_1,...,\epsilon_n)=\xi\cdot (v_1,...,v_n)^{-1}= (\xi\cdot u_1,...,\xi \cdot u_n)\,.
\end{equation}
The norm of the factor in the denominator is the dimension of the singularity at the fixed point. 

An alternative method for computing the weights is to skip the computation of the dual basis $u$'s and to instead compute determinants of the $v$'s. Consider again a vertex as above, defined by the $n$ vectors $\{v_1,...,v_n\}$. The weights are given by
\begin{equation}\label{eq:weightsdet}
    \epsilon_{i}= \frac{\det(\xi,v_{i+1},...v_n,v_1,...,v_{i-1})}{\det(v_1,....,v_n)}\, .
\end{equation}
To see why these give the same result observe that the $u$'s can be written as
\begin{equation}
    u_i^{\alpha}=\frac{1}{\det(v_1,...,v_n)}\frac{1}{(n-1)!}\epsilon^{\alpha\beta_1...\beta_{n-1}}\epsilon_{ij_1...j_{n-1}}v^{\beta_1}_{j_1}...v^{\beta_{n-1}}_{j_{n-1}}\, ,
\end{equation}
where we have used \eqref{eq:ufromv}. It then follows, by using the definition of the determinant in terms of the Levi--Civita symbol, that \eqref{eq:weightsu} and \eqref{eq:weightsdet} are equivalent.

\subsection{Toric data for \texorpdfstring{$\mathcal{O}(-p)\rightarrow S^2$}{O(-p)->S2} }\label{sec:ToricO(p)S2}

In this section we will explain how to write the toric data for the non-compact four-manifold $\mathcal{O}(-p)\rightarrow S^2$. Using the formulae from the previous section we will then be able to extract out the weights needed in the localization formulae of the main text. 

First consider a round two-sphere $S^2$ and introduce the two patches
\begin{equation}
\begin{split}
U_N&=\{(x,y,z)\in \mathbb{R}^3\big| ~|\vec{x}|^2=1, z>-|\epsilon|\}\, ,\\
U_S&=\{(x,y,z)\in \mathbb{R}^3\big|~ |\vec{x}|^2=1, z<|\epsilon|\}\, ,
\end{split}
\end{equation}
with $\epsilon$ a small (say $|\epsilon|<1/2$) parameter, which cover the $S^2$. Their intersection includes a circle on the equator at $z=0$. We can introduce polar coordinates on the $S^2$, $(\theta,\varphi)$ with $\theta\in[0,\pi]$ and $\varphi \sim\varphi +2\pi$. In addition, we take the standard volume form on the $S^2$, $\vol=\sin\theta \mskip1mu\dd\theta \wedge \dd\varphi$ which defines an orientation and symplectic structure. Near to the north pole, we introduce coordinates $\rho_N=\theta$ and $\varphi_N\equiv \varphi$, which are the standard polar coordinates on $\mathbb{R}^2_N$. Expanding around the north pole, we find that the volume form becomes $\vol=\rho_N \dd\rho_N\wedge \dd \varphi_N$, which is the standard volume form on $\mathbb{R}^2$. On the other hand, near to the south pole, we introduce $\rho_S=\pi-\theta\geq 0$ and $\varphi_S\equiv -\varphi$ which are standard polar coordinates on $\mathbb{R}^2_S$. Note that the sign of $\varphi_S$ differs with the north pole and is required for the volume form on $\mathbb{R}^2_S$ to take the correct form, namely $\vol=\rho_S\dd\phi_S \wedge \dd\varphi_S$. 
To specify the gluing of the two patches, we need to give the transition function, which, from the discussion above, must be
\begin{equation}
\varphi_N=\varphi=-\varphi_S\, .
\end{equation}

Having described $S^2$, we can now define the bundle $\mathcal{O}(-p)$ over the $S^2$. We have two patches again, which we refer to as the North and South once more. On the north patch, $\mathcal{U}_N=U_N\times \mathbb{C}$, we introduce the coordinates $(\rho_N,\varphi)\times z_N$, where $z_N$ is a complex coordinate on $\mathbb{C}$. It is useful to write the coordinate on $\mathbb{C}$ as $z_N=r_N \me^{\ii \psi_N}$. On the South patch, $\mathcal{U}_S= U_S\times \mathbb{C}$, we introduce the coordinates $(\rho_S,\varphi_S)\times z_S$, where as before $z_S$ is a complex coordinate on $\mathbb{C}$ which we write as $z_S=r_S\me^{\ii \psi_S}$. 
The total space $\mathcal{O}(-p)$ is made by gluing on the overlap according to
\begin{equation}
\begin{split}
\varphi_N&=-\varphi_S=\varphi\, ,\\
\psi_N&=\psi_S-p\mskip1mu\varphi\, .
\end{split}
\end{equation}
Note that, since all of the coordinates are $2\pi$-periodic, one must take $p\in \mathbb{Z}$ and the sign of $p$ in the last step is fixed by the definition. 

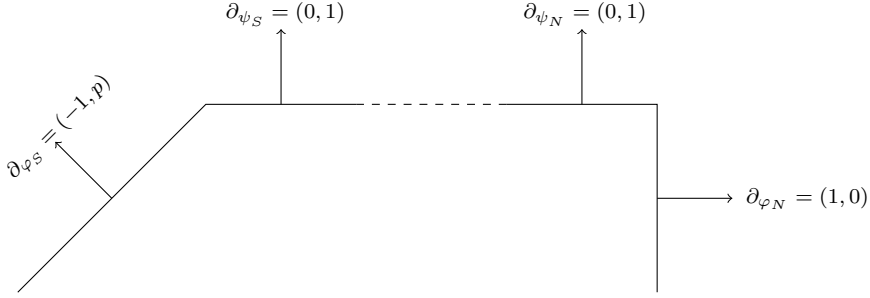
\begin{figure}[h!]
\begin{center}
\begin{tikzpicture}

\draw (0,-2.5)--(0,0)--(-2,0);
\draw[dashed] (-2,0)--(-4,0);
\draw (-4,0)--(-6,0)--(-8.5,-2.5);

\draw[->] (0,-1.25)--(1,-1.25);
\draw[->] (-1,0)--(-1,1);
\draw[->] (-5,0)--(-5,1);
\draw[->] (-7.25,-1.25)--(-8,-0.5);

\node at (2,-1.25) {\scriptsize$\partial_{\varphi_N}=(1,0)$};

\node at (-0.96,1.2) {\scriptsize$\partial_{\psi_N}=(0,1)$};

\node at (-4.94,1.2) {\scriptsize$\partial_{\psi_S}=(0,1)$};

\node at (-7.95,-0.3) [ rotate=45] {\scriptsize$\partial_{\varphi_S}=(-1,p)$};

\end{tikzpicture}
\end{center}
\caption{The toric diagram for $\mathcal{O}(-p)\rightarrow S^2$ with $p>0$.}\label{fig:OpS2}
\end{figure}

We now want to study the natural $\mathbb{T}^2$ action on this bundle. First pick a pole, without loss of generality we take the North pole. We need to construct a basis for the $\mathbb{T}^2$ action and we make the obvious choice
\begin{equation}
\begin{split}
\partial_{\phi_1}&\equiv \partial_{\varphi_N}=\partial_{\varphi}\, ,\\
\partial_{\phi_2}&\equiv \partial_{\psi_N}\, ,
\end{split}
\end{equation}
though any basis will do. It follows that the coordinates for our basis are:
\begin{equation}
\phi_1=\varphi_N=\varphi\, ,\quad \phi_2=\psi_N\, .
\end{equation}
We now want to study the $\mathbb{T}^2$ action on $\mathcal{U}_S$. We have that the coordinates on $\mathcal{U}_S$ in terms of our new basis are
\begin{equation}
\begin{split}
\varphi_S&=-\varphi_N=-\phi_1\, ,\\
\psi_S&=\psi_N+p \mskip1mu\varphi=\phi_2+p\mskip1mu \phi_1\, ,
\end{split}
\end{equation}
such that the degenerating Killing vectors are
\begin{equation}
\begin{split}
\partial_{\psi_S}=\partial_{\phi_2}\, ,\qquad \partial_{\varphi_S}=-\partial_{\phi_1}+p\mskip1mu\partial_{\phi_2}\, .
\end{split}
\end{equation}
The final step is to write these in terms of the basis $\{\partial_{\phi_1},\partial_{\phi_2}\}$,
\begin{equation}
\partial_{\psi_S}=(0,1)\cdot(\partial_{\phi_1},\partial_{\phi_2})^{T}\, ,\qquad \partial_{\varphi_S}=(-1,p)\cdot (\partial_{\phi_1},\partial_{\phi_2})^{T}\, .
\end{equation}
We read off the three vectors describing $\mathcal{O}(-p)\rightarrow S^2$ to be
\begin{equation}
    v_1=(-1,p) \, ,\quad v_2=(0,1)\, ,\quad v_3=(1,0)\,.
\end{equation}
We have drawn the polytope described by these vectors in figure \ref{fig:OpS2}.
Note that for $p<0$, the diagram is not convex (correspondingly, the total space of $\mathcal{O}(-p)$ is not holomorphically convex). Moreover, for $p=0$, the toric diagram is simply the open cup.


\subsection{Toric data for \texorpdfstring{$\mathcal{O}(-p_1,-p_2)\rightarrow S^2_1\times S^2_2$}{O->S2xS2}}\label{app:O(p1p2)S2S2}

We can now run a similar argument for $\mathcal{O}(-p_1,-p_2)\rightarrow S^2_1\times S^2_2$. The difference is simply the choices of patches. We use the same patches on the two-spheres as above, and the patches on the 6d space are
\begin{equation}
  \begin{split}
    \mathcal{U}_{NN} &= U_{1,N} \times U_{2,N} \times \mathbb{C}\, , \quad 
  \mathcal{ U}_{NS} = U_{1,N}\times U_{2,S} \times \mathbb{C}\, , \\ 
    \mathcal{U}_{SN} &= U_{1,S}\times U_{2,N} \times \mathbb{C} \, ,\quad ~
    \mathcal{U}_{SS} = U_{1,S}\times U_{2,S} \times \mathbb{C}\, ,
  \end{split}
\end{equation}
where $U_{1,N}$ is the patch covering the northern hemisphere of the first sphere, etc.
Analogously to the previous example, we introduce coordinates on all patches with labels consistent with those of the patches. We pick the obvious basis for the $\mathbb{T}^3$ action working in the patch $\mathcal{U}_{NN}$, 
\begin{equation}
  \begin{split}
    \partial_{\phi_1} \equiv\partial_{\varphi_{1,N}}\, ,\quad
    \partial_{\phi_2} \equiv \partial_{\varphi_{2,N}}\, ,\quad
    \partial_{\phi_3} \equiv\partial_{\psi_{NN}} \, .
  \end{split}
\end{equation}
This gives us three toric vectors, $v_0,v_1,v_2$ which are the canonical basis vectors on $\mathbb{R}^3$.

We now want to see how the other degenerating Killing vectors are written in terms of this basis. Consider first the degenerating Killing vector at the south pole of the second two-sphere. We need to go from the $\mathcal{U}_{NN}$ patch to the $\mathcal{U}_{NS}$ patch. The transition functions on the overlap are
\begin{equation}
  \begin{split}
    \varphi_{2,N} & = - \varphi_{2,S} \, ,\quad 
    \psi_{NN}  =  \psi_{NS} - p_2 \, \varphi_{2,N}\, .
  \end{split}
\end{equation}
Thus we find
\begin{equation}
\partial_{\varphi_{1,N}}=\partial_{\phi_1}\, ,\quad \partial_{\varphi_{2,S}} = - \partial_{\phi_2} + p_2 \, \partial_{\phi_3} \, , \quad \partial_{\psi_{NS}}= \partial_{\phi_3} \, .
\end{equation}
We therefore have the new toric vector $v_3=(0,-1,p)$.

Finally the transition function on the overlap of  $\mathcal{U}_{NN}$ and $\mathcal{U}_{SN}$ implies
\begin{equation}
  \begin{split}
    \varphi_{1,N}  = - \varphi_{1,S} \, ,\quad 
    \psi_{NN}  =  \psi_{SN} - p_1 \, \varphi_{1,N}\, ,
  \end{split}
\end{equation}
and therefore 
\begin{equation}
    \partial_{\varphi_{1,S}} = - \partial_{\phi_{1}} + p_1 \partial_{\phi_3}\, ,\quad \partial_{\varphi_2,N}=\partial_{\phi_2}\, ,\quad \partial_{\psi_{SN}}= \partial_{\phi_3}\, .
\end{equation}
One can similarly perform the same computations in the patch $\mathcal{U}_{SS}$. However, this is redundant since we have now managed to express all five of the degenerating Killing vectors in terms of our basis.

The toric data for $\mathcal{O}(-p_1,-p_2)\rightarrow S^2 \times S^2$, in our chosen basis, is
\begin{align}
    v_0 &=  (0, 0, 1)\, ,
&v_1& = (1, 0, 0)\, ,\qquad\qquad
    v_2= (0, 1, 0) \, ,\\
    v_3&= (-1, 0, p_1)\, ,
    &v_4&= (0, -1, p_2)\, .\nonumber
  \end{align}
These are then the outward pointing normals to the faces of the polytope and describe the toric manifold $\mathcal{O}(-p_1,-p_2)\rightarrow S^2 \times S^2$. This toric data will be used in section~\ref{sec:O(-p1,-p2)xS2xS2}. For $p_1=1, p_2=2$ the polytope is given in figure \ref{fig:ToricO(1,2)S2S2}.


\subsection{Toric data for \texorpdfstring{$\mathcal{O}(-p_1 )\oplus \mathcal{O}(-p_2) \rightarrow S^2$}{O(-p1)+O(-p2)->S2} }\label{app:Op1Op2S2}

Our penultimate example is $\R^4$ fibred over $S^2$.  We decompose $\R^4$ into two complex line bundles which we fibre over the $S^2$. Just like before, we use the two patches $U_N$ and $U_S$ on $S^2$, defined in section \ref{sec:ToricO(p)S2}. From these we construct the two patches $\mathcal{U}_{N}$ and $\mathcal{U}_S$ which cover the whole manifold. On each patch, we introduce the coordinates $(\rho_I,\varphi_I)\times (z_{1I} , z_{2I})$, where $I=N,S$ and $z_{iI}=r_{iI}\ex^{\ii \psi_{iI}}$. 

On the overlap of $\mathcal{U}_N$ and $\mathcal{U}_S$, the transition functions give
\begin{align}\label{eq:Op1Op2S2transition}
    \varphi_N = - \varphi_S \, ,\quad 
    \psi_{iN} = \psi_{iS} - p_i \varphi_N\, .
\end{align}
We pick the following basis for the $\mathbb{T}^3$ action working in the $\mathcal{U}_N$ patch,
\begin{align}
    \partial_{\phi_1} &\equiv \partial_{\varphi_N}\, , \quad 
    \partial_{\phi_2} \equiv \partial_{\psi_{1N}}\, ,\quad
    \partial_{\phi_3} \equiv \partial_{\psi_{2N}}\, .
\end{align}
Using the transition function \eqref{eq:Op1Op2S2transition}, on the overlap we have that the 
\begin{equation}
    \varphi_S=-\phi_1\, ,\quad \psi_{1 S}=\phi_2 +p_1\phi_1 \, ,\quad \psi_{2 S}=\phi_2+p_1 \phi_1\, .
\end{equation}
Consequently, we find
\begin{align}
    \partial_{\varphi_S} = -\partial_{\phi_1} + p_1 \partial_{\phi_2} + p_2 \partial_{\phi_3}\,, \quad 
    \partial_{\psi_{1S}}=\partial_{\phi_2}\, , \quad  \partial_{\psi_{2S}}= \partial_{\phi_3}\, . 
\end{align}
It is once again simple to extract the toric data from the above analysis, one finds
\begin{equation}
  \begin{split}
    v_1 &=  (1, 0, 0)\, ,\quad
    v_2 = (0, 1, 0)\, ,\\
    v_3 &=  (0, 0,1)\, , \quad 
    v_4 =  (-1, p_1, p_2)\, .
  \end{split}
\end{equation}
The polytope associated to $\mathcal{O}(-p_1 )\oplus \mathcal{O}(-p_2) \rightarrow S^2$ has four faces and two vertices corresponding to two fixed points. 


\subsection{Toric data for \texorpdfstring{$\mathcal{O}(-p)\rightarrow \mathbb{ CP}^2$}{O->CP2} }\label{app:CP2}
The final set of toric data we shall derive is for $\mathcal{O}(-p)\rightarrow \mathbb{CP}^2$. The computation follows the same ideas as the previous section, but is slightly more complicated since $\mathbb{CP}^2$ has three charts. The complex weighted projective space $\mathbb{CP}^2$ is defined by $\mathbb{CP}^2=\{ (z_0,z_1,z_2)\in \mathbb{C}^3 | (z_0,z_1,z_2)\equiv \lambda  (z_0,z_1,z_2)\, ,~~ \lambda \in \mathbb{C}^*\} $. We may introduce three patches on $\mathbb{CP}^2$ which cover the manifold. We define the patches as $U_\mu = \left \{ z_\mu \neq 0 \right \}$, for $\mu=\{0,1,2\}$. The coordinates on each chart $U_\mu$ are taken to be $\xi_\mu ^\nu \equiv z_\nu /z_\mu$; note that $\xi^{\mu}_{\mu}=1$. It is convenient to write the non-trivial coordinates as $\xi_\mu ^\nu = r_\mu^\nu \me^{\ii \varphi_\mu^\nu }$. On the overlap of two charts $U_\mu$ and $U_\nu$, $\mu\neq \nu$, the coordinates are related by
\begin{equation}
    \xi^\lambda_\mu = \frac{z_\nu}{z_\mu}\xi_\nu^\lambda\, .
\end{equation}

Using the patches on $\mathbb{CP}^2$ defined above, we may define three patches, $\mathcal{U}_{\mu}=U_{\mu}\times \mathbb{C}$, which cover the total space $\mathcal{O}(-p)\rightarrow \mathbb{CP}^2$. We introduce the local coordinates $(\xi^\nu_\mu , w_\mu)$ on each of the patches, and it is useful to write these coordinates as $\left (r_\mu^\nu \ex^{\ii \varphi_\mu^\nu } ,s_\mu \ex^{\ii \psi_\mu}\right)$. On the intersection of $\mathcal{U}_{\mu}$ and $\mathcal{U}_{\nu}$, with $\mu\neq \nu$ the angular coordinates are related via 
\begin{align}
    \varphi^\lambda_\mu = \varphi^\nu_\mu + \varphi^\lambda_\nu\, ,\quad 
    \psi_\mu = \psi_\nu -p\mskip1mu\varphi_\mu^\nu\, .
\end{align}
The former is a consequence of the transition functions on $\mathbb{CP}^2$, while the latter is the definition of the bundle $\mathcal{O}(-p)$. Let us now introduce a basis for the $\mathbb{T}^3$ action. Working in the patch $\mathcal{U}_0$, define the basis
\begin{equation}
    \phi_1\equiv \varphi_0^1\, ,\quad \phi_2\equiv\varphi_0^2 \, ,\quad \phi_3\equiv \psi_0\, .
\end{equation}
We have that $\partial_{\phi_1}$ degenerates at $z_1=0$, $\partial_{\phi_2}$ degenerates at $z_2=0$, and $\partial_{\phi_3}$ degenerates at $w_0=0$. To complete the data we need to understand the degenerating vector field at $z_0=0$. We choose to work in the patch $\mathcal{U}_1$, though one could equally work in the patch $\mathcal{U}_2$. On the intersection $\mathcal{U}_0\cap \mathcal{U}_1$ we have that the angular coordinates are related by
\begin{equation}
    \varphi_1^0=\varphi_0^1\, ,\quad \varphi_1^2=\varphi^2_0+\varphi_1^0\, ,\quad \psi_1=\psi_0-p\mskip1mu\varphi_1^0\,,
\end{equation}
and therefore we have
\begin{equation}
    \partial_{\varphi_1^0}=-\partial_{\phi_1}-\partial_{\phi_2}+p \mskip1mu\partial_{\psi_0}\, .
\end{equation}
We have now found all degenerating vector fields in terms of our basis and we find that the toric data for $\mathcal{O}(-p)\rightarrow \mathbb{CP}^2$ is
\begin{equation}
  \begin{split}
    v_0 =  (0,0,1)\, ,\quad
    v_1 =  (1, 0, 0)\, ,\quad 
    v_2 = (0, 1, 0)\, ,\quad 
    v_3 =  (-1, -1,p)\, .
    \end{split}
\end{equation}
Observe that the toric data for $\mathbb{CP}^2$ is simply the first two entries of $v_1,v_2$ and $v_3$ with $p=0$ and thus the toric vectors are consistent.


\section{Charge conjugated spinor}\label{app:chargeconjugate}

At the end of section \ref{sec:SUGRA}, we discussed that a geometry is supersymmetric provided  there exists a doublet of Killing spinors, $(\epsilon_1 ,\epsilon_2 )$, satisfying the Killing spinor equations \eqref{KSE}. We further made an assumption on the spinors which related the second spinor to the charge conjugate of the other spinor, that is we took $(\epsilon_1 ,\epsilon_2 )=(\epsilon, \epsilon^c )$. This was motivated by the requirement for a well-defined Lorentzian theory after Wick rotation. Throughout the paper, we have worked exclusively with the spinor $\epsilon$ and its charge conjugate has not played any role. In this appendix we show that this choice is made without loss of generality. Though intermediate results depend on the choice of $\epsilon$ or $\epsilon^c$, the final results for physical observables are independent of this choice.

\paragraph{Gauge Field Flux}
First consider the Killing spinor equations obeyed by $\epsilon$ and $\epsilon^c$. Schematically they take the form
\begin{equation}
\begin{split}
    \nabla \epsilon + \tfrac{\ii}{2}A \epsilon &= \mathcal{M}\epsilon + F \mathcal{N}\epsilon \,, \\
    \nabla \epsilon^c - \tfrac{\ii}{2}A \epsilon^c &= \mathcal{M}\epsilon^c - F \mathcal{N}\epsilon^c \, ,
\end{split}
\end{equation}
where we have kept explicit terms depending on the gauge field and its field strength.

One can see immediately that the equations are formally translated into each other on defining $A^c\equiv -A$. This means that $\epsilon^c$ is a section of $\mathcal{S}M_6\otimes \mathcal{L}^{-1/2}$ whereas $\epsilon$ is a section of $\mathcal{S}M_6\otimes \mathcal{L}^{1/2}$. Thus, in the notation introduced in section \ref{sec:gaugefield}, the magnetic charge for the charge conjugated spinor is $m^c = -m$.

\paragraph{Bilinears and invariance of the action} In addition to the sign change in the gauge field, one also finds that select bilinears  in \eqref{bilineardefs} pick up additional minus signs if we had chosen to work with $\epsilon^c$ rather than $\epsilon$.  Without loss of generality, we may work in the basis where the $\gamma_\mu$ are anti-symmetric and purely imaginary. There are two classes of bilinears that we are interested in, $\mathcal{B} = \bar{\epsilon}\gamma_{(n)}\epsilon$ and $\mathcal{V}=\bar{\epsilon}\gamma_{(n)}\gamma_7 \epsilon$. For the charge conjugate spinor one finds 
\begin{equation}
   \mathcal{B}_n^c=(-1)^{\tfrac{n(n-1)}{2}}\mathcal{B}_n\, ,\quad \mathcal{V}_n^c=(-1)^{\tfrac{n(n+1)}{2}}\mathcal{V}_n\, .
\end{equation}
In particular, this implies that $S^c =S$, $P^c=-P$, $\xi^{ c\flat} = \xi^\flat$, $Y^c = -Y$ and $\tilde{Y}^c=\tilde{Y}$. One also observes that $-\ii \gamma^{(2i-1)2i}\epsilon^c = -\sigmai \epsilon^c $ and $\gamma_7 \epsilon^c = -\chi \epsilon^c$. Substituting these results into \eqref{actionwithsigns}, one sees that the signs conspire to cancel, and the final result is independent of choosing whether to work with $\epsilon$ or $\epsilon^c$.


\section{The \texorpdfstring{$\frac{1}{2}$}{1/2} BPS black hole}\label{app:half}

In this appendix we give further details for the $1/2$ BPS black hole solution studied in section \ref{sec:hyperbolicBH}. Since the solution is $1/2$ BPS and thus preserves four supercharges, there are four distinct choices of R-symmetry vector field that one can use in the localization procedure. As we will show shortly, these four choices take the universal form
\begin{equation}\label{eq:xiH4fixed}
    \xi=\chi\left(\frac{\sigma^{(1)}}{n}\partial_{\tau}+\sigma^{(2)}\partial_{\phi_1}+\sigma^{(3)}\partial_{\phi_2}\right)\, .
\end{equation}
To exemplify the choice of signs $\sigma^{(i)}$, we will study the explicit solution, using this as a test case for the more complicated geometries where no explicit solution is known. 

The explicit solution, found in \cite{Alday:2014fsa}, has metric\footnote{Note that our $\tau$ coordinate is $2\pi$ periodic whilst the coordinate in \cite{Alday:2014fsa} is $2\pi n$ periodic. } 
\begin{equation}
\dd s^2=\frac{H(r)^{1/2}}{f(r)} \dd r^2 + \frac{9 f(r)}{2 H(r)^{3/2}}n^2 \dd \tau^2 + r^2 H(r)^{1/2} \dd s_{\mathbb{H}^4}^2 \, ,
\end{equation}
where
\begin{equation}\label{H&f1/2BH}
  H(r)= 1 + \frac{Q}{r^3}, \quad f(r)= -1 + \frac{2}{9} r^2 H(r)^2\, ,
\end{equation}
with $Q$ the charge of the black hole. We realise the hyperbolic space with a spherical slicing and put the following metric on it
\begin{equation}
     \dd s_{\mathbb{H}^4}^2 = \frac{1}{1 + q^2} \dd q^2 + q^2 \big(\dd \theta^2 +\cos^2\theta \dd \phi_1^2 + \sin^2\theta \dd \phi_2^2 \big)\, .
\end{equation} 
The gauge field supporting the solution is
\begin{equation}
    A=3 \frac{H(r)-1}{H(r)} n \dd \tau+ (1-n)\dd \tau\,,
\end{equation}
where we have fixed a pure gauge term in order to preserve the supersymmetric gauge \eqref{Asusy}. Define
\begin{equation}
    f_1(r)=\frac{r^{1/8}\sqrt{3 \sqrt{2} r^2 +2 r^3 +2 Q}}{(r^3+Q)^{3/8}}\, ,\qquad  f_2(r)=\frac{r^{1/8}\sqrt{-3 \sqrt{2} r^2 +2 r^3 +2 Q}}{(r^3+Q)^{3/8}}\, , 
\end{equation}
then with this choice of gauge the Killing spinors are
\begin{align}
    \begin{split}
        \epsilon = \frac{ \sqrt{\sqrt{q^2+1}+1}}{4}
        &\begin{pmatrix}
            f_1(r) \left(\kappa_1 \me^{-\ii \phi_2 }+\kappa_2 \me^{\ii \phi_1 }\right) \me^{-\frac{1}{2} \ii (\theta +\tau+\phi_1 -\phi_2 )}   \\
           \ii f_1(r) \left(\kappa_1 \me^{-\ii \phi_2 }-\kappa_2 \me^{\ii \phi_1 }\right) \me^{-\frac{1}{2} \ii (-\theta +\tau+\phi_1 -\phi_2 )} \\
            f_2(r) \left(\kappa_4+\kappa_3 \me^{\ii (\phi_1 -\phi_2 )}\right) \me^{-\frac{1}{2} \ii (-\theta +\tau+\phi_1 -\phi_2 )}  \\
            \ii f_2(r) \left(-\kappa_4+\kappa_3 \me^{\ii (\phi_1 -\phi_2 )}\right) \me^{-\frac{1}{2} \ii (\theta +\tau+\phi_1 -\phi_2 )} \\
            -\ii f_2(r) \left(\kappa_1 \me^{-\ii \phi_2 }+\kappa_2 \me^{\ii \phi_1 }\right) \me^{-\frac{1}{2} \ii (\theta +\tau+\phi_1 -\phi_2 )} \\
            f_2(r) \left(\kappa_1 \me^{-\ii \phi_2 }-\kappa_2 \me^{\ii \phi_1 }\right) \me^{-\frac{1}{2} \ii (-\theta +\tau+\phi_1 -\phi_2 )} \\
            -\ii f_1(r) \left(\kappa_4+\kappa_3 \me^{\ii (\phi_1 -\phi_2 )}\right) \me^{-\frac{1}{2} \ii (-\theta +\tau+\phi_1 -\phi_2 )} \\
            f_1(r) \left(-\kappa_4+\kappa_3 \me^{\ii (\phi_1 -\phi_2 )}\right) \me^{-\frac{1}{2} \ii (\theta +\tau+\phi_1 -\phi_2 )}
        \end{pmatrix}\\
        +\frac{q}{4\sqrt{\sqrt{q^2+1}+1}}
        &\begin{pmatrix}
           \ii  f_1(r) \left(\kappa_4+\kappa_3 \me^{\ii (\phi_1 -\phi_2 )}\right) \me^{-\frac{1}{2} \ii (-\theta +\tau+\phi_1 -\phi_2 )} \\
            f_1(r) \left(\kappa_4-\kappa_3 \me^{\ii (\phi_1 -\phi_2 )}\right) \me^{-\frac{1}{2} \ii (\theta +\tau+\phi_1 -\phi_2 )} \\
            -\ii f_2(r) \left(\kappa_1 \me^{-\ii \phi_2 }+\kappa_2 \me^{\ii \phi_1 }\right) \me^{-\frac{1}{2} \ii (\theta +\tau+\phi_1 -\phi_2 )} \\
            f_2(r) \left(\kappa_1 \me^{-\ii \phi_2 }-\kappa_2 \me^{\ii \phi_1 }\right) \me^{-\frac{1}{2} \ii (-\theta +\tau+\phi_1 -\phi_2 )} \\
            f_2(r) \left(\kappa_4+\kappa_3 \me^{\ii (\phi_1 -\phi_2 )}\right) \me^{-\frac{1}{2} \ii (-\theta +\tau+\phi_1 -\phi_2 )} \\
            \ii f_2(r) \left(-\kappa_4+\kappa_3 \me^{\ii (\phi_1 -\phi_2 )}\right) \me^{-\frac{1}{2} \ii (\theta +\tau+\phi_1 -\phi_2 )} \\
            -f_1(r) \left(\kappa_1 \me^{-\ii \phi_2 }+\kappa_2 \me^{\ii \phi_1 }\right) \me^{-\frac{1}{2} \ii (\theta +\tau+\phi_1 -\phi_2 )} \\
           \ii f_1(r) \left(\kappa_2 \me^{\ii \phi_1 }-\kappa_1 \me^{-\ii \phi_2 }\right) \me^{-\frac{1}{2} \ii (-\theta +\tau+\phi_1 -\phi_2 )} 
        \end{pmatrix}
    \end{split}\, ,
\end{align}
where $\kappa_a$ are four real constants giving the four independent Killing spinors. The metric admits three U$(1)$ Killing vectors: $\partial_{\tau}$, $\partial_{\phi_1}$ and $\partial_{\phi_2}$ in terms of which we can write the R-symmetry vector
\begin{equation}
    \xi=\bar{\epsilon}\gamma^{\mu}\epsilon \partial_{\mu}\, ,
\end{equation}
for each of the four Killing spinors. With the above Killing spinors one finds
\begin{equation}\label{eq:xiH4}
    \xi=\frac{(-1)^{\kappa_3+\kappa_4}}{n}\partial_{\tau} +(-1)^{\kappa_2+\kappa_4}\partial_{\phi_1} + (-1)^{\kappa_2+\kappa_3}\partial_{\phi_2}\, ,
\end{equation}
where here it is understood that one keeps only one of the $\kappa$'s non-zero, setting the non-zero one to $1$. In each case the square norm of the Killing vector field is
\begin{equation}
    \|\xi\|^2=r^2 H(r)^{1/2} q^2+\frac{9 f(r)}{2 H(r)^{3/2}}\, ,
\end{equation}
which vanishes at $q=0, r=r_h$ with $f(r_h)=0$.

We now want to understand how to extract the $\sigma^{(i)}$'s from the explicit solution. Recall that at an isolated fixed point we take our normal space to decompose as $\mathcal{N}= \mathbb{C}\oplus \mathbb{C}\oplus \mathbb{C}$ with a single Killing vector acting on a single copy of $\mathbb{C}$. We want to understand the restriction of the Killing spinor to each of these copies of $\mathbb{C}$ and the projection condition on the spinor there. A spinor on $\mathbb{C}$ can only have charge $\pm\tfrac{1}{2}$ under the action of azimuthal rotations \cite{BenettiGenolini:2023ndb}, and it is precisely this sign that we are interested in. To find this sign, we can study each of the individual Killing vectors on their four-dimensional fixed point sets. 
It is natural to work in an orthonormal frame which is invariant under $\xi$; however, the obvious frame for the metric is not invariant. Instead one finds that on each of the fixed points sets of the three Killing vectors one has the projection conditions:
\begin{equation}
\begin{split}
     \gamma_{12}\epsilon=\ii \sigma^{(1)}\epsilon& \qquad \text{for}\quad \partial_{\tau}\, ,\\
  (\cos\theta \gamma_3-\sin\theta \gamma_4)\gamma_5\epsilon=\ii \sigma^{(2)}\epsilon & \qquad \text{for}\quad   \partial_{\phi_1}\, ,\\
    - (\sin\theta \gamma_3 +\cos\theta \gamma_4)\gamma_6 \epsilon=\ii \sigma^{(3)}\epsilon &\qquad \text{for}\quad \partial_{\phi_2}\, ,
\end{split}
\end{equation}
where it is understood that this holds only on the fixed point set. For the solution one finds that the $\sigma^{(i)}$'s are precisely the signs in $ \chi \xi$, see equation \eqref{eq:xiH4} and table \ref{tab:sigmas}, and therefore we have derived \eqref{eq:xiH4fixed}.
 \begin{table}[H]
    \centering
    \begin{tabular}{@{}|c|c|c|c|c|@{}}
    \hline
        & $\kappa_1=1$ & $\kappa_2=1$ & $\kappa_3=1$ & $\kappa_4=1$ \\ \hline
        $\sigmaone$       &   $-1$      &  $-1$  &  $-1$   &    $-1$  \\ \hline
        $\sigmatwo$       &   $-1$      &   1  &   1   &    $-1$   \\ \hline
        $\sigmathree$     &    $-1$     &  1   &  $-1$   &     1  \\ \hline
        $\chi=\sigmaone\sigmatwo\sigmathree$ & $-1$ & $-1$ &  1 &  1\\ \hline
    \end{tabular}%
    \caption{The choices of projection condition for each of the four distinct Killing spinors. Here it is understood that when $\kappa_a=1$, for a particular index $a=1,\ldots,4$, the other $\kappa$'s are set to zero.}
    \label{tab:sigmas}
    \end{table}
One can now see why the final result is independent of the choice of Killing spinor with which we construct $\xi$. The combination $\sigma^{(i)}\epsilon^{i}$, which appears in the on-shell action formula, is an invariant for all four configurations. Therefore, substituting in any of these choices into \eqref{main} gives the same result
\begin{equation}
    I=\frac{(2n+1)^3}{27 n^2} F_{S^5}\, .
\end{equation}

Another observable studied in \cite{Alday:2015jsa} for the $1/2$ BPS black hole is the on-shell action of a BPS Wilson loop in the geometry. In \cite{Alday:2014bta} it was shown that the BPS Wilson loop wraps a cigar shape constructed by taking the R-symmetry circle at a fixed point to the asymptotic boundary. The string action to compute is 
\begin{equation}
    S_{\text{string}}=\int_{\Sigma_2} X^{-2}K_1 \wedge K_2 + \ii B \, .
\end{equation}
A boundary counterterm must be added to regularize the divergence arising from the infinite boundary length of the loop. However, this counterterm cancels against the boundary contribution of the integral, allowing us to disregard it. 

While the integrand of the string action above is not closed on $M_6$,  it becomes closed when restricted to a fixed point set. We may therefore equivariantly complete it and one finds the polyform
\begin{equation}
    \Phi^{\mathrm{WL}} = X^{-2}K_1 \wedge K_2 + \ii B - \frac{3}{\sqrt{2}}XS\,.
\end{equation}
One can now compute the string action for the BPS Wilson loop using the BV--AB formula and the polyform. We have
\begin{equation}
    S_{\text{string}}=- \frac{3 \pi \chi (\sigmaone \epsilon_1 + \sigmatwo \epsilon_2 + \sigmathree \epsilon_3)}{\epsilon_i}\,,
\end{equation}
and for the choice of R-symmetry vector field we find
\begin{equation}
     S_{\text{string}} =-\frac{3\pi (1+2n) }{n\epsilon_i}\, ,
\end{equation}
which agrees with the result in \cite{Alday:2014fsa}.

\section{A compendium of localization integrals}

In this final appendix, we gather various localization results for the ease of the reader, presenting the localized integrals for various observables. Since the integrals presented here are either four- or two-dimensional subspaces,  it is necessary to break the results into distinct cases depending on the type of fixed point set in $D=6$.  

We can apply the BV--AB formula to the polyform \eqref{PhiF} when there is a toric action on a two-dimensional submanifold of $M_6$ to obtain a formula for the R-symmetry gauge field flux. Denoting this submanifold $M_2$, and letting the weight of the toric action on a fixed point in $M_2$ be $\epsilon_1$, we have that 
\begin{equation}\label{mformulatoric}
    m = \frac{1}{2\pi}\int_{M_2}\Phi^{F}= - \sum_{\text{dim 0}}\frac{1}{\epsilon_1}(\sigmaone \epsilon_1 + \sigmatwo \epsilon_2 + \sigmathree \epsilon_3).
\end{equation}

Next, we want to apply \eqref{BV-AB} with the forms \eqref{eq:Phi*B} and \eqref{eq:Phi*F}. In the following, we are considering $\R^2 \rightarrow M_4$, where here the projection condition associated with $\R^2$ is $\sigmathree$ and the weights of the toric action on $M_4$ are $\epsilon_1$ and $\epsilon_2$.

First, the $\Phi^{*F}$ integral is
\begin{equation}\label{electriccharge} 
\begin{split}
    \int_{M_4}\Phi^{*F}=&~ 6\pi^2 \sum_{\text{dim } 0}\frac{\sigmaone \sigmatwo \sigmathree}{\epsilon_1 \epsilon_2}(\sigmaone \epsilon_1 + \sigmatwo \epsilon_2 + \sigmathree \epsilon_3 )^2 \\
    + &\sum_{\text{dim }2}\frac{2 \pi}{\epsilon_1}\Bigg \{-6\pi \sigmaone \sigmatwo \sigmathree (\sigmaone \epsilon_1 + \sigmatwo \epsilon_2 )\Bigg[m \\ 
    & + \frac{\sigmaone \epsilon_1 + \sigmatwo \epsilon_2}{2\epsilon_1}\int_{\Sigma_g} c_1 (L_1) \Bigg] 
     +2 \int_{\Sigma_g} C  \Bigg\} \, .
\end{split}
\end{equation}
Similarly, the integral of $\Phi^{*B}$ is
\begin{equation}\label{Bcharge}
    \begin{split}
        \int_{M_4}\Phi^{*B}= -\frac{9 \pi^2  \ii}{2} &\left \{\sum_{\text{dim }0} \frac{(\sigmaone \epsilon_1 + \sigmatwo \epsilon_2 + \sigmathree \epsilon_3 )^2}{\epsilon_1 \epsilon_2} \right .\\
        +&\left . \sum_{\text{dim 2}}\frac{\sigmaone \epsilon_1 + \sigmatwo \epsilon_2}{\epsilon_1}\left (2m +\frac{\sigmaone \epsilon_1 + \sigmatwo \epsilon_2}{\epsilon_1}\int_{\Sigma_g} c_1 (L_1 )  \right ) \right \} \, .
    \end{split}
\end{equation}
In both of these, $L_1$ is the normal bundle to $\Sigma_g$ inside $M_4$. Thus, for the example $S^2 \times \Sigma_g$, this normal bundle is a trivial bundle even if we are considering $\R^2 \rightarrow S^2 \times \Sigma_g$. For this special case, we obtain
\begin{equation} 
\begin{split}
    \int_{M_4}\Phi^{*F}= 6\pi^2 &\sum_{\text{dim } 0}\frac{\sigmaone \sigmatwo \sigmathree}{\epsilon_1 \epsilon_2}(\sigmaone \epsilon_1 + \sigmatwo \epsilon_2 + \sigmathree \epsilon_3 )^2 \\
    - &\sum_{\text{dim }2}\frac{4 \pi}{\epsilon_1}\left \{-3\pi \sigmaone \sigmatwo \sigmathree (\sigmaone \epsilon_1 + \sigmatwo \epsilon_2 )m    + \int_{\Sigma_g} C  \right \} \, ,
\end{split}
\end{equation}
and
\begin{equation}
        \int_{M_4}\Phi^{*B}= -\frac{9 \pi^2  \ii}{2} \left [\sum_{\text{dim }0} \frac{(\sigmaone \epsilon_1 + \sigmatwo \epsilon_2 + \sigmathree \epsilon_3 )^2}{\epsilon_1 \epsilon_2} \right .- \left . \sum_{\text{dim 2}}\frac{\sigmaone \epsilon_1 + \sigmatwo \epsilon_2}{\epsilon_1}2m   \right ] \, .
\end{equation}
In the special case that $M_6 = L_1\rightarrow B_4$, the $B$-charge can be calculated directly by noticing that $\Phi^{*B}|_{\mathscr{F}}=-\tfrac{9\ii}{8}F\wedge F$, and thus we have that
\begin{equation}\label{BchargeB4}
    \frac{1}{(2\pi)^2}\int_{B_4}\Phi^{*B}= -\frac{9\ii}{8}\left [2 \chi(B_4) + 3\tau(B_4)+\int_{B_4}c_1(L_1)(c_1(L_1)\mp 2\sigmaone c_1 (K_{B_4})) \right ].
\end{equation}

\bibliographystyle{JHEP} 
\bibliography{biblio}{}

\end{document}